\documentclass[twoside,11pt]{article}

\usepackage{amsmath}
\usepackage{amsthm}

\usepackage[colorlinks=true, linkcolor=blue, hypertexnames=false]{hyperref} 

\usepackage{jmlr2e}
\usepackage{concavedag-jmlr}
\usepackage{booktabs}   
\usepackage{algorithm}  
\usepackage[flushleft, online]{threeparttable} 

\def\phi{\varphi}

\def\pl{p_\lambda}
\def\pln{p_{\lambda_n}}

\def\o{\omega}

\renewcommand{\b}{\beta}

\newcommand{\hB}{\widehat{B}}
\newcommand{\hO}{\widehat{\Omega}}
\newcommand{\hP}{\widehat{\Phi}}
\newcommand{\hR}{\widehat{R}}
\newcommand{\hT}{\widehat{\Theta}}

\def\bn{\boldsymbol\nu}

\renewcommand{\vec}{{\rm vec}}

\ShortHeadings{Concave Penalized Estimation of BNs}{Aragam and Zhou}
\firstpageno{1}

\begin{document}

\title{Concave Penalized Estimation of Sparse Gaussian \\Bayesian Networks}

\author{\name Bryon Aragam \email bryon.aragam@gmail.com\\
       \name Qing Zhou \email zhou@stat.ucla.edu \\
       \addr Department of Statistics\\
       University of California, Los Angeles\\
       Los Angeles, CA 90024, USA}

\editor{[blank]}

\maketitle

\begin{abstract}
\!\!We develop a penalized likelihood estimation framework to estimate the structure of Gaussian Bayesian networks from observational data. In contrast to recent methods which accelerate the learning problem by restricting the search space, our main contribution is a fast algorithm for score-based structure learning which does not restrict the search space in any way and works on high-dimensional datasets with thousands of variables. Our use of concave regularization, as opposed to the more popular $\ell_0$ (e.g. BIC) penalty, is new. Moreover, we provide theoretical guarantees which generalize existing asymptotic results when the underlying distribution is Gaussian. Most notably, our framework does not require the existence of a so-called faithful DAG representation, and as a result the theory must handle the inherent nonidentifiability of the estimation problem in a novel way. Finally, as a matter of independent interest, we provide a comprehensive comparison of our approach to several standard structure learning methods using open-source packages developed for the \texttt{R} language. Based on these experiments, we show that our algorithm is significantly faster than other competing methods while obtaining higher sensitivity with comparable false discovery rates for high-dimensional data. In particular, the total runtime for our method to generate a solution path of 20 estimates for DAGs with 8000 nodes is around one hour.
\end{abstract}

\begin{keywords}
Bayesian networks, concave penalization, directed acyclic graphs, coordinate descent, nonconvex optimization
\end{keywords}


\section{Introduction}
\label{sec:intro}

The problem of estimating Bayesian networks (BNs) has received a significant amount of attention over the past decade, with applications ranging from medicine and genetics to expert systems and artificial intelligence. The idea of using directed graphical models such as Bayesian networks to model real-world phenomena is certainly nothing new, and while the calculus of these models has been very well-developed, the development of fast algorithms to accurately estimate these models in high-dimensions has been slow. The basic problem can be formulated as follows: Given observations from a probability distribution, is it possible to construct a directed acyclic graph (DAG) which decomposes the distribution into a sparse Bayesian network?

Based on observational data alone, it is well-known that there are many Bayesian networks that are consistent in the Markov sense with a given distribution. What we are interested in is finding the sparsest possible Bayesian network, estimated purely from i.i.d. observations without any experimental data. When the number of variables is small, there are many practical algorithms for solving this problem. Unfortunately, as the number of variables increases, this problem becomes notoriously difficult: the learning problem is nonconvex, NP-hard, and scales super-exponentially with the number of variables (\cite{chickering1996,chickering2002,robinson1977}). Since many realistic networks can have upwards of thousands or even tens of thousands of nodes---genetic networks being a prominent example of great importance---the development of new statistical methods for learning the structure of Bayesian networks is critical. 

In this work, we use a penalized likelihood estimation framework to estimate the structure of Gaussian Bayesian networks from observational data. Our framework is based on recent work by \cite{fu2013} and \cite{geer2013}, who show how these ideas lead to a family of estimators with good theoretical properties and whose estimation performance is competitive with traditional approaches. Neither of these works, however, consider the computational challenges associated with high-dimensional datasets whose dimension scales to thousands of variables, which is a key challenge in Bayesian network learning. With these computational challenges in mind, we sought to develop a score-based method that:
\begin{itemize}
\item Does not restrict or prune the search space in any way;
\item Does not assume faithfulness;
\item Does not require a known variable ordering;
\item Works on observational data (i.e. without experimental interventions);
\item Works effectively in high dimensions ($p\gg n$);
\item Is capable of handling graphs with several thousand variables.
\end{itemize}

\noindent
While various methods in the literature cover a few of these requirements, none that we are aware of simultaneously cover \emph{all} of them. The main contribution of the present work is a fast algorithm for score-based structure learning that accomplishes precisely that.

One of the key developments in our method is the application of modern regularization techniques, including both $\ell_1$ and concave penalties. Although $\ell_1$ regularization is well-understood with attractive high-dimensional and computational properties (\cite{buhlmann2011}), as we shall see, in the context of Bayesian networks many of these advantages disappear. While our approach still allows for $\ell_1$-based penalties in practice, our results will indicate that concave penalties such as the SCAD (\cite{fan2001}) and MCP (\cite{zhang2010}) offer improved performance. This is in line with recent advances in sparse learning that have highlighted the advantages of nonconvex regularization in linear and generalized linear models (\cite{lv2009,fan2010,fan2011,zhang2012,huang2012,fan2013}). Notwithstanding, both our theory and our method apply to a very general class of penalties, which can be chosen based on the application at hand.

In this light, our method also represents a major conceptual departure from existing methods in the literature on Bayesian networks through its deep involvement of recent developments in sparse regularization methods, as well as using parametric modeling via structural equations as its foundation (vs. graph theory and Markov equivalence). These techniques have long been known to be useful in regression modeling, covariance estimation, matrix factorization, and image processing, but their application to Bayesian networks, as far as we can tell, is a recent development (\cite{schmidt2007,xiang2013,fu2013,fu2014}). Finally, our method offers new insights into accelerating score-based algorithms in order to compete with hybrid and constraint-based methods which, as we will show, are generally faster and more effective. 

The organization of the rest of this paper is as follows: In the remainder of this section we review previous work and compare our contributions with the existing literature. In Section~\ref{sec:prelim}, we establish the necessary preliminaries for our approach via structural equations. In Section~\ref{sec:framework} we define and discuss the penalized estimator that is the focus of this paper. Section~\ref{sec:thy} then provides the necessary finite-dimensional theory to justify the use of our estimator. After describing this theory and establishing the necessary background material, we review recent developments towards a high-dimensional theory for score-based structure learning in Section~\ref{subsec:hdthy}. A complete description of our algorithm is outlined in Section~\ref{sec:comp}, followed by an empirical evaluation of the algorithm in Section~\ref{sec:results}. Section~\ref{sec:results} also offers a side-by-side comparison of our algorithm with four other structure learning algorithms, and Section~\ref{sec:app} provides an evaluation of these algorithms using a real-world dataset. We finally conclude with a discussion of some future directions for this research.

\subsection{Related work}
\label{subsec:related}
The idea of using sparse regularization to learn Gaussian Bayesian networks in high dimensions is a recent development, and the theoretical basis for $\ell_0$ penalization has been instigated by \cite{geer2013}. Their work relies on the interpretation of Gaussian Bayesian networks in terms of structural equation models (\cite{drton2008,drton2011}), which provides a natural interpretation of network edges in terms of coefficients of a regression model. To the best of our knowledge, the work of \cite{geer2013} is the first high-dimensional analysis of a score-based approach in the literature, and has not yet been generalized to the case of continuous $\ell_1$ or concave penalties yet. As the nontrivial and novel nature of this analysis would detract from our primary goal of addressing computational challenges, we will not pursue a corresponding high-dimensional theory here. Given this foundational work, our purpose here is to show that these ideas can be translated into a family of fast algorithms for score-based learning of Bayesian network structures.

While the traditional approach to estimating Bayesian networks uses $\ell_0$-based penalties such as BIC, \cite{fu2013} recently introduced the idea of using continuous penalties via the adaptive $\ell_1$ penalty and showed that it can be very competitive in practice. They combine a novel method of enforcing acyclicity with a block coordinate descent algorithm in order to compute an $\ell_1$-penalized maximum likelihood estimator for structure learning. Their algorithm is adapted to the case of intervention data, and does not exploit the underlying convexity of the Gaussian likelihood function. As a result, it is limited to graphs with 200 or so nodes and cannot be used on high-dimensional data. The method proposed here is essentially an adaptation of this method for use with observational, high-dimensional data, and takes explicit advantage of convexity and sparsity. We also extend these ideas to a general class of penalties which includes both $\ell_0$ and $\ell_1$ regularization as special cases. The result is an algorithm which easily handles thousands of nodes in a matter of minutes. Moreover, in contrast to the theory proposed in \cite{fu2013}, our theory does not rely on faithfulness or identifiability. 

\subsection{Review of structure learning}
\label{subsec:review}
Traditionally, there are three main approaches to learning Gaussian Bayesian networks. 

\vspace{0.5em}\noindent
\emph{Scored-based.} In the score-based approach, a scoring function is defined over the space of DAG structures, and one searches this space for a structure that optimizes the chosen scoring function. The most commonly used scoring functions are based on the a posteriori probability of a network structure (\cite{geiger2013}), while others use minimum-description length, which is equivalent to the Bayesian information criterion (\cite{lam1994}). In terms of implementation, the standard algorithmic approach is greedy hill-climbing (\cite{heckerman1995}), for which various improvements have been offered over the years (e.g. \cite{chickering2003}). Monte Carlo methods have also been used to sample network structures according to an a posteriori distribution (\cite{ellis2008,zhou2011}).

\vspace{0.5em}\noindent
\emph{Constraint-based.} In the constraint-based approach, repeated conditional independence tests are used to check for the existence of edges between nodes. The idea is to search for statistical independence between variables, which indicates that an edge cannot exist in the underlying DAG structure as long as certain assumptions are satisfied. These assumptions tend to be very strong in practice, and this constitutes the main drawback of this approach. Conversely, since the tests of independence can be very efficient, constraint-based approaches tend to be faster than score-based approaches. Two popular approaches in this spirit are the PC algorithm (\cite{spirtes1991,kalisch2007}) and the MMPC algorithm (\cite{tsamardinos2006}).

\vspace{0.5em}\noindent
\emph{Hybrid.} In the hybrid approach, constraint-based search is used to prune the search space (e.g. to find the skeleton or a moral graph representation), which is then used as an input to restrict a score-based search. By removing as many edges as possible in the first step, the second step can be significantly faster than unrestricted score-based searching. This technique has been shown to work well in practice by combining the advantages of the traditional approaches (\cite{tsamardinos2006,gamez2011,gamez2012}).

\vspace{0.5em}
As previously noted, the main issue with modern approaches to structure learning is scaling algorithms to datasets of ever-increasing sizes. \cite{tsamardinos2006} show how their hybrid MMHC algorithm scales to 5,000 variables, although the running time of 13 days left much to be desired. By assuming the underlying DAG is sparse, \cite{kalisch2007} show how exploiting sparsity in the PC algorithm leads to significant computational gains. More recently, \cite{gamez2012} have proposed modifications to hybrid hill-climbing that scale to 1000 or so variables. By taking advantage of distributed computation, \cite{scutari2014} shows how to scale constraint-based approaches to thousands of variables. Notably, none of these methods fall into the first category of score-based methods. In contrast, the method proposed in the present work is a genuine score-based method and scales efficiently to graphs with thousands of variables. To the best of our knowledge this is one of the first purely score-based methods that accomplishes this in the sense that we rely neither on significance tests (as in the constraint-based approach) nor pruning the search space (as in the hybrid approach).


\section{Preliminaries}
\label{sec:prelim}

We will develop our framework by using a multivariate Gaussian distribution as our starting point, which we will then decompose into a Bayesian network in order to define our estimator. Our approach is purely algebraic, relying on the uniqueness of the Cholesky decomposition in order to factorize a Gaussian distribution into a set of linear structural equations. In what follows, the reader may recall that the structure of a Bayesian network is completely determined by a directed acyclic graph, and hence learning the structure of a Bayesian network reduces to learning directed acyclic graphs. In order to maintain consistency and ease of translation, much of our notation is adapted from \cite{geer2013}.

\subsection{Background and notation}
We assume throughout that the data are generated from a $p$-variate Gaussian distribution,
\begin{align}\label{eq:model}
(X_1,\ldots,X_p)\sim \mathcal{N}(0,\Sigma_0),
\end{align}

\noindent
where the covariance matrix $\Sigma_0\in\R^{p\times p}$ is positive definite. Such a model can always be written as a set of Gaussian structural equations as follows (see \cite{dempster1969}):
\begin{align}\label{eq:streqn}
X_j 
&= \sum_{i=1}^p \beta_{ij}^0 X_i + \eps_j, \quad j=1,\ldots,p,
\end{align}

\noindent
where the $\eps_j$ are mutually independent with $\eps_j\sim \mathcal{N}(0,(\o^0_j)^2)$, $\eps_j$ is independent of $\Pi_j^0=\{X_i:\beta^0_{ij}\ne0\}$, and $\b^0_{jj}=0$. This decomposition is not unique, and we will let $B_0=(\beta_{ij}^0)$ denote any matrix of coefficients that satisfies \eqref{eq:streqn}. The matrix $B_0=(\b_{ij}^0)$ can then be regarded as the weighted adjacency matrix of a directed acyclic graph and represents a Bayesian network for the distribution $\mathcal{N}(0,\Sigma_0)$. Recall that a \emph{directed acyclic graph} $B$ is a directed graph containing no directed cycles. In a slight abuse of notation, we will identify a DAG $B$ with its weighted adjacency matrix, which we will also denote by $B = (\beta_{ij})$.

The nodes of $B$ are in one-to-one correspondence with the random variables $X_1,\ldots,X_p$ in our model. Following tradition, we make no distinction between random variables and nodes or vertices, and will use these terms interchangeably. We say that $X_k$ is a \emph{parent} of $X_j$ if $X_k\to X_j$, and the set of parents of $X_j$ will be denoted by $\Pi_j:=\Pi_j(B)$. We will denote the number of edges in $B$ by $s_B:=|\{\beta_{ij}\ne 0\}|$. When the underlying graph is clear from context, we will suppress the dependence on $B$ and simply denote the number of edges by $s$. For a more thorough introduction to graphical modeling concepts, see \cite{lauritzen1996}.

Unless otherwise noted, $\norm{\cdot}$ shall always mean the standard Euclidean norm and $\norm{\cdot}_F$ denotes the standard $\ell_2$ Frobenius norm on matrices.
For a general matrix $A=(a_{ij})_{n\times p}\in\R^{n\times p}$, its columns will be denoted using lowercase and single subscripts, so that 
\begin{align*}
A = [a_1\|\cdots\|a_p], \quad a_i\in\R^n \text{ for $i=1,\ldots,p$.}
\end{align*}

\noindent
The square brackets signal that $A$ is a matrix with $p$ columns given by $a_1,\ldots,a_p$. In particular, we will write $B = [\b_1\|\cdots\|\b_p]$ for an arbitrary DAG. The support of a matrix is defined by $\supp(B):=\{(i,j):\beta_{ij}\ne0\}$.

If $X=[x_1\|\cdots\|x_p]$ is an $n\times p$ data matrix of i.i.d. observations from \eqref{eq:model}, then we can rewrite \eqref{eq:streqn} as a matrix equation,
\begin{align}\label{eq:mtxmodel}
X = XB_0 + E,
\end{align}

\noindent
where $E\in\R^{n\times p}$ is the matrix of noise vectors. This model has $p(p-1)+p=p^2$ free parameters, which we encode in the two matrices $(B_0,\Omega_0)$. Here, $\Omega_0=\diag((\o^0_1)^2,\ldots,(\o^0_p)^2)$ is the matrix of error variances. We denote the matrix of error variances by $\Omega$ in order to avoid confusion with $\Sigma$, the covariance matrix of $X$.

There are thus two sets of unknown parameters in \eqref{eq:streqn}:
\begin{align*}
B &:= (\b_{ij}) \in\R^{p\times p},\\
\Omega &:= \diag(\o_1^2,\ldots,\o_p^2) \in\R^{p\times p}.
\end{align*}

\noindent
Given $n$ i.i.d. observations of the variables $(X_1,\ldots,X_p)$, the negative log-likelihood of the data $X\in\R^{n\times p}$ is easily seen to be
\begin{align}
\label{eq:loglik}
L(B, \Omega\,|\,X) 
= \sum_{j=1}^p \left[\frac{n}{2}\log(\o_j^2) + \frac{1}{2\o_j^2}\norm{x_{j}-X\beta_j}^2 \right].
\end{align}

\noindent
Observe that the function in \eqref{eq:loglik} is nonconvex; this fact will play an important role in the development of our method.

\begin{remark}
The vast majority of the literature on Bayesian networks focuses on discrete data, in contrast to our method which assumes the data are Gaussian. As the motivation for this work is to scale penalized likelihood methods for high-dimensional data, the Gaussian case is a natural starting point, as much of the high-dimensional statistical theory is tailored towards this case. Recent work has shown how to adapt our techniques to the discrete case via multi-logit regession (\cite{fu2014}). Further generalizations to more general continuous distributions remain for future work. Finally, even though our method implicitly assumes the data are Gaussian, one may naively use our algorithm on discrete data and still obtain reasonable results (see Section~\ref{sec:app}).
\end{remark}

Thus far we have viewed the distribution $\mathcal{N}(0,\Sigma_0)$ as the data-generating mechanism, rewriting this in terms of $(B_0,\Omega_0)$ by using well-known properties of the Gaussian distribution. We could just as well have gone the other way around: Given a DAG $B$ and variance matrix $\Omega=\diag(\o_1^2,\ldots,\o_p^2)$, the parameters $(B,\Omega)$ uniquely define a structural equation model as in \eqref{eq:streqn}, and this model defines a $\mathcal{N}(0,\Sigma)$ distribution. By \eqref{eq:mtxmodel}, we have for any $(B,\Omega)$, 
\begin{align}
\Sigma = (I-B)^{-T}\Omega(I-B)^{-1},
\end{align}

\noindent
and hence $\Sigma$ is uniquely determined by $(B,\Omega)$. Considering instead the inverse covariance matrix $\Theta=\Sigma^{-1}$, we can define
\begin{align}
\label{eq:invcov}
\Theta=\Theta(B,\Omega)=(I-B)\Omega^{-1}(I-B)^T. 
\end{align}

\noindent
By using \eqref{eq:invcov} and defining $S_n:=X^TX$, the negative log-likelihood in \eqref{eq:loglik} can be rewritten in terms of $\Theta=\Theta(B,\Omega)$ directly as
\begin{align}
\label{eq:invcovform}
L(\Theta\,|\,X) 
= -\frac{n}{2}\log\det\Theta + \frac{1}{2}\tr(\Theta S_n).
\end{align}

\noindent
By combining \eqref{eq:loglik} and \eqref{eq:invcovform}, we have $L(B,\Omega\|X)=L(\Theta(B,\Omega)\|X)$. This expression shows how the weighted adjacency matrix of a DAG can be considered as a reparametrization of the usual normal distribution, and gives us an explicit connection between inverse covariance estimation and DAG estimation, which will be explored further in the next subsection.

Since the decomposition of a normal distribution as a linear structural equation model (SEM) as in \eqref{eq:streqn} is not unique, we can define the following equivalence class of DAGs:
\begin{align}
\label{eq:eqclass}
\mathcal{E}(\Theta)
:=\left\{(B,\Omega):\Theta(B,\Omega) = \Theta\right\}.
\end{align}

\noindent
When $(B,\Omega)\in\mathcal{E}(\Theta)$, we shall say that $B$ \emph{represents},  or \emph{is consistent with}, $\Theta$. Two DAGs $(B,\Omega),(B',\Omega')$ will be called \emph{equivalent} if they belong to the same equivalence class $\mathcal{E}(\Theta)$.

This definition of equivalence in terms of equivalent parametrizations is indeed different from the usual definition of \emph{distributional} or \emph{Markov equivalence} that is common in the Bayesian network literature. Furthermore, while it is commonplace to assume that the true underlying distribution is faithful to the DAG $B_0$---which roughly speaking entails that $B_0$ contains exactly the same conditional independence constraints as the true distribution---we have deliberately sidestepped considerations of this hypothesis since our theory does not rely on faithfulness.

\begin{remark}
Strictly speaking, a DAG that fully encodes a Gaussian Bayesian network is specified by \emph{both} a weighted adjacency matrix $B$ and a variance matrix $\Omega$, however, we will frequently refer to a DAG simply by its adjacency matrix $B$. When there is any ambiguity one may assume that there is an assumed variance matrix $\Omega$ paired with $B$, although it may not be explicitly mentioned.
\end{remark}

\subsection{Comparison of graphical models}
\label{subsec:graphmodels}
The previous section showed how the weighted adjacency matrix of a DAG can be considered as a reparametrization of the usual normal distribution, and gave an explicit connection between inverse covariance estimation and DAG estimation: Equation \eqref{eq:invcov} shows how any DAG $(B,\Omega)$ uniquely defines an inverse covariance matrix $\Theta=\Theta(B,\Omega)$. It follows that any estimate $(\hB,\hO)$ of the true DAG yields an estimate of $\Theta_0$ given by $\widehat{\Theta}:=\Theta(\hB,\hO)$. In the context of the PC algorithm, this has been studied by \cite{rutimann2009}. As a result, one may also view our framework as defining an estimator for the inverse covariance matrix. Covariance selection and precision matrix estimation have a long history in the statistical literature (\cite{dempster1972}), with recent approaches employing regularization in various incarnations (e.g. \cite{meinshausen2006,chaudhuri2007,banerjee2008,friedman2008, ravikumar2011}). A detailed survey of recent progress in this area can be found in \cite{pourahmadi2013}. We will not pursue this connection in detail here, however, a few comments are in order.

First, while these two problems are deeply connected, estimating an inverse covariance matrix is significantly easier: the estimation problem is statistically identifiable and the parameter space is convex. This stands in stark contrast to the more difficult problem of estimating an underlying DAG, which is known to be simultaneously \emph{nonidentifiable} and \emph{nonconvex}. As a result, while the high-dimensional properties of regularized covariance estimation are well-understood, the high-dimensional properties of DAG estimation have proven much more difficult to ascertain. The only significant results we are aware of are  in \cite{geer2013} and \cite{kalisch2007}.

Second, our approach is also distinct from existing methods that directly regularize Cholesky factors (\cite{huang2006,lam2009}), as they make implicit use of an \emph{a priori} ordering amongst the variables. As such, the consistency theory in \cite{lam2009} for the sparse Cholesky decomposition does not apply directly to our method. Finally, while there are important similarities between Bayesian networks and other undirected models such as Markov random fields and Ising models, our framework has so far only been applied to the former. For applications of Bayesian networks to inferring so-called Markov blankets, see \cite{aliferis2010a,aliferis2010b}.

Part of the justification for our framework is that it produces sparse BNs that yield good fits to the true distribution, which is tantamount to producing good estimates of the inverse covariance matrix $\Theta_0$. This will be established through the theory presented in Section~\ref{sec:thy}, as well as empirically via the simulations discussed in Section~\ref{sec:results}. Because of the significance and popularity of covariance selection methods, it would of course be interesting to compare our estimate of $\Theta_0$ to the methods cited in the above discussion. As our desire is to keep the focus on estimating Bayesian networks, such comparisons are left to future work.

\subsection{Permutations and equivalence}
\label{subsec:perm}
In this section we wish to exhibit the connection between equivalent DAGs as defined in \eqref{eq:eqclass} and the choice of a permutation of the variables. Recall that a \emph{topological sort} of a directed graph is an ordering on the nodes, often denoted by $\prec$, such that the existence of a directed edge $X_k\to X_j$ implies $X_k\prec X_j$ in the ordering. A directed graph has a topological sort if and only if it is acyclic, and in general such a sort need not be unique.

When describing equivalent DAGs, it is easier to interpret an ordering in terms of a permutation of the variables. Let $\mathcal{P}$ denote the collection of all permutations of the indices $\{1,\ldots, p\}$. For an arbitrary matrix $A$ and any $\pi\in\mathcal{P}$, let us denote by $P_\pi A$ the matrix obtained by permuting the rows and columns of $A$ according to $\pi$, so that $(P_\pi A)_{ij}=a_{\pi(i)\pi(j)}$. Then a DAG can be equivalently defined as any graph whose adjacency matrix $B$ admits a permutation $\pi$ such that $P_\pi B$ is strictly triangular. When the order of the nodes in $P_\pi B$ matches a topological sort of $B$, that is if $X_k\prec X_j\implies \pi^{-1}(k) < \pi^{-1}(j)$, then the matrix $P_\pi B$ will be strictly upper triangular. For our purposes, however, it will be easier to use a \emph{lower}-triangularization, which we now describe.

A DAG $B$ will be called \emph{compatible with the permutation $\pi$} if $P_{\pi}B$ is lower-triangular, which is equivalent to saying that $X_k\to X_j$ (i.e. $X_k\prec X_j$) in $B$ implies $\pi^{-1}(k)>\pi^{-1}(j)$. Conversely, $\pi$ will also be called \emph{compatible} with $B$. Such a permutation $\pi$ may be obtained by simply reversing any topological sort for $B$, so that parents come \emph{after} their children. Formally, suppose $X_1\prec X_2\prec\cdots\prec X_p$ is a topological sort of $B$. Then the permutation
\begin{align*}
\pi(i) = p - i + 1,
\quad i=1,\ldots,p,
\end{align*}

\noindent
is compatible with $B$. Our decision to use lower-triangular matrices is for consistency with existing literature (\cite{geer2013}) and to allow a convenient interpretation of the matrix $B$ as the weighted adjacency matrix of a graph. This will also simplify the technical discussion below (e.g. compare equation \eqref{eq:invcov} above with \eqref{eq:chol} below).

Suppose $\Theta_0$ is given and $\pi\in\mathcal{P}$. Then the matrix $P_\pi\Theta_0$ represents the same covariance structure as $\Theta_0$, up to a reordering of the variables. We may use the Cholesky decomposition to write $P_\pi\Theta_0$ uniquely as 
\begin{align}
\label{eq:chol}
P_\pi\Theta_0
&= (I-L)D^{-1}(I-L)^T
= \Theta(L,D),
\end{align}

\noindent
where $L$ is strictly lower triangular and $D$ is diagonal. It follows from Lemma~\ref{lem:cholperm} in the Appendix that $P_\pi\Theta(L,D)=\Theta(P_\pi L, P_\pi D)$ for any $\pi$, so we can rewrite \eqref{eq:chol} as
\begin{align*}
\Theta_0
&= \Theta(P_{\pi^{-1}}L,P_{\pi^{-1}}D).
\end{align*}

\noindent
For each $\pi$, define
\begin{align*}
\tilde{B}_0(\pi) 
&:= P_{\pi^{-1}}L, \\
\tilde{\Omega}_0(\pi) 
&:= P_{\pi^{-1}}D.
\end{align*}

\noindent
By \eqref{eq:invcov}, this gives us the unique decomposition of $\Theta_0$ into a DAG $(\tilde{B}_0(\pi),\tilde{\Omega}_0(\pi))$ that is compatible with the permutation $\pi$. The DAGs $(\tilde{B}_0(\pi),\tilde{\Omega}_0(\pi))$ that are compatible with some permutation $\pi$ define a subset of the equivalence class $\mathcal{E}(\Theta_0)$; it is easy to check that in fact, this subset is the entire equivalence class.

\begin{lemma}
\label{lem:eqclass}
Suppose $\Sigma_0$ is a positive definite covariance matrix and let $\Theta_0:=\Sigma_0^{-1}$. Then
\begin{align*}
\mathcal{E}(\Theta_0)
&=\{(P_{\pi^{-1}}L,P_{\pi^{-1}}D):P_\pi\Theta_0=\Theta(L,D),\,\pi\in\mathcal{P}\} \\
&= \{(\tilde{B}_0(\pi),\tilde{\Omega}_0(\pi)):\emph{$\pi\in\mathcal{P}$}\}.
\end{align*}
\end{lemma}

\noindent
Note that the relationship between DAGs and permutations is not bijective: multiple permutations can lead to the same DAG. For example, the trivial DAG with no edges is compatible with all possible permutations.

The question now arises: which DAG $(\tilde{B}_0(\pi),\tilde{\Omega}_0(\pi))$ do we want to estimate? In the presence of experimental data, one may consider issues of causality, in which case each DAG represents a very different causal structure. In the absence of such data, however, we can make no such distinctions. All of the DAGs in $\mathcal{E}(\Theta_0)$ are statistically indistinguishable based on observational data alone, and the only tool at our disposal to distinguish them is by their sparsity, or number of edges. Thus a natural objective is to estimate the DAG that most parsimoniously represents the parameter $\Theta_0$ in the sense that it has the fewest number of edges. This choice can also be motivated as it represents a so-called \emph{minimal I-map}.

Under this assumption, there is an obvious connection between our approach and the \emph{sparse Cholesky factorization} problem: Given a symmetric, positive definite matrix $A$, find a permutation $\pi$ such that the Cholesky factor of $P_\pi A$ has the fewest number of nonzero entries possible. In the oracle setting in which we know $\Theta_0$, this is exactly the same problem as finding a permutation $\pi$ such that $\tilde{B}_0(\pi)$ has the fewest number of edges. This connection has been studied in much more detail in \cite{raskutti2014}. They show that in this oracle setting, there is an equivalence between $\ell_0$-penalized estimation and sparse Cholesky factorization. In contrast, here we seek to estimate $\Theta_0$ \emph{as well as} find a sparse permutation $\pi$, and in this sense we provide a non-oracular, computationally feasible alternative to searching across all $p!$ permutations when $p$ is very large. 

\begin{example}
Suppose the DAG $B_0$ has the structure $X_1\to X_2\to X_3$ with edge weights $\beta_{12}=1$ and $\beta_{23}=1$, and $\omega_j=1$ for each $j$. In this case, we have
\begin{align*}
B_0 = \begin{pmatrix}
0 & 1 & 0 \\
0 & 0 & 1 \\
0 & 0 & 0
\end{pmatrix},\quad
\Omega_0 = \begin{pmatrix}
1 & 0 & 0 \\
0 & 1 & 0 \\
0 & 0 & 1
\end{pmatrix},\quad
\Theta(B_0,\Omega_0) = \begin{pmatrix}
2 & -1 & 0 \\
-1 & 2 & -1 \\
0 & -1 & 1
\end{pmatrix}.
\end{align*}

\noindent
A topological sort for $B_0$ is $X_1\prec X_2\prec X_3$ (i.e. $B_0$ is already sorted), but $B_0$ is lower triangularized by the permutation $\pi_0=(3,2,1)$ that swaps $X_1$ and $X_3$. Thus $B_0=\tilde{B}_0(\pi_0)$.

Now consider another DAG, defined by
\begin{align*}
B_1 = \begin{pmatrix}
0 & 1/2 & 1 \\
0 & 0 & 0 \\
0 & 1/2 & 0
\end{pmatrix},\quad
\Omega_1 = \begin{pmatrix}
1 & 0 & 0 \\
0 & 1/2 & 0 \\
0 & 0 & 2
\end{pmatrix},\quad
\Theta(B_1,\Omega_1) = \begin{pmatrix}
2 & -1 & 0 \\
-1 & 2 & -1 \\
0 & -1 & 1
\end{pmatrix}.
\end{align*}

\noindent
Since $\Theta(B_1,\Omega_1) = \Theta(B_0,\Omega_0)$, the DAG $(B_1,\Omega_1)$ is equivalent to $(B_0,\Omega_0)$. Thus, according to Lemma~\ref{lem:eqclass}, there must be a permutation $\pi_1$ such that $B_1=\tilde{B}_0(\pi_1)$ and $\Omega_1=\tilde{\Omega}_0(\pi_1)$. Indeed, if we let $\pi_1=(2,3,1)$, one can check (by \eqref{eq:chol}) that these identities hold. Furthermore, if we reverse the order of the variables in $\pi_1$, we obtain a topological sort for $B_1$: $X_1\prec X_3\prec X_2$.

This example highlights two important points: (i) For the reader familiar with Markov equivalence of DAGs, it is obvious that $B_0$ and $B_1$ are not Markov equivalent, so our definition of equivalence is indeed different; and (ii) Equivalent DAGs in the sense we have defined need not have the same number of edges. This is the primary complication our framework must manage: Amongst all the DAGs which are equivalent to the true parameter $\Theta_0$, we wish to find one which has the fewest number of edges.
\end{example}

\subsection{Structural equation modeling}
\label{subsec:sem}
We have chosen to focus on the problem of structure estimation of Bayesian networks, which is not to be confused with the problem of causal inference. We view the data-generation mechanism as a multivariate Gaussian distribution as in \eqref{eq:model}. From this perspective, there are many linear structural equations \eqref{eq:streqn} that may generate \eqref{eq:model}. Our focus is on finding the most parsimonious representation of the true distribution as a set of structural equations.

Alternatively, one could view the structural equation model \eqref{eq:streqn} as the data-generating mechanism, in which case there is a \emph{particular} set of structural equations that we wish to estimate. This is the perspective commonly adopted in the social sciences and in public health, in which the structural equations model causal relationships between the variables. In this set-up, it is well-known that one cannot expect to recover the directionality of causal relationships based on observational data alone, and the issues of causality, confounding and identifiability take center stage. Since we are only considering observational data, our framework does not address these questions.


\section{The Concave Penalization Framework}
\label{sec:framework}
Now that the necessary preliminaries have been discussed, in the remainder of the paper we will develop the estimation framework thus far described at a high-level. Our approach is to use a penalized maximum likelihood estimator to estimate a sparse DAG $B_0$ that represents $\Theta_0$. Recall that the negative log-likelihood is given by $L(B, \Omega\,|\,X)$ in \eqref{eq:loglik}. This will be our loss function, however in order to promote sparsity and avoid overfitting, we will minimize a penalized loss instead. In what follows, let $\fcndef{\pl}{[0,\infty)}{\R}$ be a nonnegative and nondecreasing penalty function that depends on the tuning parameter $\lambda$ and possibly one or more additional shape parameters. Our framework is valid for a very general class of penalties, so in what follows we will allow $\pl(\cdot)$ to be arbitrary. The details of choosing the penalty function will be discussed in Section~\ref{subsec:pen}.

Once $\pl$ is chosen, one may seek to find a solution to 
\begin{align}
\label{eq:estimator}
\argmin_{B,\Omega} \big\{L(B,\Omega \| X) + n\sum_{i,j}\pl(|\b_{ij}|) : \text{$B$ is a DAG}\big\}.
\end{align}

\noindent
When $L$ is taken to be a more general scoring function such as a posterior probability, \eqref{eq:estimator} resembles most familiar score-based methods. When $\pl(\cdot)$ is taken to be the $\ell_0$ penalty, we recover the estimator discussed in \cite{geer2013}. Our approach differs from the aforementioned in two ways:
\begin{enumerate}
\item Our choice of the penalty term $\pl(\cdot)$ is different from traditional approaches and results in a continuous optimization problem,
\item Due to the nonconvexity of the loss function, we reparametrize the problem in order to obtain a convex loss function.
\end{enumerate}

\noindent
Thus, in general our estimator will not be the same as \eqref{eq:estimator}.

\begin{remark}
\label{rem:fixedpi}
If we further constrain the minimization problem in \eqref{eq:estimator} to include only DAGs which are compatible with a fixed topological sort, we can reduce the problem to a series of $p$ individual regression problems. Given a topological sort $\prec$, the parents of $X_j$ must be a subset of the variables that precede $X_j$ in $\prec$. In terms of the permutation $\pi$ described in Section~\ref{subsec:perm}, we require $\Pi^0_j\subset\{X_k:\pi^{-1}(k)>\pi^{-1}(j)\}$. The true neighbourhood of $X_j$ can then be determined by projecting $X_j$ onto this subset of nodes, which can be done via penalized least squares. Consistency in structure learning and parameter estimation can then be established through standard penalized regression theory.
\end{remark}

\subsection{Reparametrization}
\label{subsec:reparam}
One of the drawbacks of the loss in {\eqref{eq:loglik}} is that it is nonconvex, which complicates the minimization of the penalized loss. If we minimize \eqref{eq:loglik} with respect to $\Omega$ and use the adaptive Lasso penalty, we obtain the estimator described in \cite{fu2013}. By keeping the $p$ variance terms, however, we can exploit a clever reparametrization of the problem,  introduced in \cite{stadler2010}, which leads to a convex loss. 

The idea is to define new variables by $\rho_j=1/\o_j$ and $\phi_{ij} = \b_{ij}/\o_j$, which yields the reparametrized negative log-likelihood
\begin{align}
\label{eq:reloglik}
L(\Phi, R \| X)
&= \sum_{j=1}^p \left[-n\log(\rho_j) + \frac{1}{2}\norm{\rho_jx_j-X\phi_j}^2 \right],
\end{align}

\noindent
where $\Phi = [\phi_1 \| \cdots \| \phi_p]$ and $R=\diag(\rho_1,\ldots,\rho_p)$. The loss function in \eqref{eq:reloglik} is easily seen to be convex. Furthermore, if we interpret $\Phi$ as the adjacency matrix of a directed graph, then $\Phi$ has exactly the same edges and nonzero entries as $B$, and thus in particular $\Phi$ is acyclic if and only if $B$ is acyclic. 

In analogy with the parametrization $(B,\Omega)$, define
\begin{align}
\label{eq:retheta}
\Theta(\Phi,R)
&= (R-\Phi)(R-\Phi)^T,
\end{align}

\noindent
which gives a formula for the inverse covariance matrix in the parametrization $(\Phi,R)$. Note that if $\Phi=\Phi(B,\Omega)$ and $R=R(B,\Omega)$, then $\Theta(B,\Omega)=\Theta(\Phi,R)$, and hence also $L(B,\Omega)=L(\Phi,R)$.

This reparametrization is \emph{not} the same as the likelihood in \eqref{eq:invcovform}, which is well-known to lead to a convex program (see, for instance, \sec7.1 in \cite{boyd2009}). Indeed, plugging \eqref{eq:invcov} into \eqref{eq:invcovform} leads back to \eqref{eq:loglik}, which is nonconvex in the parameters $\beta_{ij}$ and $\omega_j$. To wit, the problem is convex in $\Theta$ but not in $(B,\Omega)$. The key insight from \cite{stadler2010} is to observe that one may recover convexity by switching to the alternate parametrization in terms of $\phi_{ij}$ and $\rho_j$. Unfortunately, the DAG constraint in \eqref{eq:estimator} is still nonconvex. The idea behind this reparametrization is to allow our algorithm to exploit convexity wherever possible in order to reap at least \emph{some} computational and analytical gains. As we shall see, the gains are indeed significant.

\subsection{The estimator}
\label{subsec:estimator}
We are now prepared to introduce the formal definition of the DAG estimator which is the focus of this work. 

Fix a penalty function $\pl(\cdot)$. Then given
\begin{align}
\label{eq:reestimator}
(\hP, \hR) 
&:= \argmin_{\Phi,R} \big\{L(\Phi,R \| X) + n\sum_{i,j}\pl(|\phi_{ij}|) : \text{$\Phi$ is a DAG}\big\} \\
&\,\,= \argmin_{\Phi,R} \big\{\sum_{j=1}^p \left[-n\log(\rho_j) + \frac{1}{2}\norm{\rho_jx_j-X\phi_j}^2 \right] \nonumber\\
& \hspace{12em} + n\sum_{i,j}\pl(|\phi_{ij}|) : \text{$\Phi$ is a DAG}\big\}, \nonumber
\intertext{we define our estimator to be}
\label{eq:transform}
(\hB, \hO) &= \begin{cases}
\hat{\b}_{ij}=\hat{\phi}_{ij}/\hat{\rho}_j, & i\ne j \\
\hat{\b}_{jj}=0, & \\
\hat{\o}_j^2 = 1/\hat{\rho}_j^2, & j=1,\ldots,p
\end{cases}
\end{align}

\noindent
where $\hat{\phi}_{ij}$ and $\hat{\rho}_j$ denote the respective components of $(\hP,\hR)$. When we wish to emphasize the estimator's dependence on $\lambda$, we shall denote it by $(\hP(\lambda), \hR(\lambda))$. 

There is an intuitive interpretation of the problem in \eqref{eq:reestimator}: By the identity $L(\Phi, R \| X) = L(\Theta(\Phi, R) \| X)$, it is evident that the loss function for $(\Phi, R)$ is simply the negative log-likelihood of the resulting estimate of $\Theta=\Theta(\Phi, R)$. In this sense, we are implicitly approximating the true parameter $\Theta_0$. The key ingredient, however, is the penalty term: We only penalize the edge weights $\phi_{ij}$, which has the effect of self-selecting for DAGs which are very sparse. In this way, the solution to {\eqref{eq:reestimator}} produces a sparse Bayesian network whose distribution is close to the true, underlying distribution.

\begin{remark}
\label{rem:equiv}
For most choices of the penalty, the solution to {\eqref{eq:reestimator}} is \emph{not} the same as the solution to {\eqref{eq:estimator}} since we are penalizing different terms. In the original parametrization, we penalize the coefficients $\beta_{ij}$, whereas after reparametrizing we are penalizing the rescaled coefficients $\phi_{ij}=\beta_{ij}/\omega_j$. Thus we are also penalizing choices of coefficients which overfit the data, i.e., which have very small $\omega_j$. A notable exception, however, occurs when $\pl(\cdot)$ is taken to be the $\ell_0$ penalty. In this special case, the problems in \eqref{eq:estimator} and \eqref{eq:reestimator} are the same, and thus in particular the analysis in \cite{geer2013} applies.
\end{remark}

\subsection{Choice of penalty function}
\label{subsec:pen}

The standard approach in the Bayesian network literature is to use AIC or BIC to penalize overly complex models, although $\ell_1$-based methods have been slowly gaining in popularity. Traditionally, $\ell_1$ regularization is viewed as a convex relaxation of optimal $\ell_0$ regularization, which results in a convex program that is computationally efficient to solve. Unfortunately, in our situation the constraint that $B$ is a DAG is also nonconvex, so there is little hope to recover a convex program. Thus, there is nothing lost in using concave penalties, which have more attractive theoretical properties than $\ell_1$-based alternatives. We will briefly review the details here. 

\cite{fan2001} introduce the fundamental theory of concave penalized likelihood estimation and outline three principles that should guide any variable selection procedure: unbiasedness, sparsity, and continuity. They argue that the following conditions are sufficient to guarantee that a penalized least squares estimator has these properties:
\begin{enumerate}
\item (Unbiasedness) $\pl'(t) = 0$ for large $t$;
\item (Sparsity) The minimum of $t + \pl'(t)$ is positive;
\item (Continuity) The minimum of $t + \pl'(t)$ is attained at zero.
\end{enumerate}

\noindent
Note that Condition (1) only guarantees unbiasedness for large values of the parameter; in general we cannot expect a penalized procedure to be totally unbiased. Note also that (1-3) imply that $\pl$ must be a concave function of $t$.

\begin{figure}[t]
\centering
\includegraphics[width=4in]{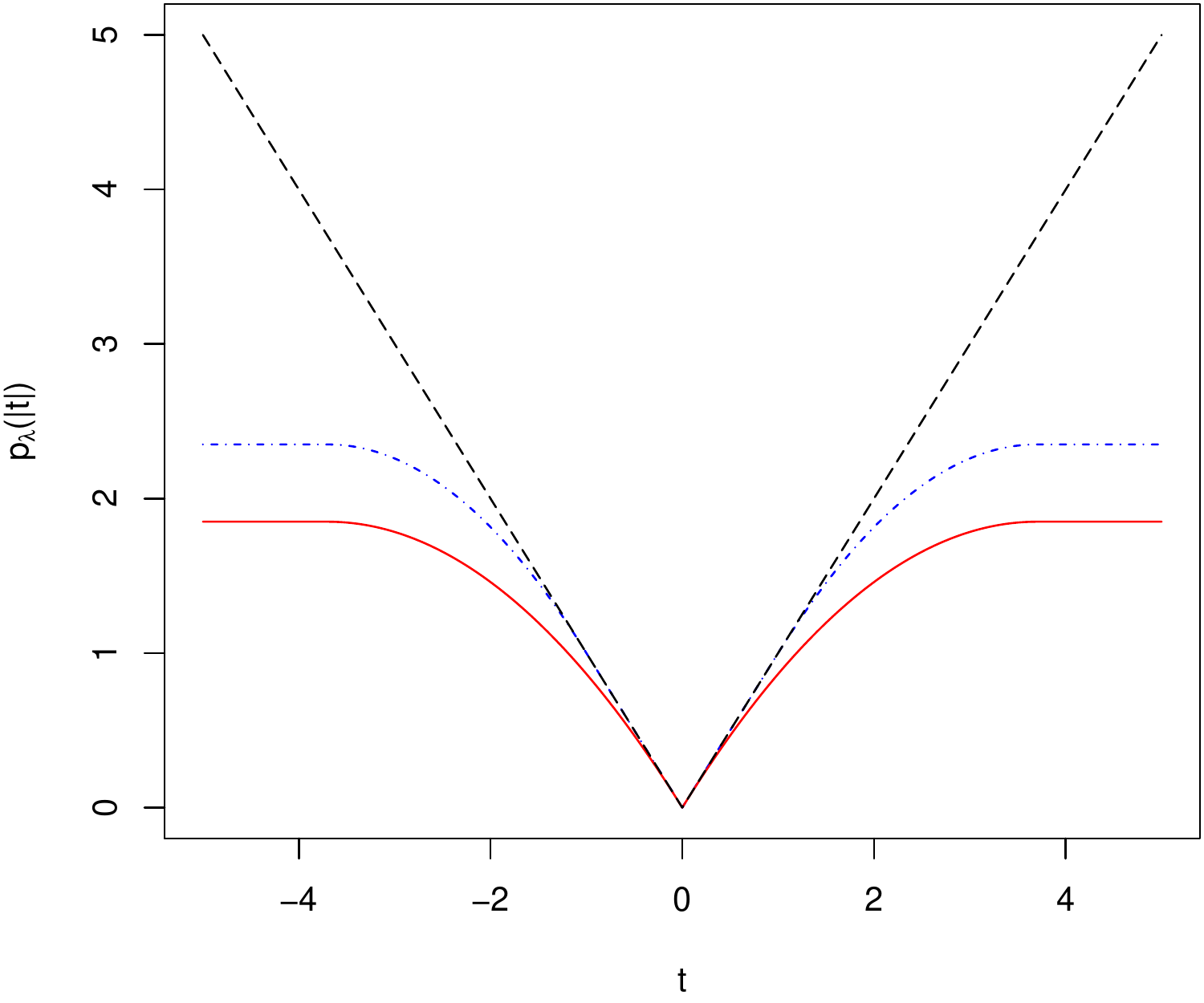}
\caption{Comparison of penalty functions. The red, solid line is the minimax concave penalty (MCP), the blue dot-dashed line is the smoothly clipped absolute deviation penalty (SCAD), and the black dashed line is the $\ell_1$ or Lasso penalty. Both the MCP and SCAD represent smooth interpolations of the $\ell_1$ and $\ell_0$ penalties and hence have better statistical properties, whereas the $\ell_1$ penalty exhibits bias due to its divergence as $t\to\infty$.}
\label{plot:pen}
\end{figure}

In the methodological developments which follow, it will not be necessary to assume that the penalty function is concave. The theory developed in Section~\ref{sec:thy} will illuminate how the properties of the penalty function influence the theoretical properties of the estimator (\ref{eq:reestimator},~\ref{eq:transform}), however, the only strict requirement on the penalty function needed for the proposed algorithm is that there exists a corresponding threshold function $S(\cdot,\lambda)$ to perform the single parameter updates (see Section~\ref{subsec:cd} for details). Examples of common penalty functions in the literature include $\ell_1$ (or Lasso, \cite{tibshirani1996}), SCAD (\cite{fan2001}) and MCP (\cite{zhang2010}). The SCAD penalty represents a smooth quadratic interpolation between the $\ell_1$ and $\ell_0$ penalties, and the MCP translates the linear part of the SCAD to the origin. See Figure~\ref{plot:pen} for a graphical comparison of these three penalties. The key difference between the $\ell_1$ penalty and SCAD or MCP is the flat part of the penalty, which helps to reduce bias.

In our computations we chose to use the MCP, defined for $t\ge 0$ by
\begin{align}
\label{eq:mcp}
\pl(t;\gamma) &:= 
\lambda\left(t-\frac{t^2}{2\lambda\gamma}\right)1(t<\lambda\gamma)
+ \frac{\lambda^2\gamma}{2}1(t\ge \lambda\gamma) \\
\nonumber&= \begin{cases}
\lambda\left(t-\frac{t^2}{2\lambda\gamma}\right), & t<\lambda\gamma, \\
\frac{\lambda^2\gamma}{2}, & t\ge \lambda\gamma.
\end{cases}
\end{align}

\noindent
The $\gamma$ parameter in the MCP controls the concavity of the penalty: As $\gamma\to0$, MCP approaches the $\ell_0$ penalty and as $\gamma\to\infty$, it approaches the $\ell_1$ penalty. In the sequel we will thus refer to $\gamma$ as the \emph{concavity parameter} and $\lambda$ as the \emph{regularization parameter}. From the above formula, MCP is easily seen to be a quadratic spline between the origin and the $\ell_0$ penalty with a knot at $t=\lambda\gamma$. To demonstrate the differences and potential advantages of a concave penalty, we also implemented our method with the $\ell_1$ penalty, $\pl(|t|) = \lambda|t|$.

As the $\ell_1$ penalty does not satisfy the unbiasedness condition (Condition (1) above), it yields biased estimates in general.  Allowing ourselves to be motivated by some recent developments in regression theory, we can say even more. There the assumptions required for consistency are rather strong and require a so-called \emph{irrepresentability condition} (\cite{zhao2006}), also known as \emph{neighbourhood stability} (\cite{meinshausen2006}). The bias issues can be circumvented by employing the adaptive Lasso (\cite{zou2006}), an idea which has been explored in \cite{fu2013}. Recent theoretical analysis of regularization with concave penalties has shown that, compared to $\ell_1$ penalties, the assumptions on the data needed for consistency can be relaxed substantially. Generalizing these ideas to Bayesian network models, we will show in Section~\ref{sec:thy} how our estimator is consistent in both parameter estimation and structure learning when concave regularization is used; with $\ell_1$ regularization we only obtain parameter estimation consistency. These theoretical results are supported by the comparisons in Section~\ref{sec:results}.

\subsection{The role of sparsity}
For a given $\Theta_0$, the equivalence class $\mathcal{E}(\Theta_0)$ will typically consist of graphs with very different numbers of edges, and in general there need not be a sparse representation $(\tilde{B}_0(\pi),\tilde{\Omega}(\pi))$ with $s_{\tilde{B}_0(\pi)}:=\tilde{s}_0(\pi)= O(p)$. Moreover, the asymptotic theory to be developed in Section~\ref{sec:thy} will not require such an assumption. When we evaluate our method in Sections~\ref{sec:comp}-\ref{sec:app}, however, we will focus our attention on the case where there exists a DAG in $\mathcal{E}(\Theta_0)$ which is sparse, that is, satisfying the condition $\tilde{s}_0(\pi)=O(p)$.

Our justification for this assumption is both practical and theoretical. In terms of the true graph, sparsity implies that we expect either (a) only a subset of the variables are truly involved, or (b) on average, each variable has only a few parents. In case (a), estimating a Bayesian network is very similar to the variable screening problem. Both of these scenarios are commonly encountered in practice, as many realistic DAG models tend to be sparse in one of these two senses. Moreover, for datasets with $p$ very large, we typically have fewer observations than variables. In fact, we expect $p\gg n$, with $p$ on the order of thousands or tens of thousands. When this happens, we can only expect to obtain reasonable results when each node has at most $n$ parents, although in practice far fewer than $n$ parents is typical. For these reasons, we chose to tailor our algorithm to the sparse, high-dimensional regime. Along with the nonconvexity of the constraint space, this is the main reason for emphasizing the use of concave penalties, whose superior performance in the $p\gg n$ regime has been already established for regression models. Furthermore, by assuming that the true graph is sparse, we can take advantage of several computational enhancements that allow our algorithm to leverage sparsity for speed. The result is an efficient algorithm when we are confident that the underlying model admits a sparse representation.


\section{Asymptotic Theory}
\label{sec:thy}

In this section we provide theoretical justification for the use of the estimator (\ref{eq:reestimator},~\ref{eq:transform}) in the finite-dimensional regime. That is, we will assume $p$ is fixed and let $n\to\infty$. The purpose of this section is not to provide novel theoretical insights, but rather simply to show that under the right conditions we can always guarantee that the estimator defined in the previous section has good estimation properties. Most importantly, we establish that these conditions can always be satisfied when the MCP is used for regularization.

In the statistics literature, a procedure which attains consistency in structure learning with high probability is sometimes referred to as \emph{model selection consistent}. This can be confusing as model selection is also used to refer to the problem of selecting the tuning parameter $\lambda$. In the sequel, we use the following conventions: (i) A procedure is \emph{structure estimation consistent} if $P(\supp(\widehat{B})=\supp(B_0))\to1$, (ii) A procedure is \emph{parameter estimation consistent} if $\norm{\widehat{B} - B_0}_F\overset{P}{\to}0$, and (iii) \emph{Model selection} will refer only to the problem of choosing $\lambda$.


\subsection{Nonidentifiability and sparsity}
\label{subsec:sketch}
Since our optimization problem is nonconvex, we must be careful when discussing ``solutions'' to \eqref{eq:reestimator}. The estimator is defined to be the global minimum of the penalized loss, but theoretical guarantees are generally only available for local minimizers. Our theory is no exception, and it is furthermore complicated by identifiability issues: Based on observational data alone, the inverse covariance matrix $\Theta_0$ is identifiable, but the DAG $(B_0,\Omega_0)$ is not. The usual theory of maximum likelihood estimation assumes identifiability, but it is possible to derive similar optimality results when the true parameter is nonidentifiable (see for instance \cite{redner1981}). 

When the model is identifiable, one establishes the existence of a consistent local minimizer for the true parameter, which is unique (e.g. \cite{fan2001}). It turns out that even if the model is nonidentifiable, we can still obtain a consistent local minimizer for each equivalent parameter. As long as there are finitely many equivalent parameters, these minimizers are unique to each parameter. In particular, in the context of DAG estimation, there are up to $p!$ equivalent parameters in the equivalence class $\mathcal{E}_0$ (Lemma~\ref{lem:eqclass}). Thus we have a finite collection of local minimizers that serve as ``candidates'' for the global minimum; the question that remains is which one of these minimizers does our estimator produce?

Each equivalent parameter has the same likelihood, so the only quantity we have to distinguish these minimizers is the penalty term. Our theory will show that by properly controlling the amount of regularization, it is possible to distinguish the \emph{sparsest} DAGs in $\mathcal{E}_0$ in the sense that they will each have strictly smaller penalized loss than their competitors. Moreover, this analysis can be transferred over to the \emph{empirical} local minimizers, so that the sparsest local minimizer has the smallest penalized loss. Because of nonconvexity, however, it is hard to guarantee that these minimizers are the \emph{only} local minimizers, and hence that the sparsest DAGs are the global minimizers. The simulations in Section {\ref{sec:results}} give us good empirical evidence that our estimator indeed approximates the sparsest DAG representation of $\Theta_0$, as opposed to another DAG with many more edges.

The remainder of this section undertakes the details of this analysis. To stay consistent with the literature, instead of minimizing the penalized loss \eqref{eq:reestimator} we will maximize the penalized log-likelihood, which is of course only a technical distinction. We begin with a discussion of the technical results and assumptions which establish the existence of consistent local maximizers before stating our main result in Section {\ref{subsec:collect}}. We also briefly discuss the high-dimensional scenario in which $p$ is allowed to depend on $n$.

\begin{remark}
For some classes of models, including nonlinear and non-Gaussian models, the DAG estimation problem considered here is known to be identifiable based on observational data alone (\cite{shimizu2006,peters2012nonlinear}), and some methods have been developed to estimate such models (\cite{hyvarinen2010,anandkumar2013}). Identifiability can also be obtained when the errors are Gaussian with equal variances (\cite{peters2012samevar}). In contrast to these developments, the main technical difficulty in our analysis is the nonidentifiability of the general Gaussian model.
\end{remark}

\subsection{Existence of local maximizers}
In the ensuing theoretical analysis, it will be easier to work with a single parameter vector (vs. the two matrices $\Phi$ and $R$), so we first transform our parameter space in this way without any loss of generality. To the end, define $U:=R+\Phi$ and let $\bn = \vec(U) = \vec(R+\Phi) \in\R^{p^2}$ to be the vectorized copy of $U$ in $\R^{p^2}$. Our parameter space is then the subset $\mathcal{D}$ of $\R^{p^2}$ such that $\bn\in\mathcal{D}$ implies $(\Phi,R)$ is a DAG, where $\bn=\vec(R+\Phi)$. In the sequel, we will refer to such a $\bn$ as a DAG. For a more in-depth treatment of the abstract framework, see Section~\ref{app:formal} in the Appendix.

The true distribution is uniquely defined by its inverse covariance matrix, $\Theta_0$. By equation \eqref{eq:retheta}, given $(\hP,\hR)$ we may consider the resulting estimate of the inverse covariance matrix $\hT=\Theta(\hP,\hR)$. By analogy, for any DAG $\bn\in\R^{p^2}$, we may define in the obvious way the matrix $\Theta(\bn)$. Thus the parameter $\bn$ is simply another parametrization of the normal distribution: For any $\Theta_0$, there exists $\bn\in\mathcal{D}$ such that $\Theta_0 = \Theta(\bn)$. Let $\mathcal{E}_0=\mathcal{E}(\Theta_0)=\{\bn\in\R^{p^2}:\Theta(\bn)=\Theta_0\}$. We will denote an arbitrary element of $\mathcal{E}_0$ by $\bn_0$ and a minimal-edge DAG in $\mathcal{E}_0$ by $\bn^*$.

As is customary, we denote the support set of a vector by $\supp(\bn):=\{j:\nu_j\ne0\}$, and likewise for matrices $\supp(B):=\{(i,j):\beta_{ij}\ne0\}$. Let $\ell_n(\bn\|X)$ be the unpenalized log-likelihood of the parameter vector $\bn$ and define
\begin{align}
\pl(\bn)
=\sum_{i\ne j}\pl(|u_{ij}|),
\end{align}

\noindent
where $u_{ij}$ denote the elements of $U$. Note that we are penalizing only the off-diagonal elements of $U$, which correspond to the elements of $\Phi$. Now let
\begin{align}
\label{eq:F}
F(\bn) 
&:= \ell_n(\bn \| X) - n\,\pln(\bn).
\end{align}

\noindent
We are interested in maximizing $F$ over $\mathcal{D}$.

For any $\bn_0\in\mathcal{E}_0$ which represents a DAG $(\Phi_0,R_0)=((\phi_{ij}^0),(\rho_j^0))$ as described above, define two sequences which depend on the choice of penalty $\pl$:
\begin{align}
a_n(\bn_0)
&:= \max\{|\pln'(|\phi^0_{ij}|)|:\phi^0_{ij}\ne0\}, \\
b_n(\bn_0)
&:= \max\{|\pln''(|\phi^0_{ij}|)|:\phi^0_{ij}\ne0\}.
\end{align}

\noindent
When it is clear from context, the dependence of $a_n$ and $b_n$ on $\bn_0$ will be suppressed. Finally, let $\tau(\lambda):=\sup_t\pl(t)$, which may be infinite. For the MCP we have $\tau(\lambda)=\lambda^2\gamma/2$ and for the $\ell_1$ penalty $\tau(\lambda)=+\infty$.

The following result, which is similar in spirit to Theorem 2 of \cite{fu2013}, guarantees the existence of a consistent local maximizer:
\begin{thm}
\label{thm:main}
Fix $p\ge 1$. If there exists $\bn_0\in\mathcal{E}_0$ with $b_n(\bn_0)\to0$, then there is a local maximizer $\widehat{\bn}_n$ of $F(\bn)$ such that
\begin{align*}
\norm{\widehat{\bn}_n-\bn_0} = O_P(n^{-1/2}+a_n(\bn_0)).
\end{align*}
\end{thm}

\noindent
When $a_n=O(n^{-1/2})$, we obtain a $n^{1/2}$-consistent estimator of $\bn_0$. Note that by Lemma~\ref{lem:eqclass}, if $\bn_0\in\mathcal{E}_0$ then $\bn_0=(\tilde{B}_0(\pi),\tilde{\Omega}_0(\pi))$ for some permutation $\pi$. For this reason, in the sequel we shall refer to the local maximizer $\widehat{\bn}_n$ as the \emph{$\pi$-local maximizer} of $F$ for the permutation $\pi$. This theorem says that as long as the curvature of the penalty at $(\tilde{B}_0(\pi),\tilde{\Omega}_0(\pi))$ tends to zero, the penalized likelihood has a $\pi$-local maximizer that converges to $(\tilde{B}_0(\pi),\tilde{\Omega}_0(\pi))$ as $n\to\infty$.

Under additional assumptions on the penalty function, we may further strengthen this result to include consistency in structure estimation when $p$ remains fixed:
\begin{thm}
\label{thm:sparsity}
Assume that the penalty function satisfies
\begin{align}
\label{eq:liminf}
\liminf_{n\to\infty}\liminf_{t\to0^+} p_{\lambda_n}'(t)/\lambda_n > 0.
\end{align}

\noindent
Assume further that $\bn_0\in\mathcal{E}_0$ satisfies $a_n(\bn_0)=O(n^{-1/2})$, $b_n(\bn_0)\to0$, and let $\widehat{\bn}_n$ be a $\pi$-local maximizer from Theorem~\ref{thm:main}. If $\lambda_n\to0$ and $\lambda_nn^{1/2}\to\infty$, then 
\begin{align}
P(\supp(\widehat{\bn}_n) = \supp(\bn_0))\to1.
\end{align}
\end{thm}

\noindent
In fact, this follows immediately from Theorem~\ref{thm:main} above and Theorem 2 in \cite{fan2001}. An obvious corollary is that $P(\hat{s}_n=s_0)\to 1$. 

We must be careful in interpreting these theorems correctly: They do not imply necessarily that the estimator defined in (\ref{eq:reestimator},~\ref{eq:transform}) is consistent. These theorems simply show that under the right conditions, there is a local maximizer of $F$ that is consistent. It remains to establish that the global maximizer of $F$ is indeed one of these local maximizers. 

\begin{remark}
\label{rem:prob}
If we assume that the conditions of Theorems~\ref{thm:main} and~\ref{thm:sparsity} hold for \emph{all} $\bn_0\in\mathcal{E}_0$, then we can conclude that every equivalent DAG has a $\pi$-local maximizer that selects the correct sparse structure. This is trivial since we assume $p$ to be fixed as $n\to\infty$, which allows us to bound the probabilities over all $p!$ choices of $\bn_0$ simultaneously. Since the number of equivalent DAGs grows super-exponentially as $p$ increases, bounding these probabilities when $p=p_n$ grows with $n$ is the main obstacle to achieving useful results in high-dimensions.
\end{remark}

The proofs of these two theorems are found in the appendix. In the course of the proofs, we will need the following lemma:
\begin{lemma}
\label{lem:unique}
If $B_1\ne B_2$ are DAGs that have a common topological sort, then for any choices of $\Omega_1$ and $\Omega_2$, we have $\Theta(B_1,\Omega_1)\ne\Theta(B_2,\Omega_2)$. A similar result holds in the parametrization $(\Phi,R)$.
\end{lemma}

\noindent
The assumption that two DAGs have a common topological sort is equivalent to each DAG being compatible with the same permutation $\pi$. The following lemma shows that the $\bn_0$ are isolated, which guarantees that $\pi$-local maximizers do not cluster around multiple $\bn_0$. For any $\eps>0$, we denote the $\eps$-neighbourhood of $\bn_0$ in $\mathcal{D}$ by $B(\bn_0,\eps):=\{\bn\in\mathcal{D}:\norm{\bn-\bn_0} < \eps\}$.

\begin{lemma}
\label{lem:sep}
For any positive definite $\Theta_0$ there exists $\eps>0$ such that $\mathcal{E}_0\cap B(\bn_0,\eps)=\{\bn_0\}$ for any $\bn_0\in\mathcal{E}_0$.
\end{lemma}

\noindent
The proofs of these lemmas are also found in the appendix.

\subsection{The main result}
\label{subsec:collect}
We will now significantly strengthen Theorems~\ref{thm:main} and~\ref{thm:sparsity} by showing that, under a concave penalty, a sparsest DAG $\bn^*\in\mathcal{E}_0$ maximizes the penalized likelihood amongst all the possible equivalent representations of the covariance matrix $\Theta_0$. Under the assumptions of Theorem~\ref{thm:main}, there is a $\pi$-local maximizer $\widehat{\bn}_n^*$ of $F(\bn)$ such that $\norm{\widehat{\bn}_n^* - \bn^*} = O_P(n^{-1/2} + a_n(\bn^*))$. Ideally, when $\bn_0$ has more edges than $\bn^*$, we would like these $\pi$-local maximizers to satisfy $F(\widehat{\bn}_n^*)>F(\widehat{\bn}_n)$ with high probability. 

Intuitively, when $a_n(\bn_0)=b_n(\bn_0)=0$, all of the nonzero coefficients lie in the flat part of the penalty where $\pln'(|\phi^0_{ij}|)=\pln''(|\phi^0_{ij}|)=0$. When this happens, the penalty ``acts'' like the $\ell_0$ penalty by penalizing all of the coefficients equally by the amount $\tau(\lambda_n)$, and any DAG with more edges than $\bn^*$ will see a heavier penalty. In order to quantify ``how close'' $\bn_0$ is to lying in the flat part of the penalty, we define 
\begin{align*}
c_n(\bn_0):=\min\{\pln(|\phi_{ij}^0|):\phi_{ij}^0\ne 0\}. 
\end{align*}

\noindent
When $c_n(\bn_0)=\tau(\lambda_n)$, the penalty mimics the $\ell_0$ penalty, and since the likelihood $\ell_n(\bn_0\|X)$ is constant for all $\bn_0$, we would then have
\begin{align*}
\pln(\bn^*) < \pln(\bn_0)
\iff
\ell_n(\bn^*\|X) - n\,\pln(\bn^*) > \ell_n(\bn_0\|X) - n\,\pln(\bn_0).
\end{align*}

\noindent
One would hope that for local maximizers $\widehat{\bn}_n$ that are sufficiently close to the $\bn_0$, the continuity of $F$ would guarantee that this intuition persists. As long as the amount of regularization grows fast enough, this is precisely the case:

\begin{thm}
\label{thm:global}
Suppose that $\pl(t)$ is nondecreasing and concave for $t\geq 0$ with $\pl(0)=0$. Assume further that the conditions for Theorem~\ref{thm:sparsity} hold for all $\bn_0\in\mathcal{E}_0$. Recall that $\tau(\lambda_n):=\sup_t\pln(t)$. If
\begin{enumerate}
\item $c_n(\bn_0) = \tau(\lambda_n)+O(n^{-1/2})$ for all $\bn_0\in\mathcal{E}_0$,
\item $\limsup_n\tau(\lambda_n)<\infty$,
\item $\tau(\lambda_n)n^{1/2}\to\infty$,
\end{enumerate}
then for any DAG $\bn_0\in\mathcal{E}_0$ with strictly more edges than $\bn^*$, $P(F(\widehat{\bn}_n^*) > F(\widehat{\bn}_n))\to1$ as $n\to\infty$.
\end{thm}

\noindent
The restriction to $\bn_0$ with strictly more edges than $\bn^*$ is necessary since $\bn^*$ may not be unique in general. Theorem~\ref{thm:global} essentially answers the question of which DAG in the true equivalence class $\mathcal{E}_0$ our estimator approximates. As we have discussed, there is a subtle technicality in which it is possible that there are \emph{other} maximizers of $F(\bn)$ besides the $\pi$-local maximizers, but this is very unlikely in practice.

These theorems provide general technical statements which can be used when weaker assumptions are necessary. By imposing all the conditions in Theorems~\ref{thm:main},~\ref{thm:sparsity}, and~\ref{thm:global} uniformly, we can combine all of the results in order to characterize the behaviour of the estimates in terms of the parametrization $(\hB,\hO)$ given by \eqref{eq:transform}. Before stating the main theorem, we will need some notation to distinguish $\pi$-local maximizers. When the conditions of Theorem {\ref{thm:main}} hold for all $\pi$, we will denote the collection of $\pi$-local maximizers by $\mathcal{M}_n$. Continuing our notation from the previous section, we also let $(B^*,\Omega^*)$ denote any graph in $\mathcal{E}_0$ with the fewest number of edges, and let $(\hB^*,\hO^*)$ be the corresponding $\pi$-local maximizer. Recall that given a DAG estimate $(\hB,\hO)$, we define $\widehat{\Theta} = \Theta(\hB,\hO)$.

\begin{thm}
\label{thm:all}
Suppose that $\pl(t)$ is nondecreasing and concave for $t\geq 0$ with $\pl(0)=0$. Fix $p\ge1$ and assume that the penalty function satisfies
\begin{align*}
\liminf_{n\to\infty}\liminf_{t\to0^+} p_{\lambda_n}'(t)/\lambda_n > 0.
\end{align*}
Assume further that $a_n(\bn_0) = O(n^{-1/2})$, $b_n(\bn_0)\to0$, and $c_n(\bn_0) = \tau(\lambda_n)+O(n^{-1/2})$ for each DAG in $\mathcal{E}_0$. If $\lambda_n\to0$, $\lambda_nn^{1/2}\to\infty$, $\limsup_n\tau(\lambda_n)<\infty$, and $\tau(\lambda_n)n^{1/2}\to\infty$, then for any permutation $\pi$, there is a local maximizer $(\hB,\hO)$ of $F$ such that
\begin{enumerate}
\item $\norm{\hB - \tilde{B}_0(\pi)}_F + \norm{\hO - \tilde{\Omega}_0(\pi)}_F = O_P(n^{-1/2})$,
\item $P(\supp(\hB)=\supp(\tilde{B}_0(\pi)))\to1$,
\item $\norm{\widehat{\Theta} - \Theta_0}_F = O_P(n^{-1/2})$.
\end{enumerate}
Furthermore,
\begin{align*}
P\left(F(\hB^*,\hO^*) = \max_{(\hB,\hO)\in\mathcal{M}_n}F(\hB,\hO)\right)
\to1.
\end{align*}
\end{thm}

\noindent
The proof of Theorem~\ref{thm:all} is immediate from the properties of the Frobenius norm and Theorems~\ref{thm:main},~\ref{thm:sparsity}, and~\ref{thm:global}.

\begin{remark}
Using an adaptive $\ell_1$ penalty, \cite{fu2013} first obtained results similar to Theorems~\ref{thm:main} and~\ref{thm:sparsity}. These results assume a weakened form of faithfulness, however, and require experimental data with interventions in order to guarantee identifiability of the true causal DAG. The results here generalize this theory to observational data without needing faithfulness. The keys to this generalization are the notion of parametric equivalence in \eqref{eq:eqclass} (as opposed to Markov equivalence) and the use of a concave penalty to rule out equivalent DAGs with too many edges. The role of concavity is highlighted by the observation that convex penalties cannot satisfy the conditions for Theorem~\ref{thm:global}.
\end{remark}

\subsection{Discussion of the assumptions}
\label{subsec:assumptions}
The general theme behind the theory described in the previous sections is that as long as the penalty is chosen cleverly enough, there will be a consistent local maximizer for the constrained penalized likelihood problem \eqref{eq:reestimator}. We pause now to discuss these conditions more carefully, and show that they can always be satisfied.

The parameters $a_n(\bn_0)$ and $b_n(\bn_0)$ measure respectively the maximum slope and concavity of the penalty function, and the conditions on these terms are derived directly from \cite{fan2001}. The idea is that as long as the concavity of the penalty is overcome by the local convexity of the log-likelihood function, our intuition from classical maximum likelihood theory continues to hold true. In order to simultaneously guarantee consistency in parameter estimation and structure learning, it is necessary that these parameters vanish asymptotically.

Furthermore, the assumptions on $a_n$ and $b_n$ in Theorems~\ref{thm:main} and \ref{thm:sparsity} highlight the advantages of concave regularization over $\ell_1$ regularization. In particular, the $\ell_1$ penalty trivially satisfies $b_n\to0$, but cannot simultaneously satisfy $a_n(\bn_0)=\lambda_n=O(n^{-1/2})$ and $\lambda_nn^{1/2}\to\infty$. Thus, for the $\ell_1$ penalty, we may apply Theorem~\ref{thm:main} to obtain a local maximizer which is consistent in \emph{parameter estimation}, but we cannot guarantee structure estimation consistency through Theorem~\ref{thm:sparsity}. In contrast, these conditions are easily satisfied by a concave penalty; in particular they are satisfied when $\pl$ is the MCP. These observations were first made in \cite{fan2001}.

The conditions on $\tau(\lambda_n)$ in Theorem~\ref{thm:global} are more interesting. When the true parameter is identifiable, there is no concern about dominating the penalized likelihood for nonsparse parameters. Since our set-up is decidedly nonidentifiable---there are up to $p!$ choices of the ``true'' graph---it is essential to control the growth of the penalty, and more specifically, how the penalty grows at the various equivalent DAGs $\bn_0\in\mathcal{E}_0$. As long as this grows at the right rate, nonsparse graphs will see the penalty term dominate, and as a result the sparsest graph $(B^*,\Omega^*)$ emerges as the best estimate of the true graph. Since $\tau(\lambda_n)=+\infty$ for any convex penalty, Theorem~\ref{thm:global} along with the remainder of this discussion do not apply to $\ell_1$ regularization.

In order to quantify the behaviour of the penalty, we need to control the growth of two different quantities: the maximum penalty $\tau(\lambda_n)$, and the rate of convergence of $c_n(\bn_0)$. By rate of convergence, we refer to the fact that the assumptions on $a_n(\bn_0)$ and $b_n(\bn_0)$ alone require that $c_n(\bn_0)=\tau(\lambda_n)+o(1)$, or equivalently $\pln(|\phi_{ij}^0|)=\tau(\lambda_n)+o(1)$ whenever $\phi_{ij}^0\ne0$. The stronger assumption that $c_n(\bn_0)=\tau(\lambda_n)+O(n^{-1/2})$ in Theorem~\ref{thm:global} shows that it is not enough that this convergence occurs at an arbitrary rate. One may think of this as a requirement on the zeroth-order convergence of $\pln$, in contrast to the first- and second-order convergence required by Theorems~\ref{thm:main} and~\ref{thm:sparsity}. In practice, it is sufficient to have $c_n(\bn_0)=\tau(\lambda_n)$ for sufficiently large $n$, and hence also $a_n=b_n=0$.

Of course, none of this is relevant if we cannot construct a penalty which satisfies all of these conditions simultaneously along with associated regularization parameters $\lambda_n$. When the penalty is chosen to be the MCP, all of the conditions required for Theorem~\ref{thm:all} are satisfied as long as  
\begin{align}
\label{eq:gammalambda}
\limsup_n \lambda_n\gamma_n < 
\min_{\bn_0\in\mathcal{E}_0}\min\{|\phi_{ij}^0|:\phi_{ij}^0 \ne 0\}
\quad\text{and}\quad
\lambda_n = O(n^{-\alpha}),\,\, 0<\alpha<1/2.
\end{align}

\begin{remark}
\label{rem:factor}
To better understand the conditions on $\tau(\lambda_n)$ in Theorems~\ref{thm:global} and \ref{thm:all}, it is instructive to consider the simplified case in which the penalty factors as $\pln(t) = \lambda_n \rho(t)$ for some function $\rho(t)$ (not to be confused with the parameters $\rho_j$ in our model). In this case, the penalty is bounded as long as $\lim_{t\to\infty}\rho(t)<\infty$ and the conditions on $\tau(\lambda_n)$ in Theorem~\ref{thm:global} reduce to $\limsup_n\lambda_n<\infty$ and $\lambda_nn^{1/2}\to\infty$. When $\lambda_n\to0$, these conditions are simply the assumptions in Theorem~\ref{thm:sparsity}. Thus, the extra conditions on $\tau(\lambda_n)$ in Theorems~\ref{thm:global} and \ref{thm:all} are redundant when the penalty factors in this way.
\end{remark}

\begin{example}
Although the usual formula for the MCP does not satisfy the factorization property in Remark~\ref{rem:factor}, we may reparametrize it so that it does. To do this, define a new penalty by
\begin{align*}
\overline{p}_\lambda(t;\delta)
:= \lambda\left(t-\frac{t^2}{2\delta}\right)1(t<\delta) + \frac{\lambda\delta}{2}1(t\ge\delta),
\qquad t\ge 0.
\end{align*}

\noindent
Then $\overline{p}_\lambda(t;\delta) = \lambda\cdot\overline{p}_{\lambda=1}(t;\delta)$, and by choosing $\delta = \lambda\gamma$ we may recover the usual formula for the MCP given by \eqref{eq:mcp}. Furthermore, the condition in \eqref{eq:gammalambda} becomes
\begin{align*}
\limsup_n \delta_n < 
\min_{\bn_0\in\mathcal{E}_0}\min\{|\phi_{ij}^0|:\phi_{ij}^0 \ne 0\},
\end{align*}

\noindent
which is independent of $\lambda_n$.
\end{example}

\subsection{Score-based theory in high-dimensions}
\label{subsec:hdthy}
The theory in this section so far has assumed that $p$ is fixed with $n>p$, the classical low-dimensional scenario. It would be interesting to obtain results for this method when $p$ is allowed to depend on $n$, and in particular the case when $p>n$. While the simulations in Section~\ref{sec:results} give good empirical evidence that our method is applicable to this scenario, formal theoretical results are not available yet. Here we take a moment to discuss some current work in this direction.

If we fix a permutation $\pi$, we have already described in Remark~\ref{rem:fixedpi} how to modify our method in order to estimate the equivalent DAG that is compatible with $\pi$, which we have denoted by $(\tilde{B}_0(\pi),\tilde{\Omega}_0(\pi))$. When the order of the variables is fixed, the problem reduces to standard multiple regression with a concave penalty, in which case Theorems~\ref{thm:main} and~\ref{thm:sparsity} can be generalized to high-dimensions, for instance using the results in \cite{fan2010}. This is very much in the spirit of similar results in the $\ell_1$ case obtained by \cite{shojaie2010}. Of course, in our set-up, we do not know in advance which permutation is optimal, so this does not tell the whole story. Theorem~\ref{thm:global} shows how our estimator selects the right permutation automatically based on the data, and eliminates the need to assume this prior knowledge.

Recently, \cite{geer2013} obtained some positive results using $\ell_0$ regularization in which it is not assumed that $\pi$ is known in advance. Under the same Gaussian framework we have adopted in this work, they show the following: When $\pl(t)=\lambda^2 1(t\ne0)$ and under certain strong regularity conditions, any \emph{global minimizer} of \eqref{eq:estimator} satisfies
\begin{align}
\label{eq:vdGBmain}
\norm{\hB-\tilde{B}_0(\hat{\pi})}_F^2 + \norm{\hO-\tilde{\Omega}_0(\hat{\pi})}_F^2
=O_P(\lambda^2s_0),
\end{align}

\noindent
where $\hat{\pi}$ is the permutation compatible with $(\hB,\hO)$. Furthermore, they establish that the estimated number of edges are all of the same order: $\hat{s}=O_P(\tilde{s}_0(\hat{\pi}))=O_P(s_0)$. These results represent the first significant analysis of score-based structure learning in high-dimensions that we know of, however, they have some drawbacks. First, they do not guarantee structure estimation consistency, and instead only give an upper bound on the number of estimated edges, which is to be of the same order as a minimal-edge DAG. With respect to computations, these results only hold for the intractable $\ell_0$ penalty, and no suggestions are made to allow computation of this estimator in practice. Furthermore, since the optimization problem is nonconvex, theoretical guarantees for global minimizers are less practical than guarantees for local minimizers. We have already observed (Remark~\ref{rem:equiv}) that the estimator defined in \cite{geer2013} is a special case of \eqref{eq:reestimator}, and so this theory applies to our framework under $\ell_0$ regularization.

A common interpretation of concave penalization is as a continuous relaxation of the discrete $\ell_0$ penalty. Our framework can thus be seen in this light. Previous work has shown that penalized likelihood estimators can have near optimal performance when compared with the $\ell_0$ estimator (\cite{zhang2012}), and thus we have good reason to believe the same holds true for our estimator. The key idea from the analysis in \cite{geer2013} is to control the behaviour of the estimates over all $p!$ possible permutations, which requires careful analysis using exponential-type concentration inequalities. Based on our preliminary work, we believe that such an analysis can be carried out for more general penalties, however, the details remain to be worked out and are expected to be technical.

Recently there has been some reported progress in high-dimensions for hybrid methods that consist of multiple learning stages. The general outline of these methods is the following:
\begin{enumerate}
\item[1.] Estimate an initial (undirected, directed, or partially directed) graph $\mathcal{G}_0$,
\item[2.] Search for an optimal DAG structure $\widehat{\mathcal{G}}$ subject to the constraint that $\widehat{\mathcal{G}}$ is a subgraph of $\mathcal{G}_0$.
\end{enumerate}

\noindent
This approach is motivated by the fact that searching for an undirected or partially directed graph in the first step can be substantially faster than searching for a DAG. In this light, \cite{loh2013} consider using inverse covariance estimation to restrict the search space, and \cite{buhlmann2013} convert the problem into three separate steps: preliminary neighbhourhood selection, order search, and maximum likelihood estimation. While they obtain some high-dimensional guarantees, these ideas do not apply directly to our framework since it consists of a single learning step.


\section{Algorithm Details}
\label{sec:comp}

Both the objective function and the constraint set in \eqref{eq:reestimator} are nonconvex, which makes traditional gradient descent algorithms for performing the necessary minimization inapplicable. One could employ naive gradient descent to find a local minimizer of \eqref{eq:reestimator}, but it would still be difficult to enforce the DAG constraint. Thus, a different approach must be taken altogether. Extending the algorithm of \cite{fu2013}, we employ a cyclic coordinate-descent based algorithm that relies on checking the DAG constraint at each update. By properly exploiting the sparsity of the estimates and the reparametrization {\eqref{eq:reloglik}}, however, we will be able to perform the single parameter updates and enforce the constraint with ruthless efficiency.

\subsection{Overview}
Before outlining the technical details of implementing our algorithm, we pause to provide a high-level overview of our approach.

The idea behind cyclic coordinate descent is quite simple: Instead of minimizing the objective function over the entire parameter space simultaneously, we restrict our attention to one variable at a time, perform the minimization in that variable while holding all others constant (hereafter referred to as a \emph{single parameter update}), and cycle through the remaining variables. This procedure is repeated until convergence. Coordinate descent is ideal in situations in which each single parameter update can be performed quickly and efficiently. For more details on the statistical perspective on coordinate descent, see \cite{wu2008,friedman2007}.

Moreover, due to acyclicity, we know \emph{a priori} that the parameters $\phi_{kj}$ and $\phi_{jk}$ cannot simultaneously be nonzero for $k\ne j$. This suggests performing the minimization in blocks, minimizing over $\{\phi_{kj},\phi_{jk}\}$ simultaneously. An immediate consequence of this is that we reduce the number of free parameters from $p^2$ to $p(p-1)/2 + p$, a substantial savings.

In order to enforce acyclicity, we use a simple heuristic: For each block $\{\phi_{kj},\phi_{jk}\}$, we check to see if adding an edge from $X_k\to X_j$ induces a cycle in the estimated DAG. If so, we set $\phi_{kj}=0$ and minimize with respect to $\phi_{jk}$. Alternatively, if the edge $X_j\to X_k$ induces a cycle, we set $\phi_{jk}=0$ and minimize with respect to $\phi_{kj}$. If neither edge induces a cycle, we minimize over both parameters simultaneously.

Before we outline the details, let us introduce some functions which will be useful in the sequel. Define
\begin{align}
\label{eq:objective}
Q(\Phi,R)
:= L(\Phi,R) + \sum_{i,j}\pl(|\phi_{ij}|)
\end{align}

\noindent
to be our objective function for coordinate descent. Note that we have suppressed the dependence of the log-likelihood on the data $X$ as well as the dependence of the penalty term on $n$. In fact, in the computations we may treat $n$ as fixed, so we can absorb this term into the penalty function $\pl$. This simply amounts to rescaling the regularization parameter $\lambda$, which causes no problems in computing $(\hP,\hR)$. Thus solving {\eqref{eq:reestimator}} is equivalent to minimizing $Q$.

Now define the single-variable functions
\begin{align}
\label{eq:Q1}
Q_1(\phi_{kj})
&= \frac{1}{2}\Norm{\rho_jx_{j}-\sum_{i=1}^p\phi_{ij}x_i}^2 + \pl(|\phi_{kj}|), \\
Q_2(\rho_j)
&= -n\log\rho_j + \frac{1}{2}\Norm{\rho_jx_{j}-\sum_{i=1}^p\phi_{ij}x_i}^2.
\end{align}

\noindent
The function $Q_1$ is $Q(\Phi,R)$ in \eqref{eq:objective} considered as a function of the single parameter $\phi_{kj}$, while holding the other $p^2 - 1$ variables fixed and ignoring terms that do not depend on $\phi_{kj}$, and $Q_2$ is the corresponding function for the parameter $\rho_j$. We express the dependence of $Q_1$ and $Q_2$ on $k$ and/or $j$ implicitly through their respective argument, $\phi_{kj}$ or $\rho_j$. 

An overview of the algorithm is given in Algorithm~\ref{alg:overview}. We use the notation $\phi_{kj}\Leftarrow 0$ to mean that $\phi_{kj}$ must be set to zero due to acyclicity, as outlined above. The remainder of this section is devoted to the details of implementing the above algorithm, which we call Concave penalized Coordinate Descent with reparametrization (CCDr).

\begin{algorithm}[t]
\begin{center}
\caption{CCDr Algorithm}
\label{alg:overview}
\end{center}
\begin{itemize}
\item[\emph{Input:}] Initial estimates $(\Phi^0, R^0)$; penalty parameters $(\lambda,\gamma)$; error tolerance $\eps>0$; maximum number of iterations $M$.
\vspace{0.5em}
\item[1.] Cycle through $\rho_j$ for $j=1,\ldots,p$, minimizing $Q_2$ with respect to $\rho_j$ at each step.
\item[2.] Cycle through the $p(p-1)/2$ blocks $\{\phi_{kj},\phi_{jk}\}$ for $j,k=1,\ldots,p$, $j\ne k$, minimizing with respect to each block:
\begin{itemize}
\item[(a)] If $\phi_{kj}\Leftarrow 0$, then minimize $Q_1$ with respect to $\phi_{jk}$ and set $(\phi_{kj},\phi_{jk}) = (0, \phi_{jk}^*)$, where $\phi_{jk}^*=\argmin Q_1(\phi_{jk})$;
\item[(b)] If $\phi_{jk}\Leftarrow 0$, then minimize $Q_1$ with respect to $\phi_{kj}$ and set $(\phi_{kj},\phi_{jk}) = (\phi_{kj}^*,0)$, where $\phi_{kj}^*=\argmin Q_1(\phi_{kj})$;
\item[(c)] If neither 2(a) nor 2(b) applies, then choose the update which leads to a smaller value of $Q$.
\end{itemize}
\item[3.] Repeat steps 1 and 2 $l$ times, until either $\max_{j,k}|\phi_{kj}^{(l-1)}-\phi_{kj}^{(l)}|<\eps$ or $l>M$.
\item[4.] Transform the final estimates $(\hP, \hR)$ back to the original parameter space  $(\hB,\hO)$ (see equation \eqref{eq:transform}) and output these values.
\end{itemize}
\end{algorithm}

\subsection{Coordinate descent}
\label{subsec:cd}
In what follows, we assume that the data have been appropriately normalized so that each column $x_j$ has unit norm, $\norm{x_j}^2=\sum_h x_{hj}^2=1$. Furthermore, although the details of the algorithm do not depend on the choice of penalty, we will focus on the MCP and $\ell_1$ penalties, as these are the methods implemented and discussed in Sections~\ref{sec:results} and \ref{sec:app}.

\subsubsection{Update for $\phi_{kj}$}
\cite{mazumder2011} show that the minimum of \eqref{eq:Q1} can be found by solving
\begin{align}\label{eq:spupls}
\argmin_\b Q^1(\beta), \quad\text{where }Q^1(\beta) := \frac{1}{2}(\beta-\tilde{\beta})^2 + \pl(|\b|).
\end{align}

\noindent
The solution to \eqref{eq:spupls} is given by a so-called threshold function which is associated to each choice of penalty. For the MCP with $\gamma>1$ this is defined by
\begin{align}
S_\gamma(\tilde{\b}, \lambda) = 
\begin{cases}
0, & |\tilde{\b}| \le \lambda, \\
\sgn(\tilde{\b})\left(\frac{|\tilde{\b}|-\lambda}{1-1/\gamma}\right), & \lambda < |\tilde{\b}|\le\lambda\gamma, \\
\tilde{\b}, & |\tilde{\b}| > \lambda\gamma.
\end{cases}
\end{align}

\noindent
For the $\ell_1$ penalty, we have
\begin{align}
S(\tilde{\b}, \lambda) = 
\begin{cases}
0, & |\tilde{\b}| \le \lambda, \\
\sgn(\tilde{\beta})(|\tilde{\beta}|-\lambda),
& |\tilde{\b}|>\lambda.
\end{cases}
\end{align}

To see how to convert \eqref{eq:Q1} into \eqref{eq:spupls}, note that 
\begin{align}
Q_1(\phi_{kj})
&= \frac{1}{2}\sum_{h=1}^n\left(\rho_jx_{hj} - \sum_{i\ne k}\phi_{ij}x_{hi} - \phi_{kj}x_{hk}\right)^2 + \pl(|\phi_{kj}|) \\
&= \frac{1}{2}\sum_{h=1}^n x_{hk}^2\left(\frac{1}{x_{hk}}r_{kj}^{(h)} - \phi_{kj}\right)^2 + \pl(|\phi_{kj}|), \nonumber
\intertext{where $r_{kj}^{(h)} := \rho_jx_{hj} - \sum_{i\ne k}\phi_{ij}x_{hi}$. Expanding the square in the last line and using $\sum_h x_{hk}^2=1$,}
Q_1(\phi_{kj})
&= \frac{1}{2}\left\{\sum_{h=1}^n (r_{kj}^{(h)})^2 - 2\phi_{kj}\sum_{h=1}^nx_{hk}r_{kj}^{(h)} + \phi_{kj} ^2\right\} + \pl(|\phi_{kj}|) \\
\label{eq:spuder1}
&=\frac{1}{2}\left(\phi_{kj} - \sum_{h=1}^nx_{hk}r_{kj}^{(h)}\right)^2 + \pl(|\phi_{kj}|) + \text{const.}
\end{align}

\noindent
The constant term in \eqref{eq:spuder1} does not depend on $\phi_{kj}$ and hence does not affect the minimization of $Q_1$. Thus minimizing $Q_1(\phi_{kj})$ is equivalent to minimizing $Q^1(\b)$ in \eqref{eq:spupls} with $\tilde{\b} = \sum_hx_{hk}r_{kj}^{(h)}$. Hence for MCP with $\gamma>1$,
\begin{align}
\label{eq:spusoln}
\argmin Q_1(\phi_{kj})
= S_\gamma\left(\sum_hx_{hk}r_{kj}^{(h)},\lambda\right),
\end{align}

\noindent
and similarly for the $\ell_1$ penalty. The existence of a closed-form solution to the single parameter update for $\phi_{kj}$ is a key ingredient to our method, and is one of the reasons we chose the MCP and $\ell_1$ penalties in our comparisons. Many other penalty functions, however, allow for closed-form solutions to \eqref{eq:spupls}, and our algorithm applies for any such penalty function. 

\subsubsection{Update for $\rho_k$}
The single parameter update for $\rho_j$ is straightforward to compute and is given by 
\begin{align}
\label{eq:spurho}
\argmin Q_2(\rho_j)
= \frac{c + \sqrt{c^2+4n}}{2},
\quad\text{ with } c = \sum_{i\ne j}\phi_{ij}\sum_h x_{hi}x_{hj}.
\end{align}

\noindent
Since $Q_2(\rho_j)$ is a strictly convex function, this is the only minimizer.

\subsection{Regularization paths}
In practice, it is difficult to select optimal choices of the penalty parameters $(\lambda,\gamma)$ in advance. Thus it is necessary to compute several models at many discrete choices of $(\lambda_i,\gamma_j)$, and then perform model selection. In testing, we observed a dependence on the concavity parameter $\gamma$, however, for simplicity we will consider $\gamma$ fixed in the sequel, and postpone further study of the method's dependence on $\gamma$ to future work.

The regularization parameter $\lambda$, on the other hand, has a strong effect on the estimates. In particular, as $\lambda\to\infty$, $\hP(\lambda)\to\mathbf{0}$, and as $\lambda\to0$ we obtain the unpenalized maximum likelihood estimates. It is thus desirable to obtain a sequence of estimates $(\hP(\lambda_i),\hR(\lambda_i))$ for some sequence $\lambda_i>\lambda_{i+1}>0$, $i=0,1,\ldots,L$. In practice, we will always choose $\lambda_0$ so that $\hP(\lambda_0)=\mathbf{0}$, with successive values of $\lambda_i$ decreasing on a linear scale. One can easily check that if we use an initial guess of $\Phi^0 = \mathbf{0}$, then the choice $\lambda_0=n^{1/2}$ ensures that the null model is a local minimizer of $Q$.

Once we have estimated a sequence of models $(\hP(\lambda_i),\hR(\lambda_i))$, $i=0,1,\ldots,L$, we must choose the best model from these $L+1$ models. This is the model selection problem, and is beyond the scope of this paper. The present work should be considered a ``proof of concept,'' showing that under the right conditions, there exists a $\lambda$ that estimates the true DAG with high fidelity. The problem of correctly selecting this parameter is left for future work, but some preliminary empirical analysis is provided in Section~\ref{subsec:modelselect}. See \cite{wang2007} for some positive results concerning the SCAD penalty, and \cite{fu2013} for a relevant discussion of some difficulties that are idiosyncratic to structure estimation of BNs. In particular, it is worth re-emphasizing here that cross-validation is suboptimal, and should be avoided.

\subsection{Implementation details}
\label{sec:speed}
As presented so far, the CCDr algorithm is not particularly efficient. Fortunately, there are several computational enhancements we can exploit to greatly improve the efficiency of the algorithm. Many of these ideas are adapted from \cite{friedman2010}, and the reader is urged to refer to this paper for an excellent introduction to coordinate descent for penalized regression problems.

In implementing the CCDr algorithm, we use warm starts and an active set of blocks as described in \cite{friedman2010, fu2013}. We also use a sparse implementation of the parameter matrix $\Phi$ to speed up internal calculations. Naive recomputation of the $n$ weighted residual factors $r_{kj}^{(h)}$ for $h=1,\ldots,n$ for every update incurs a cost of $O(np)$ operations, which is prohibitive in general, and is the main bottleneck in the algorithm. \cite{friedman2010} observe that this calculation can be reduced to $O(p)$ operations by noting that the sum in \eqref{eq:spusoln} can be written as
\begin{align}
\sum_{h=1}^n x_{hk}r_{kj}^{(h)}
&= \rho_j\dotp{x_j}{x_k}-\sum_{i\ne k}\phi_{ij}\dotp{x_i}{x_k}.
\end{align}

\noindent
The inner products above do not change as the algorithm progresses, and hence can be computed once at a cost of $O(n^2\log n)$ operations. This is a substantial improvement over several million $O(np)$ computations, which is typical for large $p$. 

Similar reasoning applies to the computation of {\eqref{eq:spurho}}, which highlights why the repara-metrization {\eqref{eq:reloglik}} is useful: the single parameter update for each $\rho_j$ only requires $O(p)$ operations, compared with $O(p^2)$ required operations for the standard residual estimate for $\omega_j^2$ in the original parametrization. Since we perform $p$ of these updates in each cycle, we reduce the total number of operations per cycle from $O(p^3)$ down to $O(p^2)$, which is a substantial savings. Moreover, by leveraging sparsity, both {\eqref{eq:spusoln}} and {\eqref{eq:spurho}} become $O(1)$ calculations when the maximum number of parents per node is bounded.

As stated, our algorithm will take a pre-specified sequence of $\lambda$-values and compute an estimate $(\hP(\lambda_i),\hR(\lambda_i))$ for all $L+1$ choices of $\lambda_i$. In general, we do not know in advance what the smallest value of $\lambda$ appropriate for the data is, and we typically choose $\lambda_L$ as some very small value. Since the model complexity (in terms of the number of edges) increases as $\lambda$ decreases, more and more time is spent computing complex models for small $\lambda$. We can exploit these facts in order to avoid wasting time on computing unnecessarily complex models. As the algorithm proceeds calculating estimates for each $\lambda_i$, if the estimated number of edges $\hat{s}_i:=s_{\hB(\lambda_i)}$ is too large, we know that we need not continue computing new models for smaller $\lambda$. We can justify this as follows: \emph{either} the true model is sparse, in which case we know that complex models with $\hat{s}_i$ large can be ignored, \emph{or} the true model is \emph{not} sparse, in which case our algorithm is less competitive. Thus, in this sense, prior knowledge or intuition of the sparsity of the true model is needed. In practice, we implement this by halting the algorithm whenever $\hat{s}_i>\alpha p$, where $\alpha > 0$ is a pre-specified parameter. While the choice of $\alpha$ should be application driven, we will use $\alpha = 3$ unless reported otherwise. In the sequel, $\alpha$ shall be referred to as the \emph{threshold parameter}.

\subsection{Full algorithm}
\label{sec:full}
A complete, detailed description of the algorithm is given in Algorithm~\ref{alg:full}, including the implementation details discussed in the previous section. We refer to steps (1-2) of Algorithm~\ref{alg:overview} as a single ``sweep'' of the algorithm (i.e. performing a single parameter update for every parameter in the active set).

\begin{algorithm}[t]
\begin{center}
\caption{Full CCDr Algorithm}
\label{alg:full}
\end{center} 
\begin{itemize}
\item[\emph{Input:}] Initial estimates $(\Phi^0_0, R^0_0)$; sequence of regularization parameters $\lambda_0>\lambda_1>\cdots>\lambda_L$; concavity parameter $\gamma > 1$; error tolerance $\eps>0$.
\vspace{0.5em}
\item[1.] Normalize the data so that $\norm{x_j}^2=1$ and compute the inner products $\dotp{x_i}{x_j}$ for all $i,j=1,\ldots,p$.
\item[2.] For each $\lambda_i$:
\begin{itemize}
\item[1.] If $i > 0$, set $(\Phi^0_i, R^0_i) = (\hP(\lambda_{i-1}),\hR(\lambda_{i-1}))$.
\item[2.] Perform a full sweep of all parameters using $(\Phi^0_i, R^0_i)$ as initial values, and identify the active set.
\item[3.] Sweep over the active set $l$ times, until either $\max_{j,k}|\phi_{kj}^{(l-1)}-\phi_{kj}^{(l)}|<\eps$ or $l>M$.
\item[4.] Repeat (2-3) $m$ times (using the current estimates as initial values) until the active set does not change, or $m>M$.
\item[5.] If $\hat{s}_i > \alpha p$, then halt the algorithm. If not, continue by computing $(\hP(\lambda_{i+1}),\hR(\lambda_{i+1}))$.
\end{itemize}
\item[3.] Transform the final estimates $(\hP(\lambda_i), \hR(\lambda_i))$ back to the original parameter space  $(\hB(\lambda_i),\hO(\lambda_i))$ (see equation \eqref{eq:transform}) and output these values.
\end{itemize}
\end{algorithm}

Finally, note that it is trivial to adapt the \emph{SparseNet} procedure from \cite{mazumder2011} to our algorithm in order to compute a \emph{grid} of estimates
\begin{align*}
(\hP(\lambda_i,\gamma_j),\hR(\lambda_i,\gamma_j)),
\quad i=0,\ldots,L,\, j=0,\ldots,J,
\end{align*}

\noindent
if one wishes to adjust the $\gamma$ parameter in addition to $\lambda$.

\section{Numerical Simulations and Results}
\label{sec:results}
In order to assess the accuracy and efficiency of the CCDr algorithm, we compared our algorithm with four other well-known structure learning algorithms: the PC algorithm (\cite{spirtes1991}), the max-min hill-climbing algorithm (MMHC; \cite{tsamardinos2006}), Greedy Equivalent Search (GES; \cite{chickering2003}), and standard greedy hill-climbing (HC). This selection was based on a pre-screening in which we compared the performance of several more algorithms in order to select those which showed the best performance in terms of accuracy and efficiency, and is by no means intended to be exhaustive. We were mainly interested in the accuracy and timing performance of each algorithm as a function of the model parameters $(p,s_0,n)$. Details on the implementations used and our experimental choices will be discussed in Section~\ref{subsec:setting}.

Our comparisons thus consist of two score-based methods (GES, HC), one constraint-based method (PC), and one hybrid method (MMHC). For brevity, in the ensuing discussion we will frequently refer to both PC and MMHC as constraint-based methods since both methods employ some form of constraint-based search whereas GES and HC do not. In order to compare the effects of regularization, we also compared each of these algorithms to two implementations of CCDr: One using MCP as the penalty (CCDr-MCP), and a second with the $\ell_1$ penalty (CCDr-$\ell_1$). This gives us a total of six algorithms overall. To offer a sense of scale, the experiments in this section total over $140,000$ individual DAG estimates for almost 1,000 ``gold-standard'' DAGs.

We begin with a comprehensive evaluation in low-dimensions ($n\ge p$) of all six algorithms using randomly generated DAGs, the main purpose of which is to show that hill-climbing and GES are significantly slower and less accurate in comparison with the other approaches. This supports our first claim that CCDr represents a clear improvement over existing score-based methods. We then move onto a similar assessment for high-dimensional data, which will show the advantages of our method over the constraint-based methods when sample sizes are limited and the number of nodes increases. Once this has been done, we show that our method scales efficiently on graphs with up to 2000 nodes as well as discuss some issues related to model selection and timing. We conclude this section with some detailed discussions about our experiments. 

In the results that follow, a general theme will emerge: CCDr is significantly faster than all the other approaches while still retaining very good estimation properties. To wit, CCDr convincingly outperforms the score-based methods in both timing and accuracy, and outperforms the constraint-based methods in timing and accuracy in high-dimensions. The fact that this is accomplished efficiently without preprocessing or constraining the search space is somewhat remarkable.

\subsection{Experimental set-up}
\label{subsec:setting}
All of the algorithms were implemented in the \texttt{R} language for statistical computing (\cite{rcore2014}). For the PC and GES algorithms, we used the \texttt{pcalg} package (version 2.0-3, \cite{kalisch2012}), and for the MMHC and HC algorithms we used the \texttt{bnlearn} package (version 3.6, \cite{scutari2010}). Both packages employ efficient, optimized implementations of each algorithm, and were updated as recently as July 2014. At the time of the experiments, these were the most up-to-date publicly available versions of either package. All of the tests were performed on a late 2009 Apple iMac with a 2.66GHz Intel Core i5 processor and 4GB of RAM, running Mac OS X 10.7.5.

For all the experiments described in this section, DAGs were randomly generated according to the Erd\"{o}s-Renyi model, in which edges are added independently with equal probability of  inclusion. In each experiment, an array of values were chosen for each of the three main parameters: $p$, $s_0$, and $n$. For every possible combination of $(p,s_0,n)$, $N$ individual tests were then run with these parameters fixed. For each test, a DAG was randomly generated using the \texttt{pcalg} function \texttt{randomDAG} with $p$ nodes and $s_0$ expected edges, and then $n$ random samples were generated using the function \texttt{rmvDAG}, according to the structural model \eqref{eq:streqn}. For tests involving different choices of the sample size, the same DAG was used for each choice of $n$ to generate datasets of different sizes. Since the edges were selected at random, the simulated DAGs did not have \emph{exactly} $s_0$ edges, but instead $s_0$ edges on average. For each simulation, the nonzero coefficients $\beta_{ij}^0$ were chosen randomly and uniformly from the interval $[0.5,2]$ and the error variances were fixed at $\omega_j^0=1$ for all $j$.

With the exception of HC and GES, each algorithm has a tuning parameter which strongly affects the accuracy of the final estimates. For CCDr, this is $\lambda$, which controls the amount of regularization, and for PC and MMHC it is $\alpha$, the significance level. In order to study the dependence of each algorithm on these parameters, we chose a sequence of parameters to use for each algorithm. For CCDr, we used a linear sequence of 20 values, starting from $\lambda_{\rm max}=n^{1/2}$. For both PC and MMHC, we used $$\alpha\in\{0.0001,0.0005,0.001,0.005,0.01,0.05\}.$$ Our choices for $\alpha$ were motivated by the recommendations in \cite{kalisch2007} and \cite{tsamardinos2006}, respectively, as well as by computational concerns: It was necessary to use a much smaller sequence for these algorithms since their running times are significantly longer than CCDr. Furthermore, we found that setting $\alpha<0.0001$ results in estimates with too few edges, and setting $\alpha>0.05$ can lead to runtimes well in excess of 24 hours. 

When using the MCP, we must also select the concavity parameter $\gamma$ in addition to $\lambda$. In order to keep our experiments constrained to a reasonable size, we elected not to study the effect of this parameter in detail. Based on the extensive evaluations in \cite{zhang2010}, we chose $\gamma=2$, which was supported by internal tests to gauge the effect of this parameter. This value represents a fair balance between convexity ($\gamma\to\infty$) and complexity ($\gamma\to0$). The CCDr algorithm also has three other user-specific parameters: $\eps$, $M$, and $\alpha$. Based on our simulations, $\eps$ and $M$ have a minimal impact on the accuracy of the estimates, and can simply be chosen to be small and large respectively. The default parameters we used in these simulations were: $\eps = 10^{-4}$, $M = p^{1/2}\vee 10$, and $\alpha = 3$. Recall that in the full algorithm (Algorithm~\ref{alg:full}), for each $\lambda_i$ there are at most $M^2=p\vee 100$ sweeps. When $p$ is very small a maximum of 100 iterations is more than enough.

\begin{remark}
\label{rem:pcorient}
Traditionally, the PC algorithm produces either a skeleton or a CPDAG, depending on how many phases of the algorithm are run (see \cite{kalisch2007} for the definition of a CPDAG and its relation to the PC algorithm). As discussed in \cite{rutimann2009}, however, it is possible to orient a DAG given its CPDAG using the function \texttt{pdag2dag} from the \texttt{pcalg} package. This works well in practice, although we found that in some cases the provided method was not able to orient the edges in the CPDAG successfully. In this case, we were able to compare skeletons but not DAGs for the PC algorithm. In the analysis, we treated this situation agnostically by ignoring such problematic estimates and entering them as missing values in the final analysis. This situation arose in less than 5\% of cases, so it was not a significant issue.
\end{remark}

\subsection{Performance metrics}
\label{subsec:metrics}
For every estimated structure, we compare both the final oriented DAG and its skeleton (i.e. the undirected graph that results by ignoring the directionality of the edges) to those of the true DAG. For a directed graph, we distinguish between \emph{true edges} (or \emph{true positives})---edges which are estimated with the correct orientation---and \emph{reversed edges}---edges which are in the skeleton but have the wrong direction. No such distinction can be made for the skeletons, of course. A \emph{false positive} is any edge---regardless of directionality---which is not in the skeleton of the true graph. 

We gauge the performance of the algorithms on the following metrics:
\begin{enumerate}
\item $P = $ number of estimated (predicted) edges,
\item $TP = $ number of true positives,
\item $R = $ number of reversed edges,
\item $FP = $ number of false positives,
\item SHD of the estimated DAG,
\item SHD of the estimated skeleton,
\item Test-data log-likelihood,
\item Test-data BIC,
\item Total and average running time in seconds.
\end{enumerate}

\noindent
SHD refers to the structural Hamming distance, which measures the number of edge reversals, additions, and/or removals necessary to convert an estimated graph into the true graph. This is a useful metric since it gives an absolute sense of ``how far'' away the estimates are from the true graph. For the precise definition of the structural Hamming distance, see \cite{tsamardinos2006}. Also, in order to compute the log-likelihood and BIC, it is necessary to estimate the parameters given the estimated structures, which we did by simple ordinary linear regression. As $p$ increases the time to compute these parameters becomes burdensome, and so comparisons of the log-likelihood and BIC were only performed for the low-dimensional experiments with $p\le 200$. While our primary concern in these evaluations is accuracy in structure learning, these two metrics give us a sense of the implied parameter estimation consistency.

We will also sometimes refer to the following common normalizations of the above metrics:
\begin{enumerate}
\item False discovery rate (FDR) $= (R+FP)/P$,
\item True positive rate (TPR) $= TP/T$,
\item False positive rate (FPR) $= (R+FP)/F$,
\end{enumerate}

\noindent
Here, $T$ is number of edges in the true graph and $F = \frac{1}{2}p(p-1)-T$ is the number of edges absent from the true graph. In some literature, the complement of the false discovery rate (i.e. $1-\text{FDR}$) is sometimes called \emph{specificity}, while TPR is also variously called \emph{recall} or \emph{sensitivity}.

Finally, when comparing the timing data it is important to recall that each algorithm computes a different number of estimates: HC and GES only produce one, the implementations of PC and MMHC used here produce exactly six, and both CCDr approaches produce up to 20 estimates. Thus it is necessary to consider both the total running time for each algorithm as well as the average time per estimate, which gives a better sense of the computational complexity of each approach. In the sequel, the \emph{total runtime} is defined as the real processor time required to run an algorithm over a full sequence of tuning parameters, and the \emph{average runtime} is defined as the total runtime divided by the number of graphs estimated, i.e., the number of tuning parameters in the sequence.

\subsection{Experiments on random graphs}
\label{subsec:exprandom}
In this section we provide detailed results comparing the performance of each algorithm on randomly generated DAGs, across a wide range of choices of $(p,s_0,n)$, using the metrics described in Section~\ref{subsec:metrics}.

In order to properly compare the algorithms, a single model needed to be selected from each sequence of estimates generated by each algorithm. To keep things simple, and since we have not considered a theoretical analysis of consistent model selection, we simply chose the most accurate model produced by each algorithm by selecting the DAG estimate with the smallest SHD. While this may seem artificial, it provides a good assessment of the potential of each approach. This choice of model selection results in DAGs with somewhat low sensitivity, but nonetheless it still provides a consistent method of comparing the performance of different algorithms. In Section~\ref{subsec:modelselect} we will discuss some interesting issues related to model selection. 


\subsubsection{Low-Dimensions}
\label{subsubsec:ldresults}
We first generated relatively small random graphs along with low-dimensional datasets according to the following settings:
\begin{itemize}
\item $p\in\{50, 100, 200\}$;
\item $s_0 / p \in\{0.2, 0.5, 1.0, 2.0\}$;
\item $n / p \in\{1, 5\}$;
\item Algorithms: CCDr-MCP, CCDr-$\ell_1$, GES, HC, MMHC, PC.
\end{itemize}

\noindent
For all combinations of $(p,s_0,n)$, we ran $N=50$ tests each. The result was 600 random DAGs, 1200 datasets, and 86,400 individual estimates across all six algorithms tested. 

\begin{figure}
\centering
\includegraphics[width=0.9\textwidth]{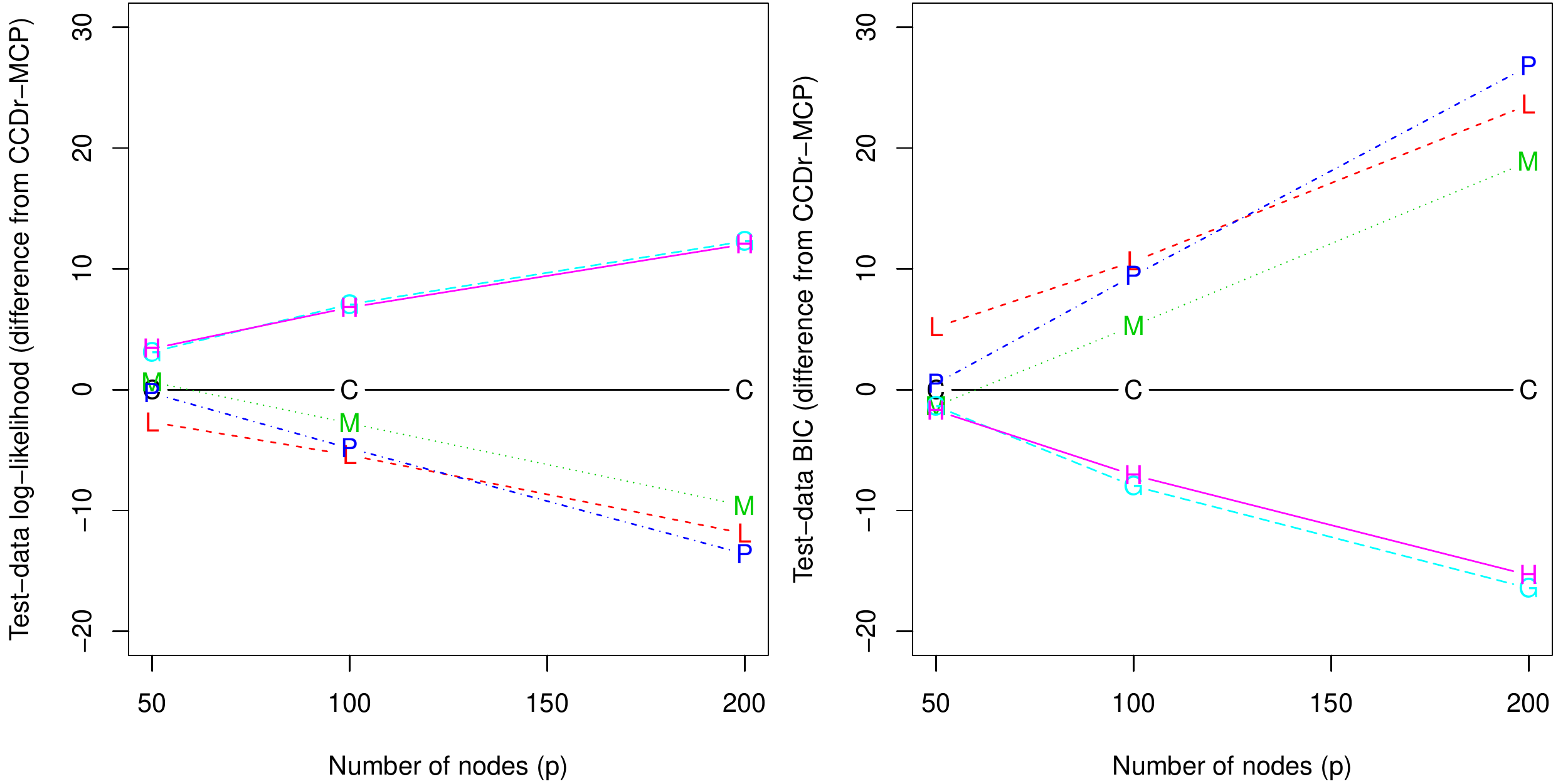}
\caption{Comparison of test-data log-likelihood and BIC scores (low dimensions). The data are presented relative to the scores for CCDr-MCP. For log-likelihood, larger scores (positive values in the plot) are better; for BIC smaller scores (negative values) are better. (C~=~CCDr-MCP, L~=~CCDr-$\ell_1$, G~=~GES, H~=~HC, M~=~MMHC, P~=~PC)}
\label{plot:loglik}
\end{figure}

The results are shown in Table~\ref{table:lowdim} and Figure~\ref{plot:loglik}. For each $p$, the results are averaged over all 50 tests and each value of $s_0$ and $n$. In the low-dimensional regime, it is expected that constraint-based algorithms will show good performance as the statistical tests on which they rely are more reliable and consistent when $n\ge p$. As expected, in our experiments, both PC and MMHC produced the most accurate results in this setting (Table~\ref{table:lowdim}). This is further substantiated by the seemingly counterintuitive observation that the performance of both algorithms improves as $p$ increases; this is explained by recalling that $n$ also increases as $p$ increases, so for larger $p$ the statistical tests also have increased power.

\begin{table}[t]
\caption{Average estimation performance of algorithms in low-dimensions.}
\vspace{0.5em}
\begin{center}
\begin{tabular}{lcccccc}
  \toprule
$p=50$, $T=46.48$ & CCDr-MCP & CCDr-$\ell_1$ & GES & HC & MMHC & PC \\ 
  \midrule
  P & 26.50 & 22.98 & 109.83 & 113.78 & 26.46 & 26.39 \\ 
  TP & 14.35 & 11.86 & \bf33.20 & 27.49 & 15.88 & 16.64 \\ 
  R & 8.38 & \bf7.96 & 8.19 & 12.29 & 9.14 & 8.26 \\ 
  FP & 3.78 & 3.15 & 68.44 & 74.00 & \bf1.44 & 1.48 \\ 
  SHD (DAG) & 35.92 & 37.77 & 81.72 & 92.99 & 32.04 & \bf31.32 \\ 
  SHD (skeleton) & 27.54 & 29.81 & 73.53 & 80.69 & \bf22.89 & 23.06 \\ 
  TPR & 0.31 & 0.26 & \bf0.71 & 0.59 & 0.34 & 0.36 \\ 
  FDR & 0.46 & 0.48 & 0.70 & 0.76 & 0.40 & \bf0.37 \\ 

  \toprule
$p=100$, $T=91.48$ & CCDr-MCP & CCDr-$\ell_1$ & GES & HC & MMHC & PC \\ 
  \midrule
  P & 67.14 & 60.32 & 241.71 & 256.20 & 60.97 & 60.33 \\ 
  TP & 36.40 & 30.85 & \bf74.30 & 60.24 & 39.03 & 39.85 \\ 
  R & 18.95 & 19.87 & \bf12.90 & 23.16 & 18.71 & 17.33 \\ 
  FP & 11.79 & 9.60 & 154.51 & 172.81 & 3.22 & \bf3.15 \\ 
  SHD (DAG) & 66.86 & 70.23 & 171.69 & 204.05 & 55.67 & \bf54.78 \\ 
  SHD (skeleton) & 47.91 & 50.36 & 158.79 & 180.88 & \bf36.95 & 37.45 \\ 
  TPR & 0.40 & 0.34 & \bf0.81 & 0.66 & 0.43 & 0.44 \\ 
  FDR & 0.46 & 0.49 & 0.69 & 0.76 & 0.36 & \bf0.34 \\ 

  \toprule
$p=200$, $T=185.06$ & CCDr-MCP & CCDr-$\ell_1$ & GES & HC & MMHC & PC \\ 
  \midrule
    P & 150.44 & 140.51 & 553.78 & 591.55 & 134.72 & 128.73 \\ 
  TP & 83.60 & 73.28 & \bf158.38 & 127.69 & 90.74 & 89.23 \\ 
  R & 39.05 & 42.58 & \bf22.35 & 45.65 & 37.59 & 34.28 \\ 
  FP & 27.79 & 24.65 & 373.06 & 418.21 & 6.39 & \bf5.22 \\ 
  SHD (DAG) & 129.24 & 136.43 & 399.74 & 475.58 & 100.70 & \bf96.69 \\ 
  SHD (skeleton) & 90.19 & 93.86 & 377.39 & 429.93 & \bf63.12 & 65.25 \\ 
  TPR & 0.45 & 0.40 & \bf0.86 & 0.69 & 0.49 & 0.48 \\ 
  FDR & 0.44 & 0.48 & 0.71 & 0.78 & 0.33 & \bf0.31 \\ 
   \bottomrule
\end{tabular}
\label{table:lowdim}
\end{center}
\end{table}

The score-based algorithms GES and HC, on the other hand, easily perform the worst in terms of structure learning: these algorithms include far too many edges and as a result obtain high sensitivity but also high false discovery rates. For example, when $p=200$ and the simulated DAGs had 185 edges on average, both HC and GES estimate well over 500 edges, almost three times the true number, and exhibit false discovery rates greater than 70\%. Notwithstanding, GES does noticeably outperform HC, which was anticipated.

Both CCDr methods fall in the middle, with CCDr-MCP outperforming CCDr-$\ell_1$ by a few edges in each case. Both methods estimate fewer edges than their score-based competitors---150 and 140 edges respectively when $p=200$---but slightly more than the constraint-based methods, which estimate 135 edges (PC) and 129 edges (MMHC). This shows that CCDr represents a clear improvement over both GES and HC, and this is even without consideration of efficiency, which we will discuss shortly (Section~\ref{subsubsec:timing}).

The results for the test-data log-likelihood and the BIC score highlight several difficulties with existing methods which the proposed methods help to overcome. GES and HC both show better log-likelihood than the others, and since the results are computed based on \emph{test data}, this cannot be attributed to overfitting. What's more, even though both methods produce far more edges than the others, they each only estimate roughly 3 edges per node, which is further evidence that these methods are not necessarily overfitting. Rather, going back to {\eqref{eq:invcovform}}, we see that the log-likelihood is a function of $\Theta$ alone, which means the test-data log-likelihood is not influenced by the accuracy of the graph structure estimated by an algorithm. This results in two distinct issues in evaluating  algorithms on the basis of test-data log-likelihood:

\begin{itemize}
\item Even if $\norm{\widehat{\Theta} - \Theta_0}_F$ is small, i.e. $\widehat{\Theta}$ is a good estimate of the true parameter, the estimated equivalence class can still be very different from the true equivalence class;
\item Even if the equivalence class is correctly estimated, the chosen representation may not be the sparsest.
\end{itemize}

This explains why GES and HC perform the best on this metric: They do a good job of estimating $\Theta_0$, as opposed to a sparse Bayesian network. By contrast, the constraint-based methods do not use the log-likelihood at all and thus exhibit the worst generalization in terms of log-likelihood. For methods which estimate approximately the same number of edges, CCDr-MCP is optimal, falling in between the score-based and constraint-based approaches (Figure~\ref{plot:loglik}). A similar discussion applies to the BIC scores, with the added complication of the BIC penalty. The fact that GES and HC still perform the best with respect to BIC---in spite of estimating far too many edges---underscores the fact that the BIC penalty is too lenient for estimating DAGs. This observation is further substantiated and discussed in more detail in Section~\ref{subsec:modelselect}.

\subsubsection{High-dimensions}
\label{subsubsec:hdresults}
In this section we use the same random set-up as in the previous section, however, our focus is now on high-dimensional estimation. Both HC and GES were omitted in this experiment because of their poor performance---both in terms of accuracy and timing---in the low-dimensional setting. This allowed us to scale up the experiments to $p=500$. In order to ensure a reasonable signal was detectable in each test, we fixed $n = 50$ for the tests. The following settings were used:
\begin{itemize}
\item $p\in\{100, 200, 500\}$;
\item $s_0 / p \in\{0.2, 0.5, 1.0, 2.0\}$;
\item $n=50$ fixed for all models;
\item Algorithms: CCDr-MCP, CCDr-$\ell_1$, MMHC, PC.
\end{itemize}

\noindent
For all combinations of $(p,s_0,n)$, we ran $N=20$ tests each, resulting in 240 tests. These tests give us a better sense of the performance of the algorithms when the sample size is small relative to $p$. 

The results are shown in Table~\ref{table:highdim}. As before, the results are presented for each value of $p$, averaged over all tests and each value of $s_0$ (note that $n$ did not change in these tests). In contrast to the low-dimensional scenario in which the constraint-based methods outperform our method, in high-dimensions we begin to see the advantages of CCDr in structure learning. As $p$ increases and $n$ remains fixed, the gap between CCDr-MCP and both PC and MMHC increases. In particular, across each value of $p$, the false discovery rates for all the methods are comparable, however, the increased sensitivity (true positive rate) and lower SHD indicates that CCDr-MCP provides a higher quality reconstruction of the true network. The numbers are illuminating: when $p=500$, for graphs which have 460 edges on average, CCDr-MCP estimates approximately 100 more edges while maintaining roughly the same false discovery rate and including 50-70 \emph{more} true edges on average.

\begin{table}[t]
\caption{Average estimation performance of algorithms in high-dimensions.}
\vspace{0.5em}
\begin{center}
\begin{tabular}{lcccc}
  \toprule
$p=100$, $T = 92.31$ & CCDr-MCP & CCDr-$\ell_1$ & MMHC & PC \\ 
  \midrule
  P & 52.74 & 43.95 & 43.02 & 43.89 \\ 
  TP & \bf27.59 & 21.48 & 23.82 & 24.12 \\ 
  R & 16.95 & 16.29 & \bf16.07 & 16.19 \\ 
  FP & 8.20 & 6.19 & \bf3.12 & 3.58 \\ 
  SHD (DAG) & 72.92 & 77.03 & \bf71.61 & 71.76 \\ 
  SHD (skeleton) & 55.98 & 60.74 & \bf55.54 & 55.58 \\ 
  TPR & \bf0.30 & 0.23 & 0.26 & 0.26 \\ 
  FDR & 0.48 & 0.51 & \bf0.45 & \bf0.45 \\ 

  \toprule
$p=200$, $T = 181.89$ & CCDr-MCP & CCDr-$\ell_1$ & MMHC & PC \\ 
  \midrule
  P & 122.05 & 97.36 & 82.71 & 86.41 \\ 
  TP & \bf65.14 & 47.40 & 44.71 & 46.70 \\ 
  R & 35.75 & 34.89 & \bf31.40 & 33.17 \\ 
  FP & 21.16 & 15.07 & 6.60 & \bf6.54 \\ 
  SHD (DAG) & \bf137.91 & 149.56 & 143.78 & 141.72 \\ 
  SHD (skeleton) & \bf102.16 & 114.67 & 112.38 & 108.55 \\ 
  TPR & \bf0.36 & 0.26 & 0.25 & 0.26 \\ 
  FDR & 0.47 & 0.51 & \bf0.46 & \bf0.46 \\ 

  \toprule
$p=500$, $T = 460.21$ & CCDr-MCP & CCDr-$\ell_1$ & MMHC & PC \\ 
  \midrule
  P & 319.94 & 252.56 & 195.07 & 202.64 \\ 
  TP & \bf172.34 & 121.75 & 101.49 & 104.33 \\ 
  R & 88.51 & 89.33 & \bf75.50 & 82.60 \\ 
  FP & 59.09 & 41.49 & 18.09 & \bf15.71 \\ 
  SHD (DAG) & \bf346.96 & 379.95 & 376.81 & 371.60 \\ 
  SHD (skeleton) & \bf258.45 & 290.62 & 301.31 & 289.00 \\ 
  TPR & \bf0.37 & 0.26 & 0.22 & 0.23 \\ 
  FDR & \bf0.46 & 0.52 & 0.48 & 0.49 \\ 
   \bottomrule
\end{tabular}
\label{table:highdim}
\end{center}
\end{table}

By comparison, CCDr-$\ell_1$ estimates fewer edges, obtaining lower sensitivity, and more closely mirrors the performance of PC and MMHC. This discrepancy in the performance of concave and $\ell_1$ regularization in high dimensions highlights the advantages of concave regularization and supports the conclusions in the literature on sparse regression. This is not altogether surprising since our framework is closely tied to the Gaussian linear model and regression analysis.

Comparing Tables~\ref{table:lowdim}~and~\ref{table:highdim} when $p=100,200$, we also see that the CCDr methods are more robust to smaller sample sizes. When $p=200$, for example, the increase in SHD from low- to high-dimensions is roughly 9 edges for CCDr-MCP, 13 edges for CCDr-$\ell_1$, 43 edges for MMHC, and 45 edges for PC. Similar patterns are observed for $p=100$, and for other metrics as well. This confirms what we already know about constraint-based methods: they are more reliable when sample sizes are large. Moreover, in spite of the fact that GES and HC were omitted from the high-dimensional experiments, we of course do not expect \emph{improved} performance when $n$ decreases. These observations confirm our expectations that regularization can improve the performance of structure learning algorithms in high-dimensions, with concave regularization providing a noticeable improvement upon $\ell_1$ regularization.

\subsubsection{Timing comparison}
\label{subsubsec:timing}
\begin{figure}
\centering
\includegraphics[width=0.9\textwidth]{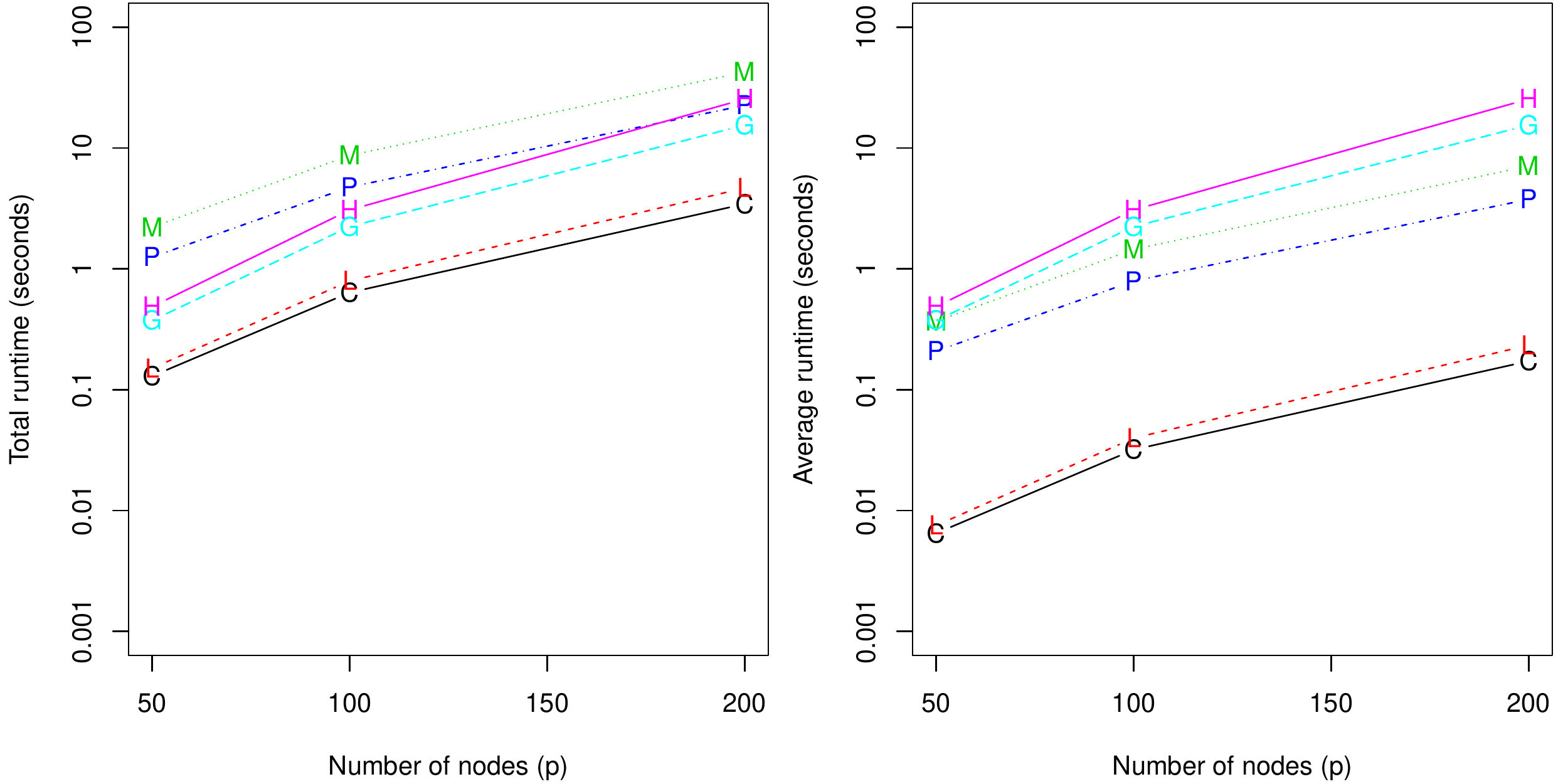}
\vspace{-0.5em}
\caption{Timing comparison in low dimensions for all six algorithms (C~=~CCDr-MCP, L~=~CCDr-$\ell_1$, G~=~GES, H~=~HC, M~=~MMHC, P~=~PC).}
\label{plot:timinglow}

\vspace{1em}
\includegraphics[width=0.9\textwidth]{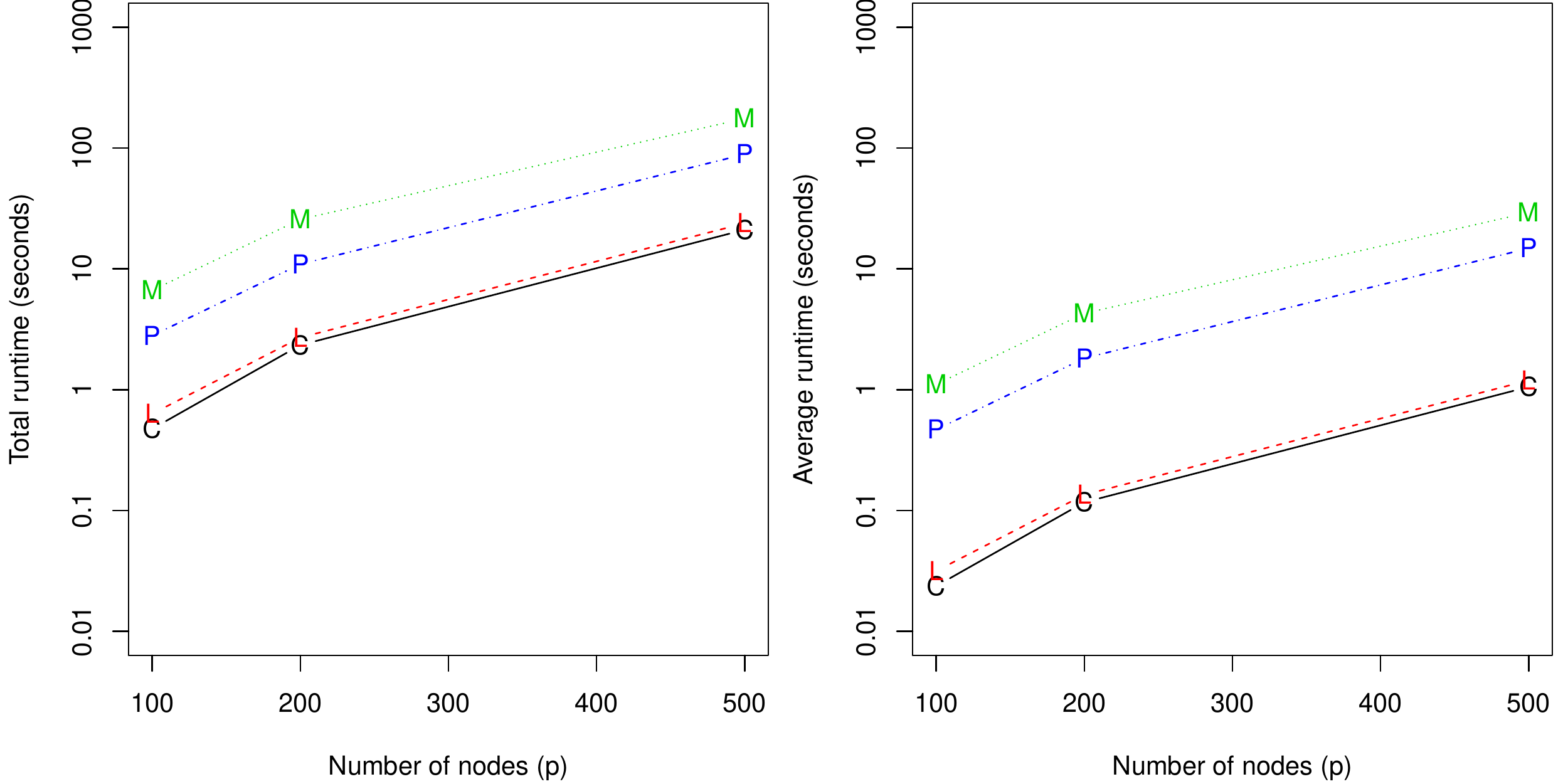}
\vspace{-0.5em}
\caption{Timing comparison in high dimensions, excluding GES and HC (C~=~CCDr-MCP, L~=~CCDr-$\ell_1$, M~=~MMHC, P~=~PC).}
\label{plot:timinghigh}
\end{figure}

A comparison of the total and average runtimes for all the algorithms is provided by Figures~\ref{plot:timinglow}~and~\ref{plot:timinghigh}. The results are displayed graphically here; detailed tables can be found in the Supplementary Materials (Tables~S1 and S2). In both settings, the CCDr methods are significantly faster than the other approaches, particularly the score-based approaches.

In low-dimensions, both GES and HC produce a single DAG estimate and take 15s and 25s, respectively, to estimate graphs with 200 nodes. This is compared with 3-5s for both CCDr-MCP and CCDr-$\ell_1$, in which time both methods compute approximately 20 estimates. Amongst all the compared methods, the fastest alternative is the PC algorithm, however, the difference in timing is still roughly an order of magnitude: When $p=200$, PC takes a little less than 4s on average for a single estimate, whereas CCDr takes approximately \emph{one-fifth of a second} per estimate. This translates to a total runtime of less than 4s for 20 CCDr estimates---faster than the time to compute a single model, on average, for the PC algorithm. Furthermore, CCDr-MCP is slightly faster than CCDr-$\ell_1$, although the difference is small (less than 2s in total runtime).

Similar observations continue to hold in high-dimensions up to the tested limit of $p=500$. For the largest graphs tested, the CCDr methods are still the fastest, taking 21s (CCDr-MCP) and 23s (CCDr-$\ell_1$) to estimate a full solution path on average. MMHC takes almost twice as long as the PC algorithm, requiring almost 3 minutes of total runtime versus 88s for PC. In terms of \emph{average} runtime, the differences are still over an order of magnitude: one second for each CCDr method versus 14.8s (PC) and 29.3s (MMHC). Interestingly, both PC and MMHC are significantly faster in high-dimensions than in low-dimensions (see Tables~S1 and S2 in the Supplementary Materials), which we suspect is due to how these algorithms scale with $n$: datasets with more samples require more time to process (see Section~\ref{subsec:disc} for more details).

Combined with the improved performance in high-dimensions (Section~\ref{subsubsec:hdresults}), these results support our claim that CCDr is an improvement in both timing and accuracy over existing methods for high-dimensional data when $p\le 500$. To see how CCDr performs when $p>500$, we will show in the next subsection that the CCDr algorithm scales efficiently to high-dimensional problems with thousands of variables with almost no loss in reconstruction accuracy.


\subsection{Large graphs}
\label{subsec:timing}
The previous section offered a detailed assessment of the performance of the CCDr algorithm when $p\le 500$. In order to test how our algorithm scales as the number of nodes increases, we ran further tests up to $p=2000$ using CCDr-MCP. The purpose of these tests is to show how the proposed method scales as $p$ increases in terms of timing and accuracy. Since the timing is acutely dependent on the relationship between the dimension, the sparsity of the true graph, and the number of samples, we opted to compare the timing over random choices of the latter two parameters. This also gives us a sense of how the algorithm performs when faced with a more realistic scenario in which the relationship between $p$, $s_0$, and $n$ can be unpredictable. Specifically, we ran $N=20$ tests with the following parameters:
\begin{itemize}
\item $p\in\{100, 200, 500, 1000, 1500, 2000\}$;
\item $s_0 / p \in\{0.2,0.3,0.4,\ldots, 2\}$;
\item $n / p \in\{0.1,0.2,0.3,\ldots, 5\}$.
\end{itemize}

\begin{figure}
\centering
\includegraphics[width=5.5in]{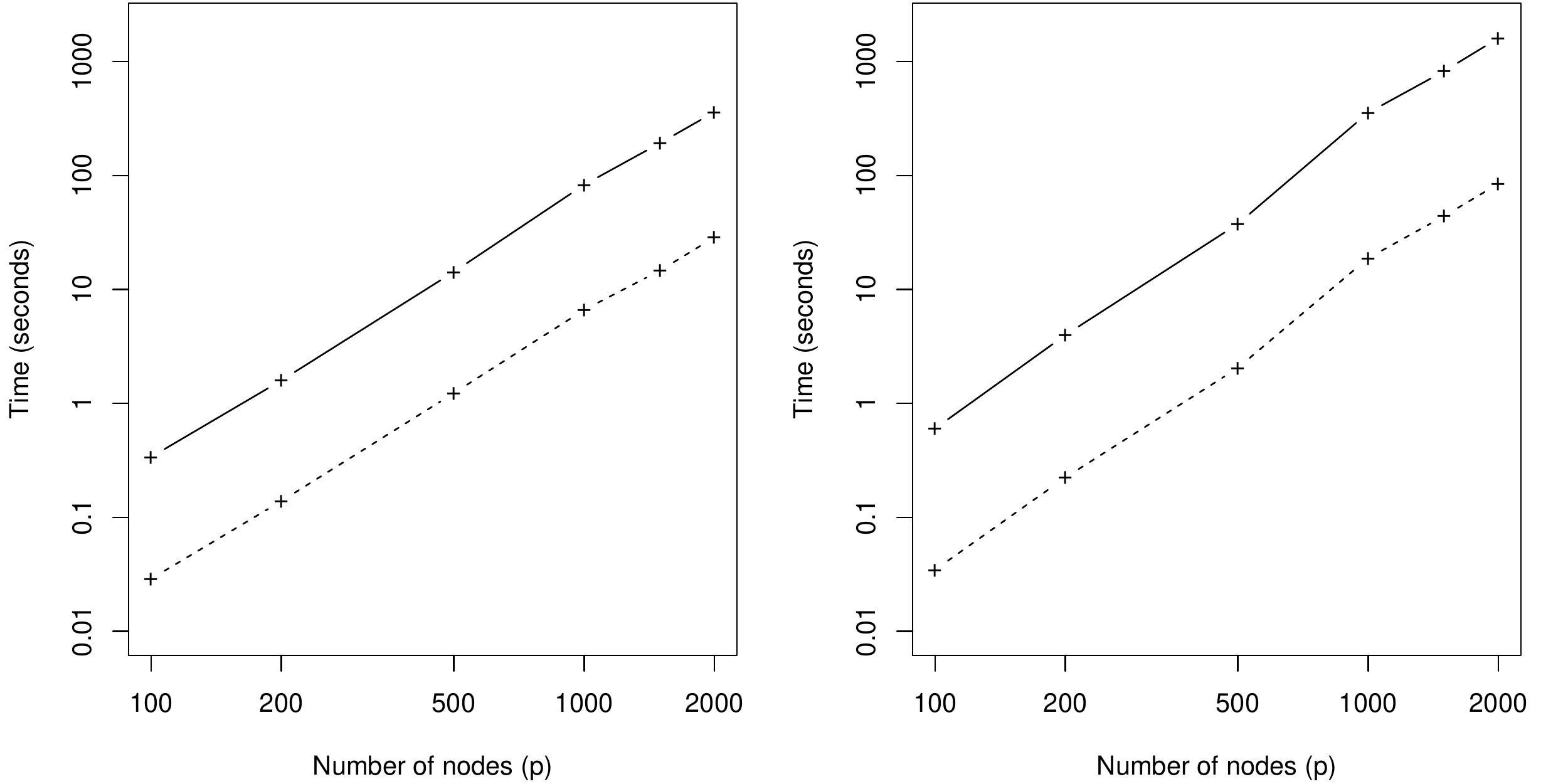}
\caption{Timing data for CCDr-MCP up to $p=2000$. The solid line is the total runtime and the dashed line is the average runtime. (left) Time to estimate graphs with at most $p$ edges; (right) Full runtime with edge threshold $\alpha=3$.}
\label{plot:timing}
\end{figure}

\noindent
The parameters $s_0$ and $n$ were chosen randomly from the above sets in each test, which resulted in an average sparsity level of $s_0/p=1.06$. The results are displayed in Table~\ref{table:timing} and Figure~{\ref{plot:timing}}. Since the timing of the algorithm depends crucially on the total number of models estimated, and also on the threshold parameter $\alpha$, we have plotted both the total and average runtimes for two scenarios: The time it took to estimate DAGs with up to $p$ edges, and then the full running time with the edge threshold set at $\alpha=3$. When $p=1000$, the total running time is just under six minutes, with an average time per model of about 20 seconds. When $p=2000$, the total running time is just under thirty minutes, with an average time per model of about 85 seconds. 

\begin{table}
\caption{Average estimation performance of CCDr-MCP from Section~\ref{subsec:timing}, averaged over $N=20$ random choices of $s_0$ and $n$ for each $p$.}
\begin{center}
\begin{tabular}{lcccccc}
  \toprule
Number of nodes ($p$) & 100 & 200 & 500 & 1000 & 1500 & 2000 \\ 
  \midrule
  Number of samples ($n$) & 114 & 190 & 520 & 1280 & 1470 & 2260 \\ 
  T & 83.15 & 237.15 & 538.15 & 1186.35 & 1550.15 & 2057.95 \\ 
  P & 66.15 & 191.90 & 488.30 & 1082.20 & 1434.20 & 1926.90 \\ 
  TP & 36.15 & 111.50 & 279.80 & 636.70 & 854.25 & 1156.10 \\ 
  R & 20.75 & 46.45 & 115.80 & 226.45 & 323.75 & 447.90 \\ 
  FP & 9.25 & 33.95 & 92.70 & 219.05 & 256.20 & 322.90 \\ 
  SHD (DAG) & 56.25 & 159.60 & 351.05 & 768.70 & 952.10 & 1224.75 \\ 
  SHD (skeleton) & 35.50 & 113.15 & 235.25 & 542.25 & 628.35 & 776.85 \\ 
  TPR & 0.43 & 0.47 & 0.52 & 0.54 & 0.55 & 0.56 \\ 
  FDR & 0.45 & 0.42 & 0.43 & 0.41 & 0.40 & 0.40 \\
   \bottomrule
\end{tabular}
\label{table:timing}
\end{center}
\end{table}

In terms of accuracy, Table~\ref{table:timing} shows that the results are comparable to those in Section~\ref{subsec:exprandom}. Furthermore, as $p$ increases we notice that TPR increases while FDR decreases, which is likely due to the increased number of samples (on average) as $p$ increases; when $p=100$, there were $n=114$ samples on average vs. $n=2260$ when $p=2000$. Combined with the timing data in Figure~\ref{plot:timing}, this confirms that CCDr scales efficiently in terms of both $n$ and $p$ when the underlying graph is sparse.

After these experiments in this work were completed, the performance of our method was further improved, so that the total runtime for $p=2000$ is now less than five minutes.\footnote{A comprehensive comparison of the updated implementation vs. the numbers reported here can be found in Figure~S1 in the Supplementary Materials.} These changes were made to the underlying codebase, and \emph{not} to the algorithm, thus the improvements were purely in terms of code efficiency. Using this updated implementation, we can report that our method has been successfully tested on graphs with up to 8000 nodes, with comparable accuracy to the results exhibited in Table~\ref{table:timing}. The total runtime for 20 estimates was 75 minutes, which may be compared with the 13 days reported for MMHC on a graph with $p=5000$ in \cite{tsamardinos2006}. Regarding the internal implementation of our method, we did not make use of an internal cache, memoization, or efficient data structures (i.e. besides standard vectors), all of which are common strategies used in existing methods. It stands to reason that an optimized implementation would yield even faster results. For instance, we perform the acyclicity check statically with each edge addition; one could imagine a more sophisticated strategy such as incremental topological sorting would lead to significant performance enhancements.

\subsection{Model selection}
\label{subsec:modelselect}
Thus far, we have used the ``best estimate'' according to distance from the true graph, measured by SHD, in order to select models from the estimated solution paths for CCDr, MMHC, and PC. This choice provides a consistent comparison, but results in relatively sparse estimates since missing edges are penalized equally against false positives. One of the advantages of CCDr is that it is able to estimate models with higher sensitivity much more efficiently than PC or MMHC. Alternatively, one could use empirical model selection techniques such as BIC or cross-validation. It has already been noted that these empirical model selection techniques are suboptimal in high-dimensions, particularly for graphical models. This has been previously reported in the literature, see for instance \cite{fu2013}. Here we briefly discuss the results of some tests to confirm this behaviour for our method.

Using both conventional BIC and the extended BIC for high-dimensional problems developed in \cite{foygel2010}, we selected the tuning parameters for CCDr-MCP, CCDr-$\ell_1$, PC, and MMHC. The results confirm that BIC tends to select models with too many edges by insufficiently penalizing the model complexity, consistent with Figure~\ref{plot:loglik}. One may ask if all the algorithms suffer equally, and the answer is no. For the reasons already discussed, we were not able to test the performance of either PC or MMHC for $\alpha>0.05$, which is the regime in which more edges tend to be selected. Thus, in using BIC to select the significance level, the maximum value of $\alpha=0.05$ was over-represented. We suspect that if we had run PC and MMHC with $\alpha>0.05$ in order to produce estimates with extraneous edges, BIC would also select these models. As a result of these limitations, in selecting models based on BIC, CCDr appeared to perform worse relative to either PC or MMHC than reported in previous sections.

To correct for this, we ran the same model selection test using BIC as the selection criterion, but this time restricting the set of CCDr candidates to those with at most as many edges as the most produced by either the PC algorithm or the MMHC algorithm. Using the same data as in Section~\ref{subsubsec:ldresults}, the results resemble those previously reported (Table S3 in the Supplementary Materials). Across the board, graphs with more edges were selected, but the qualitative observations between CCDr and PC / MMHC remain the same.

\subsection{Further discussion}
\label{subsec:disc}

The experiments and results described already, while providing a general overview of the performance of the algorithms tested, also raise several questions which we address briefly in this section.

While we tested a variety of sparsity levels in Section~\ref{subsec:exprandom}, we did not provide a detailed assessment of how the performance of the algorithms varied as the sparsity increases or decreases. An analysis of the effect of sparsity shows that the same qualitative behaviour observed in Sections~\ref{subsubsec:ldresults}~and~\ref{subsubsec:hdresults} persists (see Figures~S3~and~S4 in the Supplementary Materials). Generally speaking, even after controlling for sparsity, CCDr-MCP is the most accurate in high-dimensions and the constraint-based methods are more accurate in low-dimensions. As before, CCDr-MCP still outperforms CCDr-$\ell_1$ after controlling for sparsity. We do observe a small decrease in reconstruction accuracy for the CCDr methods when the graph is more dense ($s_0/p=2$); improving our method when the true graph is more dense remains for future work.

For the CCDr algorithm, in order to provide a reasonable balance of complexity and efficiency in the resulting estimation problem, we fixed $\gamma = 2$. Nonetheless, this parameter was observed to have a non-negligible effect on the results and a more in-depth study in the future would account for the effect of this parameter. Another parameter which we have not discussed is the maximum neighbourhood size in the true graph, which we controlled in our simulations by controlling the expected neighbourhood size. Keeping the neighbourhoods small is critical for keeping the running time of the PC algorithm reasonable. Further simulations in which we allowed each node to have arbitrarily many parents showed that the running time of the CCDr algorithm does not depend on this parameter. Moreover, restricting the maximum size of the conditioning sets used in the conditional independence tests in the PC algorithm, as suggested by the work of \cite{anandkumar2012}, also had a negligible effect. Finally, both PC and MMHC show relatively poor computational complexity with respect to the sample size $n$, with more instances requiring more time to process. Our tests indicate that the complexity of CCDr is essentially independent of $n$---the only dependence on sample size enters through the computation of the correlation matrix in the first step. 

\section{Real Networks}
\label{sec:app}
While the random set-up in the previous section provided a convenient setting to test many random structures quickly and efficiently, random graphs may not be good representatives of realistic network structures. For this reason, we augmented these experiments with tests on real network structures, using both simulated and scientific (unsimulated) data. Our first experiment uses network structures from the Bayesian Network Repository,\footnote{\texttt{http://www.cs.huji.ac.il/site/labs/compbio/Repository/}} a standardized collection of networks which is commonly used as a benchmark for structure learning methods, as well as a simulated scale-free network. In order to assess the impact of these methods on actual scientific data, we also compare the performance of the algorithms on the well-known flow cytometry dataset (\cite{sachs2005}). 

\subsection{Bayesian network repository}
\label{subsec:bnrep}

All of the networks examined in this experiment were loaded using the \texttt{bnlearn} package.\footnote{A mirror of the repository used by the \texttt{bnlearn} package can be found at: \texttt{http://www.bnlearn.com/bnrepository/}.} We then used the graph structures to generate  data according to a structural equation model, as in the previous section. Furthermore, in order to keep the focus on high-dimensional estimation, we fixed the number of samples at $n=50$, which narrowed the choice of networks to those that satisfy $p>50$. Seven such network structures were tested, to which we added one randomly generated scale-free structure with 200 nodes. The scale-free network was created using the \texttt{igraph} package. For each network, we generated random coefficients in the interval $[0.5,1]$ for each edge as before and generated a single random dataset with unit variances for testing. This procedure was replicated $N=50$ times, and the number of true positives and false positives were tracked for each algorithm. We also increased the length of the regularization path used for the CCDr methods to 50 estimates while keeping both PC and MMHC fixed at six estimates for each graph. Based on the results in the previous section---particularly with respect to timing---both HC and GES were excluded from these tests.

\begin{figure}
\includegraphics[width=\textwidth]{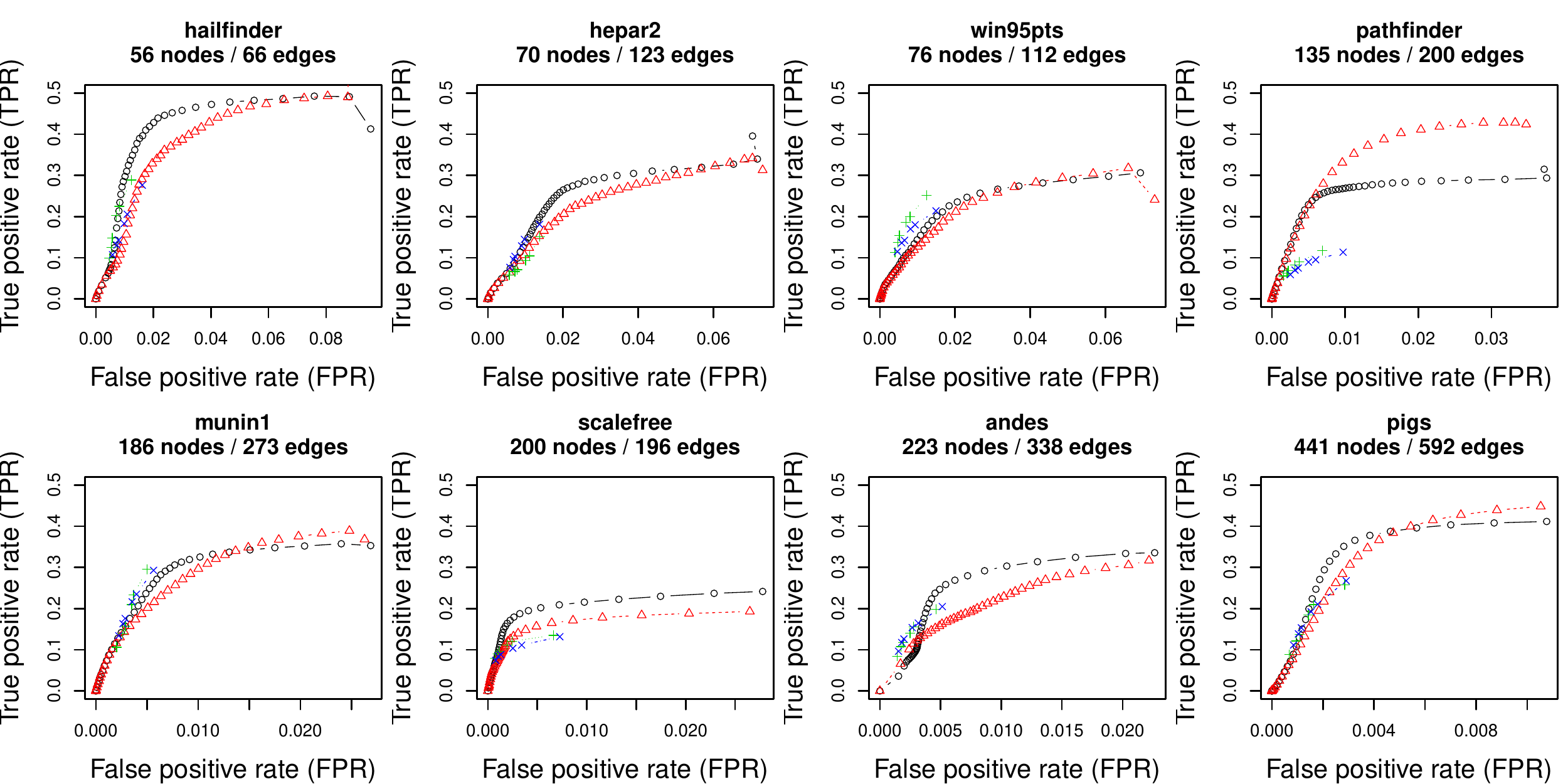}
\caption{ROC curves for real networks (black $\bigcirc$~=~CCDr-MCP, red $\bigtriangleup$~=~CCDr-$\ell_1$, blue $\times$~=~MMHC, green $+$~=~PC).}
\label{plot:bnrep}
\end{figure}

We have already observed in Section~{\ref{subsec:modelselect}} how traditional model selection techniques such as BIC and cross-validation perform very poorly. For this reason, we chose to present the results graphically by their ROC curves in order to compare the true positive rate against the false positive rate as a function of the tuning parameters. The resulting ROC curves are displayed in Figure~\ref{plot:bnrep}.

In terms of reconstruction accuracy, with only one exception, we see that the CCDr methods perform as well or better than the other methods in these experiments. Consistent with the previously reported experiments on random graphs, the CCDr methods tend to show higher sensitivity with comparable false positive rates in high dimensions. In some cases the improvements are dramatic---for instance, \texttt{pathfinder}, \texttt{scalefree}, and \texttt{pigs}. The one exception is the \texttt{win95pts} network, in which the PC algorithm attains slightly higher sensitivity and lower FDR compared with the CCDr methods as well as MMHC. These results further highlight the tradeoffs in learning between each approach and confirm the patterns observed previously in the literature: constraint-based methods tend to miss edges in the true skeleton, resulting in lower false discovery rates and lower sensitivity, whereas regularization tends to increase overall sensitivity with the risk of higher false positive rates if the amount of regularization is not calibrated properly.

More interesting is the comparison between CCDr-MCP and CCDr-$\ell_1$. Compared with the simulation results in Section~\ref{sec:results}, there is a more pronounced difference between the performance of concave vs $\ell_1$ regularization, with the former outperforming the latter. This is most visible in the \texttt{hailfinder} and \texttt{pigs} networks, where both methods show comparable sensitivity but CCDr-MCP exhibits lower false positive rates. The only network in which $\ell_1$ regularization is preferable is \texttt{pathfinder}, where CCDr-$\ell_1$ obtains higher sensitivity later in the solution path.

Consistent with the previous experiments, however, the main advantages of CCDr come in the form of efficiency: When averaged across all 8 networks, the total runtime of CCDr-MCP was 8.5x faster than PC and 15x faster than MMHC. In terms of average runtime, CCDr-MCP is roughly 58x and 100x faster, respectively. See Figure~S2 in the Supplementary Materials for the detailed runtime comparisons. Recall that for these networks, unlike in the previous experiments, the estimated solution path for the CCDr methods is 2.5 times longer, with up to 50 estimates per solution path. For example, on average the \texttt{pathfinder} network with $p=135$ nodes took 110x and 150x longer per estimate to compute, respectively, for PC and MMHC. At the other end of the spectrum, the hardest graph to reconstruct was the \texttt{pigs} network, which took 39s for CCDr-MCP, 29s for CCDr-$\ell_1$, 71s for PC, and 147s for MMHC. In both cases CCDr-MCP easily did the best job reconstructing the true networks.

\subsection{Application to real data}

We analyzed the well-known flow cytometry dataset, generated by \cite{sachs2005}, which has been previously analyzed by \cite{fu2013,shojaie2010,friedman2008} among others. The dataset contains $n=7466$ measurements of $p=11$ continuous variables corresponding to proteins and phospholipids in human immune system cells. The underlying network, constructed through a careful series of biological experiments, has $s_0=20$ edges, and represents a gold-standard for comparison currently accepted by the biology community. Hereafter, we regard this consensus network as the true network in order to assess the algorithms. While this dataset is hardly high-dimensional, it represents one of the few continuous datasets for which we have oracular knowledge of the true underlying DAG \emph{as well as} real data from which to infer the true structure.

The original dataset contains a mixture of both observational and experimental data. Since the methods presented here assume the data are normally distributed, we first tested the original continuous variables for normality, and much as expected the data were highly non-normal. To correct for this, we applied a logarithm transform, which produced variables that were much closer to Gaussian. This dataset was used for our tests on continuous data.

We also analyzed a discretized version of the dataset containing $n=5400$ measurements, created by transforming the continuous data into three nonnegative levels which correspond to \emph{high}, \emph{medium}, and \emph{low}, so that magnitudes were partially preserved (\cite{sachs2005}).  This dataset is especially interesting for a number of reasons. First, it represents a test of model misspecification: Our method was developed for continuous data, but nothing prevents us from naively feeding this dataset into the algorithm. By treating the three levels as numeric values (\emph{high} = 2, \emph{medium} = 1, \emph{low} = 0), we can compute the correlation matrix and proceed with the second and third steps in Algorithm~\ref{alg:full}. Since the data are clearly not Gaussian, the results of this test give us a sense of how well our method performs on discrete, non-Gaussian data. Second, as a result of postprocessing to clean up the data as well as the discretization itself, it is much less noisy than the original dataset, which provides an interesting side-by-side comparison.

A few changes were made to the set-up used in previous experiments. First, since the number of variables was small, it was feasible to run the constraint-based methods on a longer sequence of significance levels. Thus, we used a sequence of 10 levels: $$\alpha\in\{10^{-6}, 5\times10^{-6},10^{-5}, 5\times10^{-5},10^{-4}, 5\times10^{-4},10^{-3}, 5\times10^{-3},0.01,0.05\}.$$ Furthermore, in a majority of the tests we ran, the PC algorithm was unable to orient all the edges in the final step, leading to a partially directed graph (formally a CPDAG, see Remark~\ref{rem:pcorient}). As a result, we had to modify our metrics to allow for undirected edges. We did this favourably for the PC algorithm by counting an undirected edge as a true edge as long as the same edge exists in the skeleton of the true graph. Any edge that was successfully oriented by the PC algorithm was treated as a directed edge. Finally, we split each dataset in half in order to obtain a testing dataset on which to compute the log-likelihood of the estimated models. Since the PC algorithm was not able to estimate DAGs, log-likelihood scores could not be computed for the continuous dataset.

\begin{figure}
\centering
\includegraphics[width=0.3\textwidth]{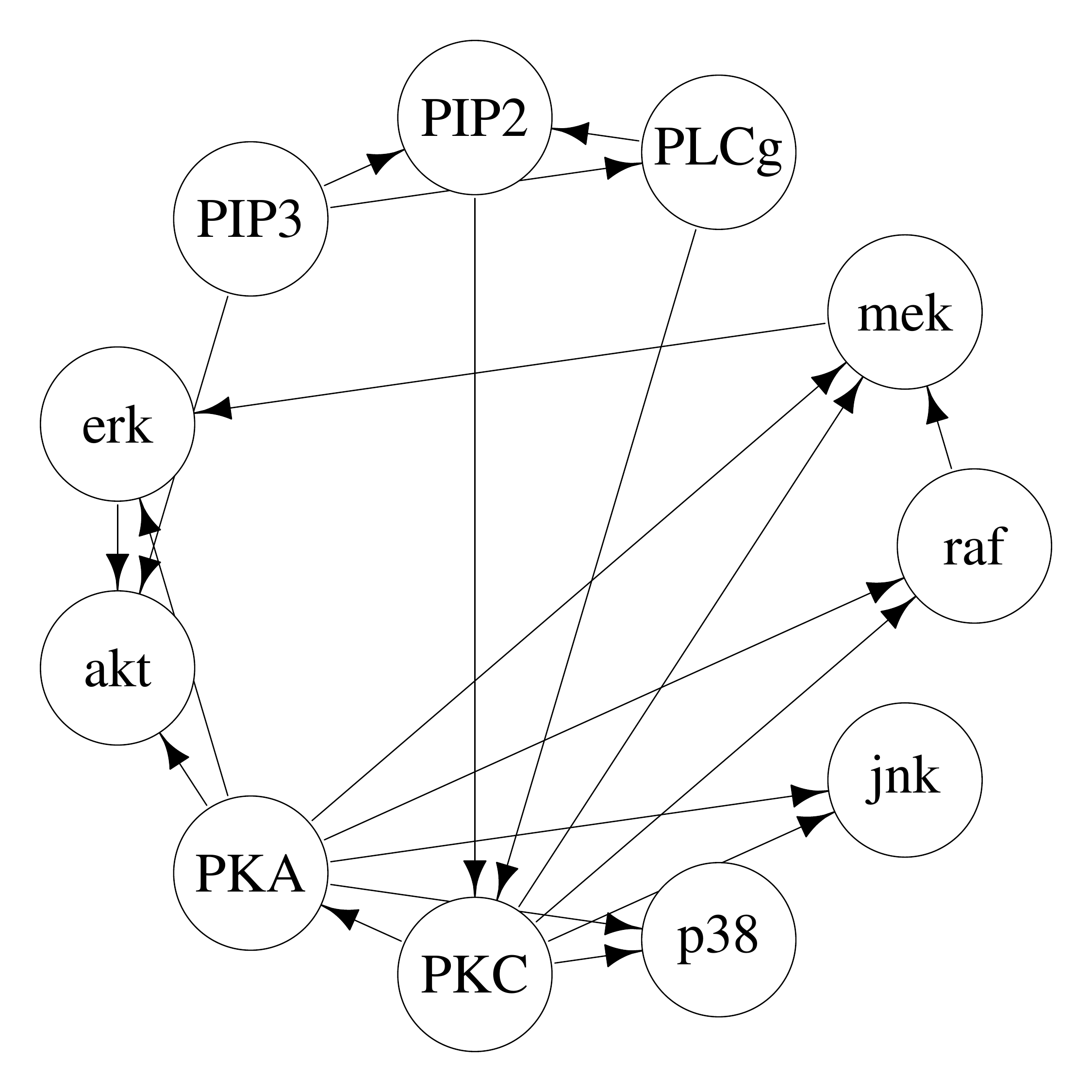}
\includegraphics[width=0.3\textwidth]{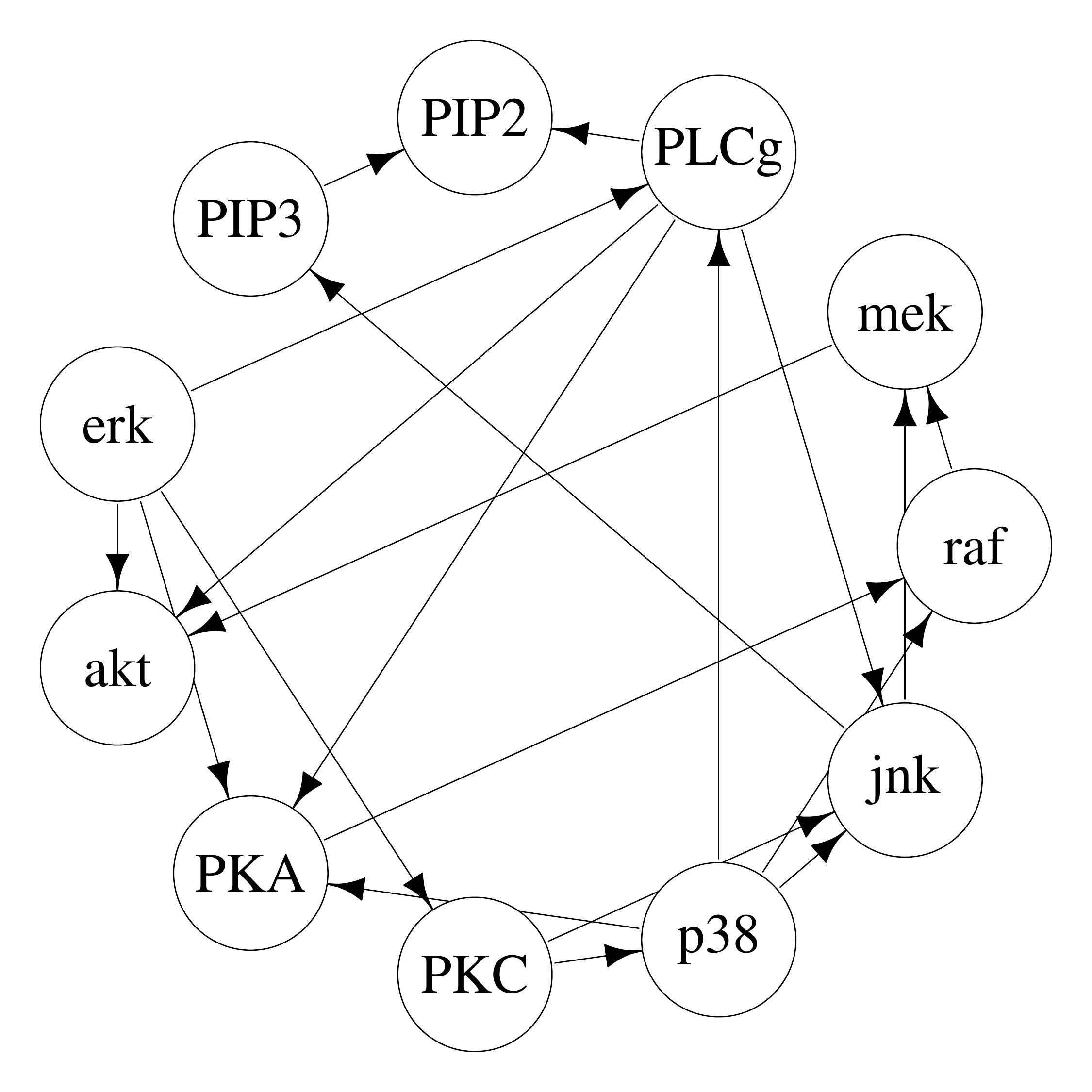}
\includegraphics[width=0.3\textwidth]{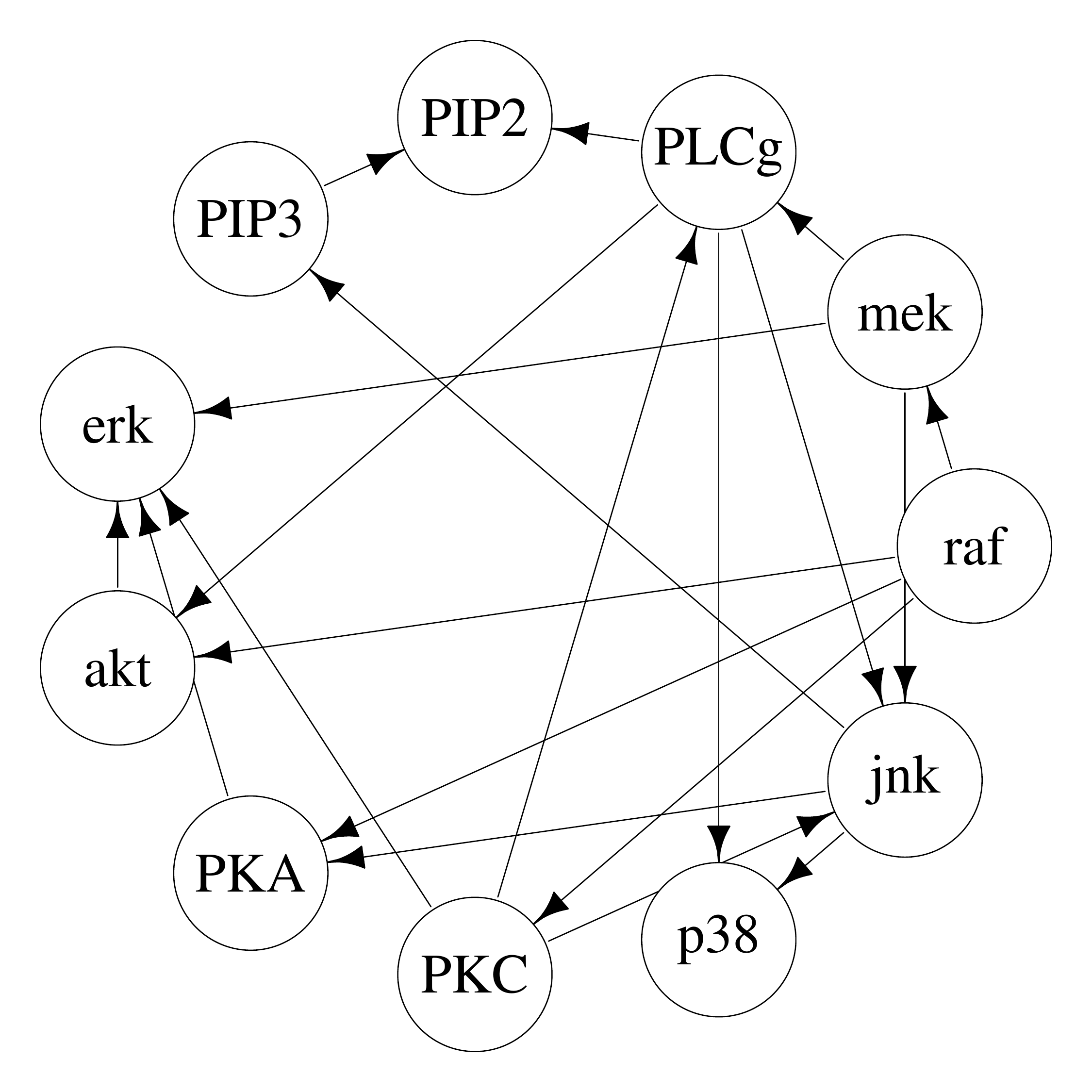}
\caption{Comparison of the consensus network (left) against the DAGs estimated by the CCDr-MCP algorithm for both datasets: (middle) Log-transformed continuous dataset; (right) Discretized dataset.}
\label{fig:cytodags}
\end{figure}

Tables~\ref{table:cytologcts}~and~\ref{table:cytodiscrete} summarize the results for a sample run, which are indicative of the general behaviour when different random splits are tested. Instead of selecting the best estimates as in Section~\ref{sec:results}, we chose estimates with comparable numbers of edges, selected to match the true graph as closely as possible with $s_0=20$. The results for CCDr-MCP are visualized in Figure~\ref{fig:cytodags}. Both GES and HC consistently estimated too many edges, which matches the behaviour observed in Section~\ref{subsubsec:ldresults}. For the continuous dataset, CCDr-MCP and MMHC perform the best with almost identical metrics, while for the discrete dataset CCDr-MCP is clearly optimal with fewer false positives and smaller SHD across the board. This indicates that even though this method was developed with continuous Gaussian data in mind, it can still be applied to discrete data with reasonable results.

\begin{table}
\caption{Structure estimation performance for all algorithms using the log-transformed continuous cytometry data.}
\centering
\begin{tabular}{lcccccc}
  \toprule
$p=11$, $T=20$ & CCDr-MCP & CCDr-$\ell_1$ & GES & HC & MMHC & PC \\ 
  \midrule
  P & 20 & 20 & 41 & 38 & 20 & 20 \\ 
  TP & 7 & 7 & 9 & \bf10 & 7 & 7 \\ 
  R & 2 & \bf1 & 7 & 6 & 2 & 2 \\ 
  FP & \bf11 & 12 & 25 & 22 & \bf11 & \bf11 \\ 
  SHD (DAG) & \bf24 & 25 & 36 & 32 & \bf24 & 25 \\ 
  SHD (skeleton) & \bf22 & 24 & 29 & 26 & \bf22 & \bf22 \\ 
  Test Log-likelihood & -2.05 & -2.19 & \bf-0.34 & -1.09 & -2.03 & --- \\ 
   \bottomrule
\end{tabular}
\label{table:cytologcts}

\caption{Structure estimation performance for all algorithms using discretized cytometry data.}
\centering
\begin{tabular}{lcccccc}
  \toprule
$p=11$, $T=20$ & CCDr-MCP & CCDr-$\ell_1$ & GES & HC & MMHC & PC \\ 
  \midrule
  P & 20 & 20 & 43 & 35 & 20 & 20 \\ 
  TP & 6 & 3 & \bf13 & 7 & 3 & 6 \\ 
  R & 5 & 6 & 4 & 7 & 5 & \bf2 \\ 
  FP & \bf9 & 11 & 26 & 21 & 12 & 12 \\ 
  SHD (DAG) & \bf23 & 28 & 33 & 34 & 29 & 26 \\ 
  SHD (skeleton) & \bf18 & 22 & 29 & 27 & 24 & 24 \\ 
  Test Log-likelihood & -0.68 & -1.86 & \bf-0.10 & 0.18 & -2.32 & -2.01 \\ 
   \bottomrule
\end{tabular}
\label{table:cytodiscrete}
\end{table}

Due to the small size of the graph with only $p=11$ nodes, the differences in timing are largely negligible, taking fractions of a second to complete. Because of this, the processor time is subject to fluctuations in low-level bottlenecks most likely unrelated to the core algorithms themselves, and so we do not report exact times here. At a high level we did observe that HC and GES show much improved performance relative to PC and MMHC, however, the CCDr methods are still consistently the fastest.

\section{Conclusion}
\label{sec:conclusion}
We have introduced a general penalized likelihood framework for estimating sparse Bayesian networks, along with a fast algorithm that is easily implemented on a personal computer. In the finite dimensional scenario, the resulting estimator has good theoretical properties. Through a series of tests designed to test the limits of this new algorithm, we have shown that our approach accurately estimates networks with 2000 nodes while scaling efficiently to handle networks with up to 8000 nodes. The proposed method is compatible with high-dimensional data where $p\gg n$, and outperforms many existing methods in both speed and accuracy in this regime. Tests on real networks have validated the performance and applicability of this method in a variety of domains.

The central theme of exploiting convexity to solve nonconvex problems is an intriguing prospect for the development of new algorithms in statistics and machine learning. Indeed, the main difficulties with nonconvex regularization are computational in nature. Although recent progress has broken this barrier in the case of least squares regression, to our knowledge the algorithm presented here is one of the first to approximate this type of nonconvex optimization problem---whose dimension scales quadratically---when $p$ is in the thousands. Moreover, since our method revolves around a continuous optimization problem, we avoid approaches that rely on individual edge additions and removals, which are intrinsically discrete. Such approaches tend to scale very poorly as the number of nodes increases due to the combinatorial nature of the algorithms. Unlike these methods, future advances in nonconvex optimization will directly affect how we solve the maximum likelihood problem presented here.

We have already indicated that the performance of the method has been improved even further without any changes to the core algorithm itself. Notwithstanding, there are several potential improvements to the algorithm that remain promising, such as adaptive and stochastic coordinate descent. It also remains to incorporate prior knowledge either via whitelists and blacklists, or through a more sophisticated hybrid Bayesian approach. As research into nonconvex optimization is a rapidly developing area, the methods presented here merely scratch the surface of how these techniques can be applied to the structure learning problem for Bayesian networks.

\subsection*{Acknowledgements}
We would like to thank the referees for their helpful comments. We would also like to thank Marco Scutari and Sara van de Geer for their thoughtful discussions, as well as Damon Alexander for his assistance and suggestions in implementing the algorithm. This work was supported by NSF grants DMS-1055286 and DMS-1308376 and NSF graduate research fellowships DGE-1144087 and DGE-0707424.


\setcounter{section}{1}
\renewcommand{\thesection}{\Alph{section}}
\section*{Appendix}

\subsection{Formal preliminaries}
\label{app:formal}
Conceptually our theory is quite simple: we have a function $F$ on $\R^{p^2}$ which we would like to maximize over a subset defined by the space of DAGs, $\mathcal{D}$. In order to properly specify a topology for this space, and to ensure that the translation between our statistical model for $(B,\Omega)$ and the mathematical model for $\bn$ is coherent, we carefully outline the mathematical set-up here.

Given a DAG $(B,\Omega)$, consider the reparametrization $(\Phi,R)$ given by 
\begin{align}
\Phi
&= B\Omega^{-1/2} \\
R 
&= \Omega^{-1/2}.
\end{align}

\noindent
This is of course just the matrix version of the reparametrization that leads to \eqref{eq:reloglik}. Now define the following function which maps $(\Phi,R)\in\R^{p\times p}\times\R^{p\times p}$ into $\R^{p^2}$:
\begin{align*}
\bn(\Phi,R)
&= \vec(U)
= (u_1,\ldots,u_p),
\quad\text{where $U = [u_1\|\ldots\|u_p] = R+\Phi$}.
\end{align*}

\noindent
Recall that $\Phi$ has zeroes on the diagonal and $R$ is a diagonal matrix, so that the sum $U:=R+\Phi$ has the same number of nonzero entries as $R$ and $\Phi$ separately. Furthermore, the sparsity pattern of the off-diagonal elements of $U$ exactly matches that of $\Phi$. 

In the proofs, when there is no confusion we will simply write $\bn=U=(\Phi,R)=(B,\Omega)$ to mean that these are all equivalent representations of the same DAG in various parametrizations. In particular, for any $\bn_0\in\mathcal{E}_0$, we have $\bn_0=U_0=(\Phi_0,R_0)=(B_0,\Omega_0)$. Mathematically, we will work with $\bn$, however, our results should always be interpreted in terms of the original model $(B,\Omega)$.

We formally define the space of DAGs as follows:
\begin{align*}
\mathcal{D}
:=\left\{\bn = \bn(\Phi,R)\in\R^{p^2}:\Phi\in\R^{p\times p} \text{ is a DAG},\,\rho_j>0 \text{ for all $j$}\right\}.
\end{align*}

\noindent
This space inherits its topology from the ambient space $\R^{p^2}$, and it is this space on which we wish to maximize the function $F(\bn)=\ell_n(\bn)-n\pln(\bn)$. 

\subsection{Proof of Theorem \ref{thm:main}}
We begin by formalizing some of the background material on the Cholesky decomposition used in Section~\ref{subsec:perm}, which will also be used in the proof of Lemma~\ref{lem:unique}. First recall the following standard result:

\begin{lemma}
\label{lem:cholperm}
For any symmetric positive definite matrix $A\in\R^{p\times p}$ and permutation $\pi\in\mathcal{P}$, the Cholesky decomposition $A=LDL^T$ satisfies
\begin{align*}
P_\pi A=(P_\pi L)(P_\pi D) (P_\pi L)^T,
\end{align*}

\noindent
where $L$ is lower triangular and $D$ is a diagonal matrix.
\end{lemma}


\noindent
Now suppose $\Theta$ is given and use the Cholesky decomposition to write $\Theta = \Theta(L,D)$ as in \eqref{eq:chol}. Then, taking $A=\Theta(L, D)$ in Lemma~\ref{lem:cholperm}, we obtain $P_\pi\Theta(L,D)=\Theta(P_\pi L, P_\pi D)$. Alternatively, suppose $(B,\Omega)\in\mathcal{E}(\Theta)$ and suppose $\pi\in\mathcal{P}$ is compatible with $(B,\Omega)$. Since $P_\pi B$ is lower-triangular, by taking $A=\Theta(P_\pi B,P_\pi \Omega)$, we may similarly deduce
\begin{align*}
P_{\pi^{-1}}\Theta(P_\pi B,P_\pi \Omega)
&= \Theta(B,\Omega)
\implies
\Theta(P_\pi B,P_\pi \Omega)
= P_{\pi}\Theta(B,\Omega).
\end{align*}

\noindent
This proves the following lemma, which will be useful:

\begin{lemma}
\label{lem:dagperm}
Let $(B,\Omega)$ be a DAG. For any permutation $\pi\in\mathcal{P}$ that is compatible with $(B,\Omega)$, we have
\begin{align*}
P_\pi \Theta(B,\Omega) = \Theta(P_\pi B,P_\pi \Omega).
\end{align*}
\end{lemma}

We now prove Lemma~\ref{lem:unique}, which will be used in the proof of the Theorem~\ref{thm:main}.

\vspace{1em}\noindent
{\bf Proof of Lemma~\ref{lem:unique}}
We only prove this for the original parametrization $(B,\Omega)$; the reparametrized case is similar.

Since $B_1$ and $B_2$ have a common topological sort, there is a permutation $\pi$ of the vertices that orders $B_1$ and $B_2$ simultaneously, so that $P_\pi B_1$ and $P_\pi B_2$ are both strictly lower triangular. Suppose then that $\Theta(B_1,\Omega_1)=\Theta(B_2,\Omega_2):=\tilde{\Theta}$, so that (using Lemma~\ref{lem:dagperm} above)
\begin{align*}
&P_\pi\Theta(B_1,\Omega_1)
= P_\pi\Theta(B_2,\Omega_2) \\
\iff
&\Theta(P_\pi B_1,P_\pi\Omega_1)
= \Theta(P_\pi B_2,P_\pi\Omega_2) \\
\iff
&(I-P_\pi B_1)(P_\pi \Omega_1)^{-1}(I-P_\pi B_1)^T
= (I-P_\pi B_2)(P_\pi \Omega_2)^{-1}(I-P_\pi B_2)^T.
\end{align*}

\noindent
The last expression is equal to $P_\pi\tilde{\Theta}$, which is a symmetric positive definite matrix. By the uniqueness of the Cholesky factorization, we must have
\begin{align*}
I-P_\pi B_1
&= I-P_\pi B_2 \\
(P_\pi \Omega_1)^{-1} 
&= (P_\pi \Omega_2)^{-1},
\intertext{which implies}
B_1=B_2, &\quad
\Omega_1=\Omega_2.
\end{align*}

\noindent
Since $B_1$ was assumed to be distinct from $B_2$, this contradiction establishes the desired result.
\endprooff

\noindent
{\bf Proof of Theorem~\ref{thm:main}}
Suppose $\bn_0\in\mathcal{E}_0$ with $b_n(\bn_0)\to0$. It suffices to check Conditions (A)-(C) from \cite{fan2001}, which are simply the standard regularity conditions for asymptotic efficiency of ordinary maximum likelihood estimates. Model identifiability is not an issue since the same analysis can be carried out for any equivalent parameter (see Section~\ref{subsec:sketch}). Since the densities $f(\cdot\|\bn)$ are Gaussian, the only condition that needs to be checked is that the Fisher information is positive definite at $\bn_0$ restricted to the DAG space $\mathcal{D}$. Theorem~\ref{thm:main} will then follow immediately from Theorem 1 in \cite{fan2001}.

Let $I(\bn_0)$ denote the usual Fisher information matrix at this point; we will show that $I(\bn_0)$ is positive definite. Since $f$ is always a Gaussian density, it will suffice to show that $f(\cdot\|\bn) \ne f(\cdot\|\bn_0)$ for $\bn$ in a sufficiently small neighbourhood of $\bn_0$.

Now suppose $\bn=(\Phi,R)$ is in an arbitrarily small neighbourhood of $\bn_0=(\Phi_0,R_0)$. Then it must hold that $\phi_{ij}\phi_{ij}^0>0$ whenever $\phi_{ij}^0\ne0$. Indeed, otherwise
\begin{align*}
\norm{\Phi-\Phi_0}^2
&\ge (\phi_{ij}-\phi_{ij}^0)^2
\ge |\phi_{ij}^0|^2.
\end{align*}

\noindent
Thus, $\phi_{ij}^0\ne0$ implies $\phi_{ij}\ne0$, or $i\to j$ in $\Phi_0$ implies $i\to j$ in any DAG close to $\Phi_0$. In particular, $\Phi$ contains all the edges (including orientation) in $\Phi_0$, with the possible addition of extra edges. That is, $\Phi_0$ is a subgraph of $\Phi$. It follows that there is an ordering of the vertices that is compatible with $\Phi$ and $\Phi_0$ simultaneously. Since $\Phi\ne \Phi_0$, it follows from Lemma~\ref{lem:unique} that $\Theta(\bn)\ne\Theta(\bn_0)$, whence $f(\cdot\|\bn) \ne f(\cdot\|\bn_0).$ 
\endprooff

\noindent
{\bf Proof of Lemma~\ref{lem:sep}}
Note that Lemma~\ref{lem:eqclass} implies that the equivalence class $\mathcal{E}_0$ is finite. Set $\eps=\min_{\bn_0\in\mathcal{E}_0}\min_{i,j}\{|\phi_{ij}^0|^2:\phi_{ij}^0\ne0\}>0$. Then if $\norm{\Phi-\Phi_0}\le\norm{\bn-\bn_0}<\eps$, the arguments in the proof of Theorem~\ref{thm:main} guarantee the existence of an ordering that is compatible with $\Phi$ and $\Phi_0$, and the result follows from Lemma~\ref{lem:unique}.
\endprooff

\subsection{Proof of Theorem \ref{thm:global}}
Instead of directly proving Theorem~\ref{thm:global}, we will prove a slightly more general statement under weaker assumptions. Theorem~\ref{thm:global} will then follow as a special case. 

The following technical lemmas ensure that the objective function $F(\bn)$ is well-behaved with respect to taking limits. The first is a standard application of the uniform law of large numbers (see, for example, \sec16 in \cite{ferguson1996}) and the second is a direct consequence of concavity.

\begin{lemma}\label{lm:loglikas}
Fix $\bn_0$ and suppose $\bn_n$ is a sequence with $\norm{\bn_n-\bn_0}=o(1)$. If the empirical log-likelihood $\ell_n(\bn)$ is continuous for all $n$, then 
\begin{align*}
P\left(\lim_{n\to\infty}\frac{1}{n}\ell_n(\bn_n) 
= \lim_{n\to\infty}\frac{1}{n}\ell_n(\bn_0)\right)
=1.
\end{align*}
\end{lemma}

\begin{lemma}
\label{lemma:pen}
Suppose that $\pl(t)$ is nondecreasing and concave for $t\geq 0$ with $\pl(0) = 0$. If $\limsup_n\tau(\lambda_n)<\infty$, then for any $x_0>0$ there exists a constant $C$, depending only on $x_0$, such that
\begin{align*}
|\pln(x)-\pln(x_0)|\le C|x-x_0|
\quad\text{for all $x\ge0$ and all $n$.}
\end{align*}
\end{lemma}

Recall that $f(n)=\omega(g(n))\iff g(n)=o(f(n))$, that is, for every $C>0$,
\begin{align*}
f(n) \ge Cg(n) \quad\text{for all large $n$}.
\end{align*}

\noindent
As in Section~\ref{sec:thy}, we use $\widehat{\bn}_n$ and $\widehat{\bn}^*_n$ to denote the local maximizers close to $\bn_0$ and $\bn^*$, respectively, whose existence is guaranteed by Theorem~\ref{thm:main}.

\begin{thm}
\label{thm:genglobal}
Suppose that $\pl(t)$ is nondecreasing and concave for $t\geq 0$ with $\pl(0)=0$. Let $\bn_0\in\mathcal{E}_0$ be a DAG with strictly more edges than $\bn^*$. Assume further that the conditions for Theorem~\ref{thm:sparsity} hold for both $\bn_0$ and $\bn^*$. If
\begin{enumerate}
\item $c_n(\bn^*) = \tau(\lambda_n)+O(n^{-1/2})$ and $c_n(\bn_0) = \tau(\lambda_n)+O(n^{-1/2})$,
\item $\limsup_n\tau(\lambda_n)<\infty$,
\item $\tau(\lambda_n) = \omega(n^{-1/2})$,
\end{enumerate}

\noindent
then for every $\eps>0$,
\begin{align*}
P\left(\ell_n(\widehat{\bn}_n^*) - n\,\pln(\widehat{\bn}_n^*) > \ell_n(\widehat{\bn}_n) - n\,\pln(\widehat{\bn}_n)\right)
\ge 1-\eps
\quad\text{for sufficiently large $n$.}
\end{align*}
\end{thm}

\begin{prooff}
Since we assume Theorem~\ref{thm:sparsity} holds for both $\bn_0$ and $\bn^*$, we may assume without loss of generality that $\supp(\widehat{\bn}^*_n) = \supp(\bn^*)$ and $\supp(\widehat{\bn}_n) = \supp(\bn_0)$. 

Since $\ell_n$ is continuous for each $n$, $\norm{\widehat{\bn}_n - \bn_0} = O_P(n^{-1/2})$, and $\norm{\widehat{\bn}_n^* - \bn^*} = O_P(n^{-1/2})$, Lemma~\ref{lm:loglikas} implies that 
\begin{align*}
\frac{1}{n}(\ell_n(\widehat{\bn}_n) - \ell_n(\widehat{\bn}_n^*))\to0
\end{align*}
almost surely. It is easy to show that in fact $n^{-1}(\ell(\widehat{\bn}_n) - \ell(\widehat{\bn}_n^*))=O_P(n^{-1/2})$.

It will suffice to show that for any $\eps>0$, there exists an $N$ such that for all $n>N$, we have
\begin{align*}
P\left(\pln(\widehat{\bn}_n) - \pln(\widehat{\bn}_n^*) - \frac{1}{n}(\ell_n(\widehat{\bn}_n) - \ell_n(\widehat{\bn}_n^*))
>0\right) \ge 1-\eps.
\end{align*}
Given $\eps>0$, there exists $M>0$ such that
\begin{align*}
P\left(\frac{1}{n}(\ell_n(\widehat{\bn}_n) - \ell_n(\widehat{\bn}_n^*)) \le Mn^{-1/2}\right)
\ge 1-\eps,
\end{align*}
so that it suffices to check that $\pln(\widehat{\bn}_n) - \pln(\widehat{\bn}_n^*) > Mn^{-1/2}$ for sufficiently large $n$.

Lemma~\ref{lemma:pen} implies that for each $\phi_{ij}^0\ne0$,
\begin{align*}
|\pln(\widehat{\phi}_{ij}^0) - \pln(\phi_{ij}^0)|
\le C|\widehat{\phi}_{ij}^0-\phi_{ij}^0|
=O(n^{-1/2}),
\end{align*}
and similarly for all $\phi_{ij}^*\ne0$. Thus we can write $\pln(\widehat{\bn}_n)=\pln(\bn_0)+O_P(n^{-1/2})$ and similarly for $\widehat{\bn}^*$. It thus suffices to show that
\begin{align*}
&\pln(\bn_0) - \pln(\bn^*)=\omega(n^{-1/2}).
\end{align*}
Now, using Condition 1, 
\begin{align*}
\pln(\bn_0) - \pln(\bn^*)
&=\sum_{\phi_{ij}^0\ne 0}\pln(|\phi^0_{ij}|) - \sum_{\phi_{ij}^*\ne 0}\pln(|\phi^*_{ij}|)\\
&\ge s_0c_n(\bn_0) - s^*\tau(\lambda_n) + s^*\tau(\lambda_n) - \sum_{\phi_{ij}^*\ne 0}\pln(|\phi^*_{ij}|)\\
&= (s_0-s^*)\tau(\lambda_n) + O(n^{-1/2}) + \sum_{\phi_{ij}^*\ne 0}(\tau(\lambda_n)-\pln(\phi^*_{ij}))\\
&\ge (s_0-s^*)\tau(\lambda_n) + O(n^{-1/2}).
\end{align*}
Since $\tau(\lambda_n)=\omega(n^{-1/2})$ (Condition 3), it follows that
$\pln(\bn_0) - \pln(\bn^*)\ge \omega(n^{-1/2})$,
from which the claim follows.
\end{prooff}

\noindent
{\bf Proof of Theorem~\ref{thm:global}}
Condition 3 in Theorem~\ref{thm:genglobal} is equivalent to $\tau(\lambda_n)/n^{-1/2}\to\infty$, and Theorem~\ref{thm:global} follows as a special case since the equivalence class $\mathcal{E}_0$ is finite.
\endprooff


\bibliography{concavedagbib}

\begin{thebibliography}{72}
\providecommand{\natexlab}[1]{#1}
\providecommand{\url}[1]{\texttt{#1}}
\expandafter\ifx\csname urlstyle\endcsname\relax
  \providecommand{\doi}[1]{doi: #1}\else
  \providecommand{\doi}{doi: \begingroup \urlstyle{rm}\Url}\fi

\bibitem[Aliferis et~al.(2010{\natexlab{a}})Aliferis, Statnikov, Tsamardinos,
  Mani, and Koutsoukos]{aliferis2010a}
Constantin~F Aliferis, Alexander Statnikov, Ioannis Tsamardinos, Subramani
  Mani, and Xenofon~D Koutsoukos.
\newblock Local causal and {M}arkov blanket induction for causal discovery and
  feature selection for classification {P}art {I}: {A}lgorithms and empirical
  evaluation.
\newblock \emph{The Journal of Machine Learning Research}, 11:\penalty0
  171--234, 2010{\natexlab{a}}.

\bibitem[Aliferis et~al.(2010{\natexlab{b}})Aliferis, Statnikov, Tsamardinos,
  Mani, and Koutsoukos]{aliferis2010b}
Constantin~F Aliferis, Alexander Statnikov, Ioannis Tsamardinos, Subramani
  Mani, and Xenofon~D Koutsoukos.
\newblock Local causal and {M}arkov blanket induction for causal discovery and
  feature selection for classification {P}art {II}: {A}nalysis and extensions.
\newblock \emph{The Journal of Machine Learning Research}, 11:\penalty0
  235--284, 2010{\natexlab{b}}.

\bibitem[Anandkumar et~al.(2012)Anandkumar, Tan, Huang, and
  Willsky]{anandkumar2012}
Animashree Anandkumar, Vincent~YF Tan, Furong Huang, and Alan~S Willsky.
\newblock High-dimensional {G}aussian graphical model selection: {W}alk
  summability and local separation criterion.
\newblock \emph{The Journal of Machine Learning Research}, 13\penalty0
  (1):\penalty0 2293--2337, 2012.

\bibitem[Anandkumar et~al.(2013)Anandkumar, Hsu, Javanmard, and
  Kakade]{anandkumar2013}
Animashree Anandkumar, Daniel Hsu, Adel Javanmard, and Sham Kakade.
\newblock Learning linear {B}ayesian networks with latent variables.
\newblock In \emph{Proceedings of The 30th International Conference on Machine
  Learning}, pages 249--257, 2013.

\bibitem[Banerjee et~al.(2008)Banerjee, El~Ghaoui, and
  d'Aspremont]{banerjee2008}
Onureena Banerjee, Laurent El~Ghaoui, and Alexandre d'Aspremont.
\newblock Model selection through sparse maximum likelihood estimation for
  multivariate {G}aussian or binary data.
\newblock \emph{The Journal of Machine Learning Research}, 9:\penalty0
  485--516, 2008.

\bibitem[Boyd and Vandenberghe(2009)]{boyd2009}
Stephen Boyd and Lieven Vandenberghe.
\newblock \emph{Convex optimization}.
\newblock {C}ambridge {U}niversity {P}ress, 2009.

\bibitem[B{\"u}hlmann and van~de Geer(2011)]{buhlmann2011}
Peter B{\"u}hlmann and Sara van~de Geer.
\newblock \emph{Statistics for high-dimensional data: {M}ethods, theory and
  applications}.
\newblock Springer, 2011.

\bibitem[B{\"u}hlmann et~al.(2013)B{\"u}hlmann, Peters, and
  Ernest]{buhlmann2013}
Peter B{\"u}hlmann, Jonas Peters, and Jan Ernest.
\newblock {CAM}: {C}ausal {A}dditive {M}odels, high-dimensional order search
  and penalized regression.
\newblock \emph{arXiv preprint arXiv:1310.1533}, 2013.

\bibitem[Chaudhuri et~al.(2007)Chaudhuri, Drton, and Richardson]{chaudhuri2007}
Sanjay Chaudhuri, Mathias Drton, and Thomas~S Richardson.
\newblock Estimation of a covariance matrix with zeros.
\newblock \emph{Biometrika}, 94\penalty0 (1):\penalty0 199--216, 2007.

\bibitem[Chickering(1996)]{chickering1996}
David~Maxwell Chickering.
\newblock Learning {B}ayesian networks is {NP}-complete.
\newblock In \emph{Learning from data}, pages 121--130. Springer, 1996.

\bibitem[Chickering(2003)]{chickering2003}
David~Maxwell Chickering.
\newblock Optimal structure identification with greedy search.
\newblock \emph{The Journal of Machine Learning Research}, 3:\penalty0
  507--554, 2003.

\bibitem[Chickering and Meek(2002)]{chickering2002}
David~Maxwell Chickering and Christopher Meek.
\newblock Finding optimal {B}ayesian networks.
\newblock In \emph{Proceedings of the Eighteenth conference on Uncertainty in
  artificial intelligence}, pages 94--102. Morgan Kaufmann Publishers Inc.,
  2002.

\bibitem[Dempster(1972)]{dempster1972}
Arthur~P Dempster.
\newblock Covariance selection.
\newblock \emph{Biometrics}, 28\penalty0 (1):\penalty0 157--175, 1972.

\bibitem[Dempster(1969)]{dempster1969}
Arthur~Pentland Dempster.
\newblock \emph{Elements of continuous multivariate analysis}, volume 388.
\newblock Addison-Wesley Reading, Mass., 1969.

\bibitem[Drton and Richardson(2008)]{drton2008}
Mathias Drton and Thomas~S Richardson.
\newblock Graphical methods for efficient likelihood inference in {G}aussian
  covariance models.
\newblock \emph{The Journal of Machine Learning Research}, 9:\penalty0
  893--914, 2008.

\bibitem[Drton et~al.(2011)Drton, Foygel, and Sullivant]{drton2011}
Mathias Drton, Rina Foygel, and Seth Sullivant.
\newblock Global identifiability of linear structural equation models.
\newblock \emph{The Annals of Statistics}, 39\penalty0 (2):\penalty0 865--886,
  2011.

\bibitem[Ellis and Wong(2008)]{ellis2008}
Byron Ellis and Wing~Hung Wong.
\newblock Learning causal {B}ayesian network structures from experimental data.
\newblock \emph{Journal of the American Statistical Association}, 103\penalty0
  (482), 2008.

\bibitem[Fan and Li(2001)]{fan2001}
Jianqing Fan and Runze Li.
\newblock Variable selection via nonconcave penalized likelihood and its oracle
  properties.
\newblock \emph{Journal of the American Statistical Association}, 96\penalty0
  (456):\penalty0 1348--1360, 2001.

\bibitem[Fan and Lv(2010)]{fan2010}
Jianqing Fan and Jinchi Lv.
\newblock A selective overview of variable selection in high dimensional
  feature space.
\newblock \emph{Statistica Sinica}, 20\penalty0 (1):\penalty0 101, 2010.

\bibitem[Fan and Lv(2011)]{fan2011}
Jianqing Fan and Jinchi Lv.
\newblock Nonconcave penalized likelihood with {NP}-dimensionality.
\newblock \emph{Information Theory, IEEE Transactions on}, 57\penalty0
  (8):\penalty0 5467--5484, 2011.

\bibitem[Fan and Lv(2013)]{fan2013}
Yingying Fan and Jinchi Lv.
\newblock Asymptotic equivalence of regularization methods in thresholded
  parameter space.
\newblock \emph{Journal of the American Statistical Association}, 108\penalty0
  (503):\penalty0 1044--1061, 2013.

\bibitem[Ferguson(1996)]{ferguson1996}
Thomas~Shelburne Ferguson.
\newblock \emph{A course in large sample theory}, volume~38.
\newblock CRC Press, 1996.

\bibitem[Foygel and Drton(2010)]{foygel2010}
Rina Foygel and Mathias Drton.
\newblock Extended {B}ayesian information criteria for {G}aussian graphical
  models.
\newblock In \emph{Advances in Neural Information Processing Systems}, pages
  604--612, 2010.

\bibitem[Friedman et~al.(2007)Friedman, Hastie, H{\"o}fling, and
  Tibshirani]{friedman2007}
Jerome Friedman, Trevor Hastie, Holger H{\"o}fling, and Robert Tibshirani.
\newblock Pathwise coordinate optimization.
\newblock \emph{The Annals of Applied Statistics}, 1\penalty0 (2):\penalty0
  302--332, 2007.

\bibitem[Friedman et~al.(2008)Friedman, Hastie, and Tibshirani]{friedman2008}
Jerome Friedman, Trevor Hastie, and Robert Tibshirani.
\newblock Sparse inverse covariance estimation with the {G}raphical {L}asso.
\newblock \emph{Biostatistics}, 9\penalty0 (3):\penalty0 432--441, 2008.

\bibitem[Friedman et~al.(2010)Friedman, Hastie, and Tibshirani]{friedman2010}
Jerome Friedman, Trevor Hastie, and Rob Tibshirani.
\newblock Regularization paths for generalized linear models via coordinate
  descent.
\newblock \emph{Journal of statistical software}, 33\penalty0 (1):\penalty0 1,
  2010.

\bibitem[Fu and Zhou(2013)]{fu2013}
Fei Fu and Qing Zhou.
\newblock Learning sparse causal {G}aussian networks with experimental
  intervention: {R}egularization and coordinate descent.
\newblock \emph{Journal of the American Statistical Association}, 108\penalty0
  (501):\penalty0 288--300, 2013.

\bibitem[Fu and Zhou(2014)]{fu2014}
Fei Fu and Qing Zhou.
\newblock Penalized estimation of sparse directed acyclic graphs from
  categorical data under intervention.
\newblock \emph{arXiv preprint arXiv:1403.2310}, 2014.

\bibitem[G{\'a}mez et~al.(2011)G{\'a}mez, Mateo, and Puerta]{gamez2011}
Jos{\'e}~A G{\'a}mez, Juan~L Mateo, and Jos{\'e}~M Puerta.
\newblock Learning {B}ayesian networks by hill climbing: {E}fficient methods
  based on progressive restriction of the neighborhood.
\newblock \emph{Data Mining and Knowledge Discovery}, 22\penalty0
  (1-2):\penalty0 106--148, 2011.

\bibitem[G{\'a}mez et~al.(2012)G{\'a}mez, Mateo, and Puerta]{gamez2012}
Jos{\'e}~A G{\'a}mez, Juan~L Mateo, and Jos{\'e}~M Puerta.
\newblock One iteration {CHC} algorithm for learning {B}ayesian networks: {A}n
  effective and efficient algorithm for high dimensional problems.
\newblock \emph{Progress in Artificial Intelligence}, 1\penalty0 (4):\penalty0
  329--346, 2012.

\bibitem[Geiger and Heckerman(2013)]{geiger2013}
Dan Geiger and David Heckerman.
\newblock Learning {G}aussian networks.
\newblock \emph{arXiv preprint arXiv:1302.6808}, 2013.

\bibitem[Heckerman et~al.(1995)Heckerman, Geiger, and
  Chickering]{heckerman1995}
David Heckerman, Dan Geiger, and David~M Chickering.
\newblock Learning {B}ayesian networks: {T}he combination of knowledge and
  statistical data.
\newblock \emph{Machine learning}, 20\penalty0 (3):\penalty0 197--243, 1995.

\bibitem[Huang et~al.(2012)Huang, Breheny, and Ma]{huang2012}
Jian Huang, Patrick Breheny, and Shuangge Ma.
\newblock A selective review of group selection in high-dimensional models.
\newblock \emph{Statistical science: a review journal of the Institute of
  Mathematical Statistics}, 27\penalty0 (4), 2012.

\bibitem[Huang et~al.(2006)Huang, Liu, Pourahmadi, and Liu]{huang2006}
Jianhua~Z Huang, Naiping Liu, Mohsen Pourahmadi, and Linxu Liu.
\newblock Covariance matrix selection and estimation via penalised normal
  likelihood.
\newblock \emph{Biometrika}, 93\penalty0 (1):\penalty0 85--98, 2006.

\bibitem[Hyv{\"a}rinen et~al.(2010)Hyv{\"a}rinen, Zhang, Shimizu, and
  Hoyer]{hyvarinen2010}
Aapo Hyv{\"a}rinen, Kun Zhang, Shohei Shimizu, and Patrik~O Hoyer.
\newblock Estimation of a structural vector autoregression model using
  non-{G}aussianity.
\newblock \emph{The Journal of Machine Learning Research}, 11:\penalty0
  1709--1731, 2010.

\bibitem[Kalisch and B{\"u}hlmann(2007)]{kalisch2007}
Markus Kalisch and Peter B{\"u}hlmann.
\newblock Estimating high-dimensional directed acyclic graphs with the
  {PC}-algorithm.
\newblock \emph{The Journal of Machine Learning Research}, 8:\penalty0
  613--636, 2007.

\bibitem[Kalisch et~al.(2012)Kalisch, M{\"a}chler, Colombo, Maathuis, and
  B{\"u}hlmann]{kalisch2012}
Markus Kalisch, Martin M{\"a}chler, Diego Colombo, Marloes~H Maathuis, and
  Peter B{\"u}hlmann.
\newblock Causal inference using graphical models with the {R} package pcalg.
\newblock \emph{Journal of Statistical Software}, 47\penalty0 (11):\penalty0
  1--26, 2012.

\bibitem[Lam and Fan(2009)]{lam2009}
Clifford Lam and Jianqing Fan.
\newblock Sparsistency and rates of convergence in large covariance matrix
  estimation.
\newblock \emph{The Annals of Statistics}, 37\penalty0 (6B):\penalty0 4254,
  2009.

\bibitem[Lam and Bacchus(1994)]{lam1994}
Wai Lam and Fahiem Bacchus.
\newblock Learning {B}ayesian belief networks: {A}n approach based on the {MDL}
  principle.
\newblock \emph{Computational intelligence}, 10\penalty0 (3):\penalty0
  269--293, 1994.

\bibitem[Lauritzen(1996)]{lauritzen1996}
Steffen~L Lauritzen.
\newblock \emph{Graphical models}.
\newblock Oxford University Press, 1996.

\bibitem[Loh and B{\"u}hlmann(2013)]{loh2013}
Po-Ling Loh and Peter B{\"u}hlmann.
\newblock High-dimensional learning of linear causal networks via inverse
  covariance estimation.
\newblock \emph{arXiv preprint arXiv:1311.3492}, 2013.

\bibitem[Lv and Fan(2009)]{lv2009}
Jinchi Lv and Yingying Fan.
\newblock A unified approach to model selection and sparse recovery using
  regularized least squares.
\newblock \emph{The Annals of Statistics}, 37\penalty0 (6A):\penalty0
  3498--3528, 2009.

\bibitem[Mazumder et~al.(2011)Mazumder, Friedman, and Hastie]{mazumder2011}
Rahul Mazumder, Jerome~H Friedman, and Trevor Hastie.
\newblock Sparse{N}et: {C}oordinate descent with nonconvex penalties.
\newblock \emph{Journal of the American Statistical Association}, 106\penalty0
  (495):\penalty0 1125--1138, 2011.

\bibitem[Meinshausen and B{\"u}hlmann(2006)]{meinshausen2006}
Nicolai Meinshausen and Peter B{\"u}hlmann.
\newblock High-dimensional graphs and variable selection with the {L}asso.
\newblock \emph{The Annals of Statistics}, 34\penalty0 (3):\penalty0
  1436--1462, 2006.

\bibitem[Peters and B{\"u}hlmann(2012)]{peters2012samevar}
Jonas Peters and Peter B{\"u}hlmann.
\newblock Identifiability of {G}aussian structural equation models with same
  error variances.
\newblock \emph{arXiv preprint arXiv:1205.2536}, 2012.

\bibitem[Peters et~al.(2012)Peters, Mooij, Janzing, and
  Sch{\"o}lkopf]{peters2012nonlinear}
Jonas Peters, Joris Mooij, Dominik Janzing, and Bernhard Sch{\"o}lkopf.
\newblock Identifiability of causal graphs using functional models.
\newblock \emph{arXiv preprint arXiv:1202.3757}, 2012.

\bibitem[Pourahmadi(2013)]{pourahmadi2013}
Mohsen Pourahmadi.
\newblock \emph{High-dimensional covariance estimation}.
\newblock John Wiley \& Sons, 2013.

\bibitem[{R Core Team}(2014)]{rcore2014}
{R Core Team}.
\newblock \emph{{R}: {A} Language and Environment for Statistical Computing}.
\newblock R Foundation for Statistical Computing, Vienna, Austria, 2014.
\newblock URL \url{http://www.R-project.org}.

\bibitem[Raskutti and Uhler(2014)]{raskutti2014}
Garvesh Raskutti and Caroline Uhler.
\newblock Learning directed acyclic graphs based on sparsest permutations.
\newblock \emph{arXiv preprint arXiv:1307.0366}, 2014.

\bibitem[Ravikumar et~al.(2011)Ravikumar, Wainwright, Raskutti, and
  Yu]{ravikumar2011}
Pradeep Ravikumar, Martin~J Wainwright, Garvesh Raskutti, and Bin Yu.
\newblock High-dimensional covariance estimation by minimizing
  $\ell_1$-penalized log-determinant divergence.
\newblock \emph{Electronic Journal of Statistics}, 5:\penalty0 935--980, 2011.

\bibitem[Redner(1981)]{redner1981}
Richard Redner.
\newblock Note on the consistency of the maximum likelihood estimate for
  nonidentifiable distributions.
\newblock \emph{The Annals of Statistics}, 9\penalty0 (1):\penalty0 225--228,
  1981.

\bibitem[Robinson(1977)]{robinson1977}
Robert~W Robinson.
\newblock Counting unlabeled acyclic digraphs.
\newblock In \emph{Combinatorial mathematics V}, pages 28--43. Springer, 1977.

\bibitem[R{\"u}timann and B{\"u}hlmann(2009)]{rutimann2009}
Philipp R{\"u}timann and Peter B{\"u}hlmann.
\newblock High dimensional sparse covariance estimation via directed acyclic
  graphs.
\newblock \emph{Electronic Journal of Statistics}, 3:\penalty0 1133--1160,
  2009.

\bibitem[Sachs et~al.(2005)Sachs, Perez, Pe'er, Lauffenburger, and
  Nolan]{sachs2005}
Karen Sachs, Omar Perez, Dana Pe'er, Douglas~A Lauffenburger, and Garry~P
  Nolan.
\newblock Causal protein-signaling networks derived from multiparameter
  single-cell data.
\newblock \emph{Science}, 308\penalty0 (5721):\penalty0 523--529, 2005.

\bibitem[Schmidt et~al.(2007)Schmidt, Niculescu-Mizil, and Murphy]{schmidt2007}
Mark Schmidt, Alexandru Niculescu-Mizil, and Kevin Murphy.
\newblock Learning graphical model structure using {L1}-regularization paths.
\newblock In \emph{AAAI}, volume~7, pages 1278--1283, 2007.

\bibitem[Scutari(2010)]{scutari2010}
Marco Scutari.
\newblock Learning {B}ayesian networks with the bnlearn {R} package.
\newblock \emph{Journal of Statistical Software}, 35\penalty0 (i03), 2010.

\bibitem[Scutari(2014)]{scutari2014}
Marco Scutari.
\newblock Bayesian network constraint-based structure learning algorithms:
  {P}arallel and optimised implementations in the bnlearn {R} package.
\newblock \emph{arXiv preprint arXiv:1406.7648}, 2014.

\bibitem[Shimizu et~al.(2006)Shimizu, Hoyer, Hyv{\"a}rinen, and
  Kerminen]{shimizu2006}
Shohei Shimizu, Patrik~O Hoyer, Aapo Hyv{\"a}rinen, and Antti Kerminen.
\newblock A linear non-{G}aussian acyclic model for causal discovery.
\newblock \emph{The Journal of Machine Learning Research}, 7:\penalty0
  2003--2030, 2006.

\bibitem[Shojaie and Michailidis(2010)]{shojaie2010}
Ali Shojaie and George Michailidis.
\newblock Penalized likelihood methods for estimation of sparse
  high-dimensional directed acyclic graphs.
\newblock \emph{Biometrika}, 97\penalty0 (3):\penalty0 519--538, 2010.

\bibitem[Spirtes and Glymour(1991)]{spirtes1991}
Peter Spirtes and Clark Glymour.
\newblock An algorithm for fast recovery of sparse causal graphs.
\newblock \emph{Social Science Computer Review}, 9\penalty0 (1):\penalty0
  62--72, 1991.

\bibitem[St{\"a}dler et~al.(2010)St{\"a}dler, B{\"u}hlmann, and Van
  De~Geer]{stadler2010}
Nicolas St{\"a}dler, Peter B{\"u}hlmann, and Sara Van De~Geer.
\newblock $\ell_1$-penalization for mixture regression models.
\newblock \emph{Test}, 19\penalty0 (2):\penalty0 209--256, 2010.

\bibitem[Tibshirani(1996)]{tibshirani1996}
Robert Tibshirani.
\newblock Regression shrinkage and selection via the lasso.
\newblock \emph{Journal of the Royal Statistical Society. Series B
  (Methodological)}, pages 267--288, 1996.

\bibitem[Tsamardinos et~al.(2006)Tsamardinos, Brown, and
  Aliferis]{tsamardinos2006}
Ioannis Tsamardinos, Laura~E Brown, and Constantin~F Aliferis.
\newblock The max-min hill-climbing {B}ayesian network structure learning
  algorithm.
\newblock \emph{Machine Learning}, 65\penalty0 (1):\penalty0 31--78, 2006.

\bibitem[van~de Geer and B{\"u}hlmann(2013)]{geer2013}
Sara van~de Geer and Peter B{\"u}hlmann.
\newblock $\ell_0$-penalized maximum likelihood for sparse directed acyclic
  graphs.
\newblock \emph{The Annals of Statistics}, 41\penalty0 (2):\penalty0 536--567,
  2013.

\bibitem[Wang et~al.(2007)Wang, Li, and Tsai]{wang2007}
Hansheng Wang, Runze Li, and Chih-Ling Tsai.
\newblock Tuning parameter selectors for the {S}moothly {C}lipped {A}bsolute
  {D}eviation method.
\newblock \emph{Biometrika}, 94\penalty0 (3):\penalty0 553--568, 2007.

\bibitem[Wu and Lange(2008)]{wu2008}
Tong~Tong Wu and Kenneth Lange.
\newblock Coordinate descent algorithms for {L}asso penalized regression.
\newblock \emph{The Annals of Applied Statistics}, pages 224--244, 2008.

\bibitem[Xiang and Kim(2013)]{xiang2013}
Jing Xiang and Seyoung Kim.
\newblock A* {L}asso for learning a sparse {B}ayesian network structure for
  continuous variables.
\newblock In \emph{Advances in Neural Information Processing Systems}, pages
  2418--2426, 2013.

\bibitem[Zhang(2010)]{zhang2010}
Cun-Hui Zhang.
\newblock Nearly unbiased variable selection under minimax concave penalty.
\newblock \emph{The Annals of Statistics}, 38\penalty0 (2):\penalty0 894--942,
  2010.

\bibitem[Zhang and Zhang(2012)]{zhang2012}
Cun-Hui Zhang and Tong Zhang.
\newblock A general theory of concave regularization for high-dimensional
  sparse estimation problems.
\newblock \emph{Statistical Science}, 27\penalty0 (4):\penalty0 576--593, 2012.

\bibitem[Zhao and Yu(2006)]{zhao2006}
Peng Zhao and Bin Yu.
\newblock On model selection consistency of {L}asso.
\newblock \emph{The Journal of Machine Learning Research}, 7:\penalty0
  2541--2563, 2006.

\bibitem[Zhou(2011)]{zhou2011}
Qing Zhou.
\newblock Multi-domain sampling with applications to structural inference of
  {B}ayesian networks.
\newblock \emph{Journal of the American Statistical Association}, 106\penalty0
  (496):\penalty0 1317--1330, 2011.

\bibitem[Zou(2006)]{zou2006}
Hui Zou.
\newblock The adaptive {L}asso and its oracle properties.
\newblock \emph{Journal of the American Statistical Association}, 101\penalty0
  (476):\penalty0 1418--1429, 2006.

\end{thebibliography}
\label{bib:bib}


\newpage

%
%
\setcounter{topnumber}{3}
\setcounter{figure}{0}
\setcounter{table}{0}

%
%
\renewcommand{\thefigure}{S\arabic{figure}}
\renewcommand{\thetable}{S\arabic{table}}


\maketitle
\vspace*{4em}
\begin{center}
\Large\emph{Supplementary Material}
\end{center}
\vspace{2em}


\subsection*{1 Supplementary timing figures and tables.}
\begin{table}[h]
\centering
\caption{Total runtime (top) and average runtime (bottom) in seconds for all six algorithms from Section~\ref{subsubsec:ldresults}.}
\begin{tabular}{lcccccc}
  \toprule
$p$ & CCDr-MCP & CCDr-$\ell_1$ & GES & HC & MMHC & PC \\ 
  \midrule
50 & 0.13 & 0.15 & 0.37 & 0.49 & 2.21 & 1.25 \\ 
  100 & 0.64 & 0.79 & 2.24 & 3.08 & 8.70 & 4.76 \\ 
  200 & 3.45 & 4.67 & 15.50 & 25.47 & 42.55 & 22.59 \\ 
%
  \toprule
$p$ & CCDr-MCP & CCDr-$\ell_1$ & GES & HC & MMHC & PC \\ 
  \midrule
50 & 0.01 & 0.01 & 0.37 & 0.49 & 0.37 & 0.21 \\ 
  100 & 0.03 & 0.04 & 2.24 & 3.08 & 1.45 & 0.79 \\ 
  200 & 0.17 & 0.23 & 15.50 & 25.47 & 7.09 & 3.77 \\ 
   \bottomrule
\end{tabular}
\end{table}

\begin{table}[h]
\centering
\caption{Total runtime (top) and average runtime (bottom) in seconds for the four algorithms from Section~\ref{subsubsec:hdresults}.}
\begin{tabular}{lcccccc}
  \toprule
$p$ & CCDr-MCP & CCDr-$\ell_1$ & MMHC & PC \\ 
  \midrule
100 & 0.47 & 0.63 & 6.64 & 2.80 \\ 
  200 & 2.34 & 2.70 & 25.71 & 10.93 \\ 
  500 & 21.09 & 23.84 & 175.54 & 88.85 \\ 
%
  \toprule
$p$ & CCDr-MCP & CCDr-$\ell_1$ & MMHC & PC \\ 
  \midrule
100 & 0.02 & 0.03 & 1.11 & 0.47 \\ 
  200 & 0.12 & 0.13 & 4.29 & 1.82 \\ 
  500 & 1.05 & 1.19 & 29.26 & 14.81 \\ 
   \bottomrule
\end{tabular}
\end{table}

\begin{figure}[h]
\centering
\includegraphics[width=0.9\textwidth]{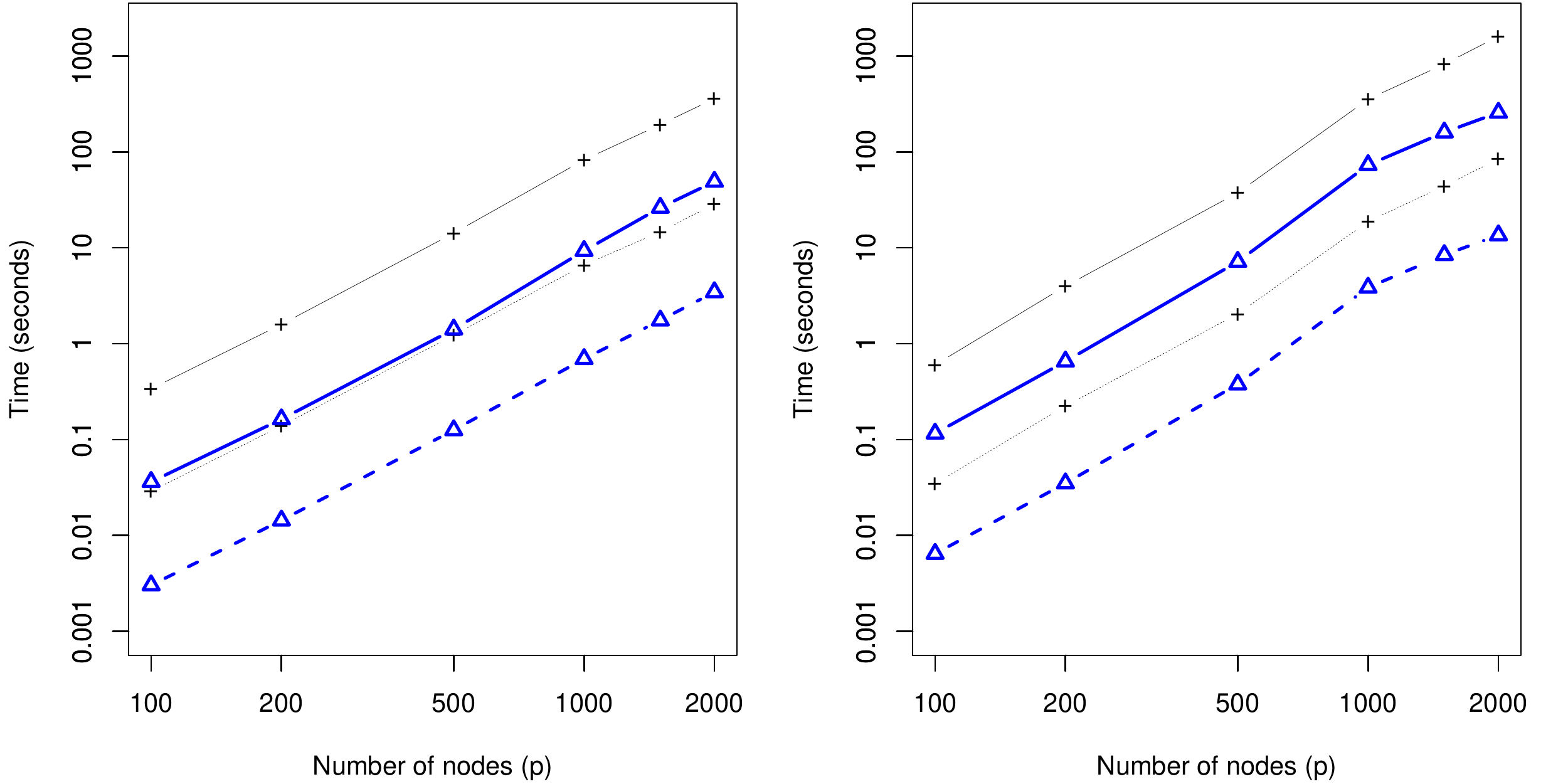}
\caption{Comparison of the updated implementation of CCDr against the implementation presented in the main text. The triangles representing the new implementation are overlaid on top of Figure~\ref{plot:timing} from the main text, based on duplicating the test in Section~\ref{subsec:timing} using the same graphs.  The solid line is the total runtime and the dashed line is the average runtime. (left) Time to estimate graphs with at most $p$ edges, (right) Full runtime with edge threshold $\alpha=3$.}
\label{splot:newtiming}
\end{figure}

\begin{figure}[h]
\centering
\includegraphics[width=\textwidth]{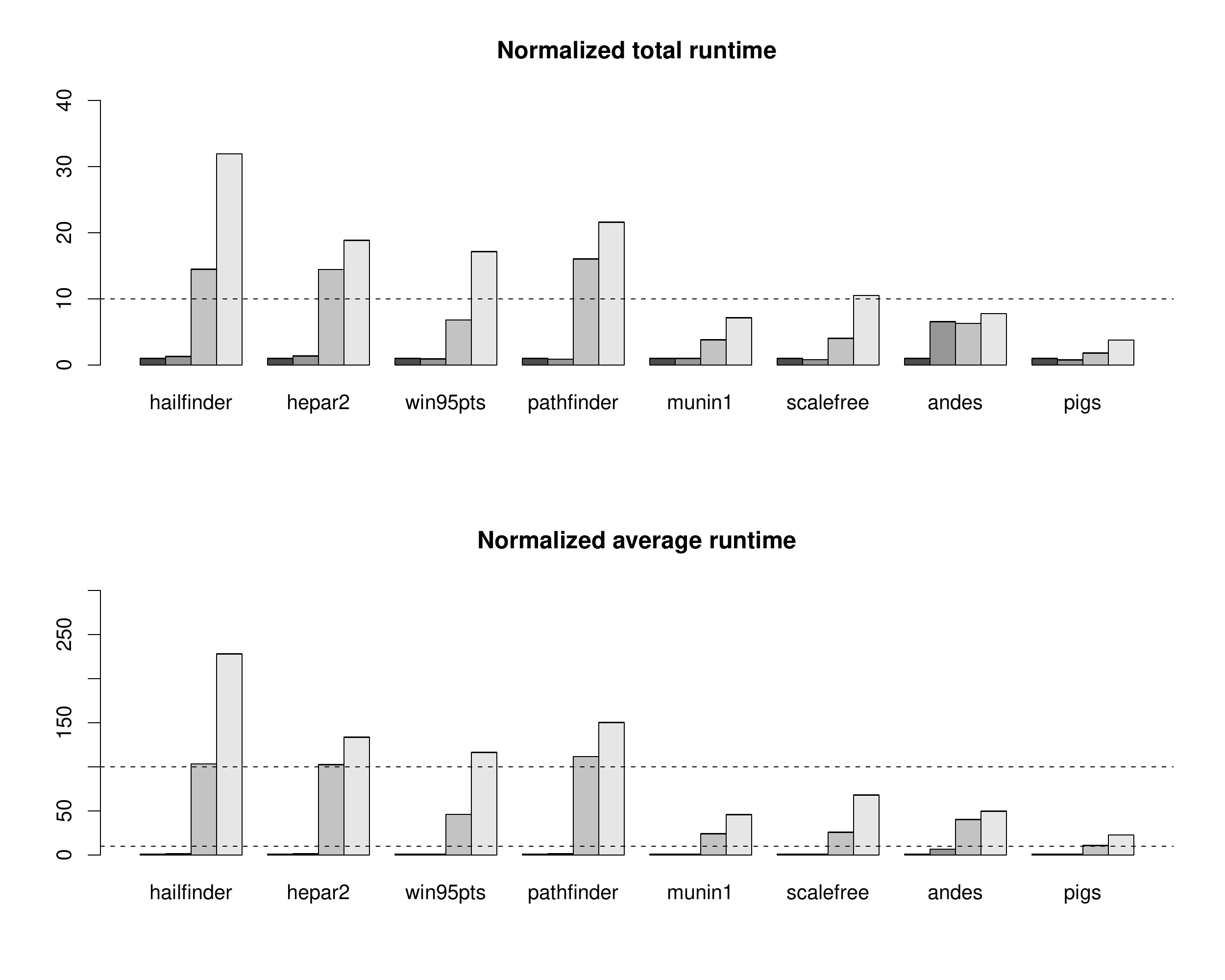}
\caption{Comparison of the total and average runtime for the four algorithms tested in Section~\ref{subsec:bnrep}, normalized by the respective runtimes for CCDr-MCP. From left to right (also, darkest to lightest), the bars represent CCDr-MCP, CCDr-$\ell_1$, PC, MMHC.}
\label{splot:bnreptiming}
\end{figure}

\clearpage
\subsection*{2 Supplementary model selection tables.}

\begin{table}[h]
\caption{Average estimation performance of algorithms in low-dimensions using BIC as model selection criteria.}
\vspace{0.5em}
\begin{center}
\begin{tabular}{lcccccc}
  \toprule
$p=50$, $T=46.48$ & CCDr-MCP & CCDr-$\ell_1$ & GES & HC & MMHC & PC \\ 
  \midrule
  P & 29.82 & 29.59 & 109.83 & 113.78 & 29.91 & 28.93 \\ 
  TP & 14.20 & 12.85 & 33.20 & 27.49 & 16.68 & 16.85 \\ 
  R & 9.65 & 10.20 & 8.19 & 12.29 & 9.76 & 9.11 \\ 
  FP & 5.96 & 6.54 & 68.44 & 74.00 & 3.47 & 2.97 \\ 
  SHD (DAG) & 38.24 & 40.17 & 81.72 & 92.99 & 33.27 & 32.60 \\ 
  SHD (skeleton) & 28.59 & 29.98 & 73.53 & 80.69 & 23.51 & 23.50 \\ 
  TPR & 0.31 & 0.28 & 0.71 & 0.59 & 0.36 & 0.36 \\ 
  FDR & 0.52 & 0.57 & 0.70 & 0.76 & 0.44 & 0.42 \\ 
%
  \toprule
$p=100$, $T=91.48$ & CCDr-MCP & CCDr-$\ell_1$ & GES & HC & MMHC & PC \\ 
  \midrule
  P & 68.48 & 69.51 & 241.71 & 256.20 & 68.25 & 65.57 \\ 
  TP & 34.85 & 32.16 & 74.30 & 60.24 & 40.94 & 40.96 \\ 
  R & 20.02 & 22.09 & 12.90 & 23.16 & 19.27 & 18.02 \\ 
  FP & 13.60 & 15.25 & 154.51 & 172.81 & 8.04 & 6.59 \\ 
  SHD (DAG) & 70.23 & 74.57 & 171.69 & 204.05 & 58.58 & 57.12 \\ 
  SHD (skeleton) & 50.20 & 52.48 & 158.79 & 180.88 & 39.30 & 39.10 \\ 
  TPR & 0.38 & 0.35 & 0.81 & 0.66 & 0.45 & 0.45 \\ 
  FDR & 0.49 & 0.54 & 0.69 & 0.76 & 0.40 & 0.38 \\ 
%
  \toprule
$p=200$, $T=185.06$ & CCDr-MCP & CCDr-$\ell_1$ & GES & HC & MMHC & PC \\ 
  \midrule
  P & 156.38 & 160.66 & 553.78 & 591.55 & 153.24 & 138.90 \\ 
  TP & 82.45 & 76.45 & 158.38 & 127.69 & 94.96 & 90.95 \\ 
  R & 40.74 & 45.44 & 22.35 & 45.65 & 38.00 & 35.51 \\ 
  FP & 33.18 & 38.77 & 373.06 & 418.21 & 20.28 & 12.45 \\ 
  SHD (DAG) & 135.79 & 147.38 & 399.74 & 475.58 & 110.38 & 102.20 \\ 
  SHD (skeleton) & 95.05 & 101.93 & 377.39 & 429.93 & 72.38 & 69.44 \\ 
  TPR & 0.45 & 0.41 & 0.86 & 0.69 & 0.51 & 0.49 \\ 
  FDR & 0.47 & 0.52 & 0.71 & 0.78 & 0.38 & 0.35 \\ 
   \bottomrule
\end{tabular}
\label{table:modelselect}
\end{center}
\end{table}


\clearpage
\subsection*{3 Supplementary comparisons for sparsity levels $s_0$.}

\begin{figure}[h]
\centering
\includegraphics[width=\textwidth]{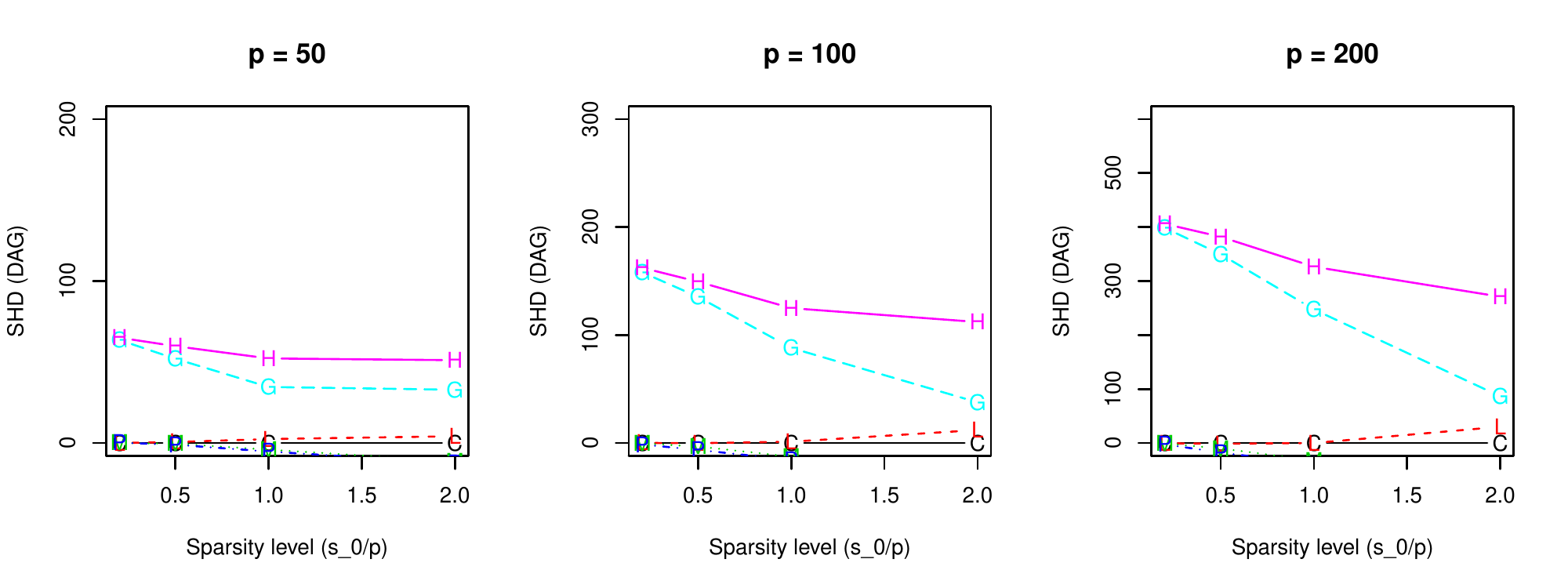}
\vspace{-1em}
\caption{Effect of sparsity on SHD in low dimensions for all six algorithms (C~=~CCDr-MCP, L~=~CCDr-$\ell_1$, G~=~GES, H~=~HC, M~=~MMHC, P~=~PC).}
\label{plot:sparselow}

\vspace{1em}
\includegraphics[width=\textwidth]{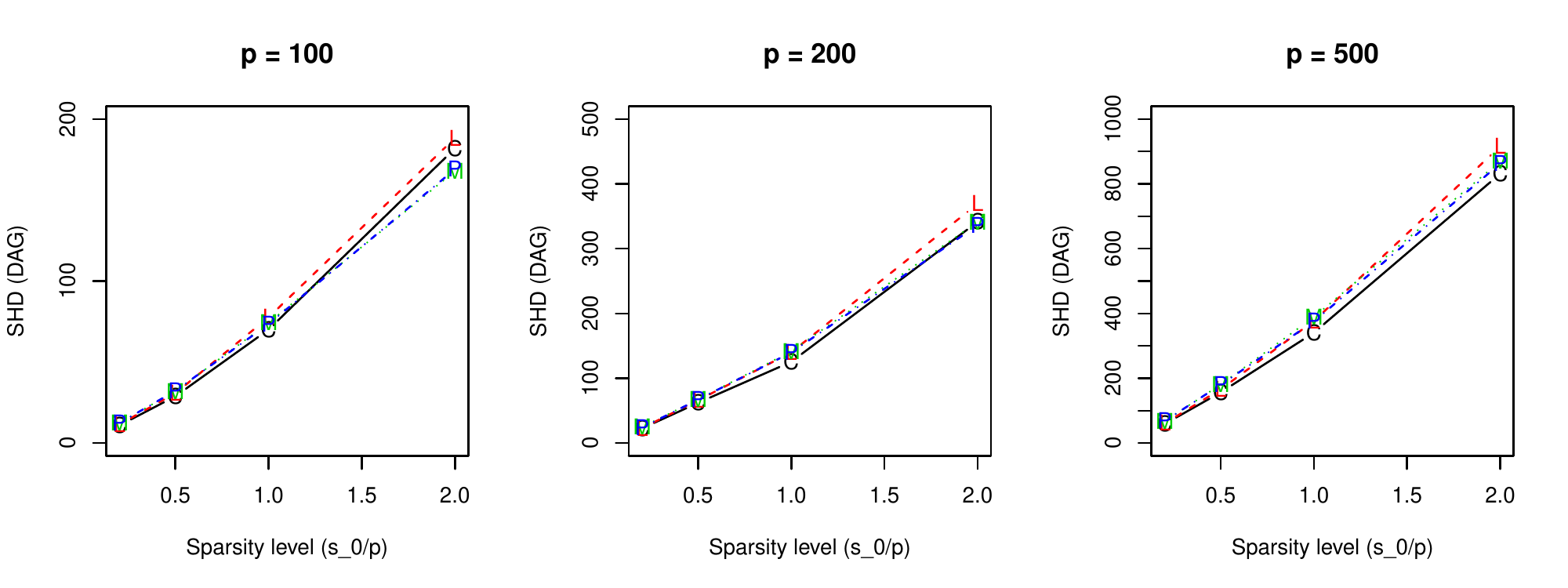}
\vspace{-1em}
\caption{Effect of sparsity on SHD in high dimensions, excluding GES and HC (C~=~CCDr-MCP, L~=~CCDr-$\ell_1$, M~=~MMHC, P~=~PC).}
\label{plot:sparsehigh}
\end{figure}

\end{document}